\documentclass{aastex61}
\shorttitle{MHOs toward HMOs}
\begin{document}

\title{MHOs toward HMOs: A Search for Molecular Hydrogen emission-line Objects toward High-Mass Outflows }

\correspondingauthor{Grace Wolf-Chase}
\email{gwolfchase@adlerplanetarium.org}

\author[0000-0002-9896-331X]{Grace Wolf-Chase} 
\affiliation{Astronomy Department \\
Adler Planetarium \\
1300 S. Lake Shore Drive \\
Chicago, IL 60605}
\affiliation{Dept. of Astronomy \& Astrophysics \\
University of Chicago \\
5640 S. Ellis Ave. \\
Chicago, IL 60637}

\author{Kim Arvidsson}
\affiliation{Trull School of Sciences and Mathematics \\ 
Schreiner University \\
2100 Memorial Blvd. \\ 
Kerrville, TX 78028}

\author{Michael Smutko}
\affiliation{Center for Interdisciplinary Exploration and Research in Astrophysics (CIERA) \\
\& Dept. of Physics \& Astronomy \\
Northwestern University \\
2131 Tech Drive \\ 
Evanston, IL 60208}

\begin{abstract}

We present the results of a narrow-band near-infrared imaging survey for Molecular Hydrogen emission-line Objects (MHOs) toward 26 regions containing high-mass protostellar candidates and massive molecular outflows.  We have detected a total of 236 MHOs, 156 of which are new detections, in 22 out of the 26 regions. We use H$_2$ 2.12-$\mu$m/H$_2$ 2.25-$\mu$m flux ratios, together with morphology, to separate the signatures of fluorescence associated with photo-dissociation regions (PDRs) from shocks associated with outflows in order to identify the MHOs. PDRs have typical low flux ratios of $\sim$ 1.5 - 3, while the vast majority of MHOs display flux ratios typical of C-type shocks ($\sim$ 6-20). A few MHOs exhibit flux ratios consistent with expected values for J-type shocks ($\sim$ 3-4), but these are located in regions that may be contaminated with fluorescent emission.  Some previously reported MHOs have low flux ratios, and are likely parts of PDRs rather than shocks indicative of outflows. We identify a total of 36 outflows across the 22 target regions where MHOs were detected. In over half these regions, MHO arrangements and fluorescent structures trace features present in CO outflow maps, suggesting the CO emission traces a combination of dynamical effects, which may include gas entrained in expanding PDRs as well as bipolar outflows. Where possible, we link MHO complexes to distinct outflows and identify candidate driving sources.  

\end{abstract}


\section{INTRODUCTION}

Outflows play an important role in the formation of stars across the entire stellar mass spectrum (e.g., Bally 2016); however, the nature of that role in the formation of  stars $>$ 8 M$_{\odot}$ remains poorly understood, since massive stars form in highly clustered environments that contain a mix of objects in different evolutionary stages. Zhang et al. (2005; hereafter, ZHB05) detected a total of 39 molecular outflows towards 69 luminous {\it IRAS} point sources associated with dense molecular gas and far-infrared luminosities indicative of massive star formation. From observations of the CO J=2$\rightarrow$1 transition with a resolution of $\sim$ 30$^{\prime\prime}$, they presented maps and derived physical parameters for 35 of these outflows, showing that their mass, momentum, and energy range from one to a few orders of magnitude larger than typical values for outflows associated with low-mass young stellar objects (YSOs). Although nearly half of the outflows mapped show spatially resolved bipolar lobes, the remainder show complicated lobe morphology or little evidence of bipolarity, and in several cases, the outflow centroid is clearly offset from the {\it IRAS} point source position.

The H$_2$ 1-0 S(1) line at 2.12 $\micron$ is a powerful tracer of shocks in molecular outflows (Davis et al. 2010 and references therein; Smith 2012). It can be used to identify collimated outflows and candidate driving sources even in massive star-forming regions (e.g., Varricatt et al. 2010; hereafter VDR10), with the caveat that H$_2$ emission produced by shocks (Molecular Hydrogen emission-line Objects or MHOs) is typically interspersed with fluorescent emission from photo-dissociation regions (PDRs) associated with expanding \ion{H}{2} regions. Although morphology of the H$_2$ emission is typically used to distinguish MHOs, H$_2$  2.12 $\micron$/2.25 $\micron$ flux ratios provide a more robust method for separating the effects of shocks and fluorescence (Wolf-Chase et al. 2013; hereafter, WAS13). To investigate the relationship between massive CO outflows, MHOs, and their driving sources, we acquired 4.$^{\prime}$6 $\times$ 4.$^{\prime}$6 narrow-band near-infrared images at a resolution of $\sim$ 1$^{\prime\prime}$ toward 25 of the CO outflows mapped by ZHB05 and one outflow mapped by Zhang et al. (2007).

This paper is organized as follows. Details of the observations and data reduction are presented in \S 2. Table 1 lists the observations for all targets.  Results and discussion of the survey are presented in \S3. Table 2 indicates whether targets contain detected MHOs, catalogued \ion{H}{2} regions, and catalogued massive young stellar objects (MYSOs). Table 3 lists the positions and fluxes of all MHOs for each target where MHOs were detected. Table 4 lists candidate outflows and driving sources. Our summary and conclusions are presented in \S4. Images, results and discussion of individual regions containing MHOs are presented in Appendix A, except for Mol 121 \& Mol 160, which were published previously (WAS13; Wolf-Chase et al. 2012, hereafter WSS12). 

\section{OBSERVATIONS AND DATA REDUCTION}

We obtained H$_2$ 2.12-$\micron$, H$_2$ 2.25-$\micron$ and H$_2$r 2.13-$\micron$ (narrow-band continuum) images of 26 regions thought to contain massive YSOs, using the Near-Infrared Camera and Fabry-Perot Spectrometer (NICFPS: Vincent et al. 2003; Hearty et al. 2004) on the Astrophysical Research Consortium (ARC) 3.5-m telescope at the Apache Point Observatory (APO) in Sunspot, NM. All targets contain energetic molecular outflows identified through CO observations and thought be associated with massive YSOs (ZHB05 Table 2; Zhang et al. 2007). Table 1 lists  the targets by {\it IRAS} designation (column 1) and Molinari number (Molinari et al. 1996, 1998, 2000, 2002), center position (columns 2 \& 3), filter (column 4), date (column 5), total exposure time (column 6), and resolution (column 7) as determined by seeing conditions at the time of the observations. We used the narrowband filter centered on 2.13 $\micron$ to allow for continuum subtraction in the final images. All images were acquired with a 4.$^{\prime}$6 $\times$ 4.$^{\prime}$6 FOV and pixel scale of 0.$^{\prime\prime}$273 pixel$^{-1}$. Our basic data acquisition, reduction, and calibration methods are described in WSS12, and the methods we used to perform irregular aperture photometry on extended emission are presented in WAS13. 

The target regions have varying brightness and morphologies, as well as varying background emission, which makes it impossible to apply one criterium for the spatial extent of all suspected MHOs. Rather, the morphologies of the regions used for aperture photometry were determined by signal-to-noise with respect to the background in the continuum subtracted images. To reduce the effect of the patchy nature of the extended background emission in continuum subtracted images, the images were smoothed using a 3-pixel Gaussian and the resulting background noise measured. The regions were then chosen to trace contours of multiples of the smoothed background noise. This could range from 10 times the background noise in the fainter regions to 60 times inside some of the brighter emission, but for most regions, 20 to 40 times the background noise contours were used. Whenever possible, the regions were chosen to trace emission that has approximately the same morphology in both 2.12 and 2.25-$\micron$ images. 

\startlongtable
\begin{deluxetable}{lllllcc}
\tablecolumns{7}
\tabletypesize{\footnotesize}
\tablenum{1}
\tablewidth{0pt}
\tablecaption{Observation Log}
\tablehead{ \colhead{Source} &  \colhead{$\alpha$(J2000)}  &  \colhead{$\delta$(J2000)}  & \colhead{Filter}   & \colhead{UT date} & \colhead{Exposure} & \colhead{FWHM(2.12 $\micron$)}\\
\colhead{ IRAS (Mol)}  &  \colhead{h m s}  &  \colhead{$^{\circ}$ $^{\prime}$ $^{\prime\prime}$}  &  \colhead{} &  \colhead{(yymmdd)} &  \colhead{Time (s)} &  \colhead{(arcsec)}}
\startdata
00117$+$6412 (2)  &  00 14 27.7  & +64 28 46 & H$_2$ 2.12 $\micron$ & 071218 & 1800 &  0.85  \\
&  &  & H$_2$ 2.25 $\micron$ & 071218 & 1800 &  \\
&  &  &  H$_2$r 2.13 $\micron$ & 071218 & 1800 &  \\
00420$+$5530 (3) & 00 44 57.6 &  +55 47 18 &  H$_2$ 2.12 $\micron$  & 071119 & 1800 &  0.87 \\
&  &  & H$_2$ 2.25 $\micron$ & 071119  & 1800 &  \\
&  &  &  H$_2$r 2.13 $\micron$ & 071119 & 1800 &  \\
05137$+$3919 (8) &  05 17 13.3 & +39 22 14 & H$_2$ 2.12 $\micron$  & 071203;070227;070204 & 4800 &  0.94 \\
&  &  & H$_2$ 2.25 $\micron$ & 071203 & 1800 &  \\
&  &  &  H$_2$r 2.13 $\micron$ & 071203;070227;070204 & 4800 &  \\
05168$+$3634 (9) & 05 20 16.2 &  +36 37 21 & H$_2$ 2.12 $\micron$  & 071218  & 1200 &  0.79  \\
&  &  & H$_2$ 2.25 $\micron$ & 071218 &  1200 &  \\
&  &  &  H$_2$r 2.13 $\micron$ & 071218 & 1200 &  \\
05274$+$3345 (10) &  05 30 48.0 &  +33 47 52 & H$_2$ 2.12 $\micron$  & 080115 & 1800 & 1.48  \\
&  &  & H$_2$ 2.25 $\micron$ & 080115  & 1800 &  \\
&  &  &  H$_2$r 2.13 $\micron$ & 080115 & 1800 &  \\
05345$+$3157 (11) & 05 37 47.8 & +31 59 24 & H$_2$ 2.12 $\micron$  & 051224 & 3900 &  1.17 \\
&  &  & H$_2$ 2.25 $\micron$ &   061013 & 1800 &  \\
&  &  &  H$_2$r 2.13 $\micron$ &  051224 & 3900  &  \\
05373$+$2349 (12) & 05 40 24.4 & +23 50 54 & H$_2$ 2.12 $\micron$  & 080319 & 1680 & 1.03  \\
&  &  & H$_2$ 2.25 $\micron$ & 080319  & 1680 &  \\
&  &  &  H$_2$r 2.13 $\micron$ & 080319  & 1680 &  \\
06056$+$2131 (15) & 06 08 41.0  & +21 31 01 & H$_2$ 2.12 $\micron$  & 061010;061013 & 3000 &  0.84  \\
&  &  & H$_2$ 2.25 $\micron$ & 061013 & 1200 &  \\
&  &  &  H$_2$r 2.13 $\micron$ &  061013 & 3000 &  \\
06584$-$0852 (28)  & 07 00 51.0 & -08 56 29 & H$_2$ 2.12 $\micron$  & 070227;070327;080319 & 5400 &  1.21 \\
&  &  & H$_2$ 2.25 $\micron$ & 070327;080319  & 3600 &  \\
&  &  &  H$_2$r 2.13 $\micron$ & 070227;070327;080319 & 5400 &  \\
18532$+$0047 (78)  & 18 55 50.6 & +00 51 22 & H$_2$ 2.12 $\micron$  & 060612;060613 & 4500 & 1.38  \\
&  &  & H$_2$ 2.25 $\micron$ & 080611;080915 &  1800 &  \\
&  &  &  H$_2$r 2.13 $\micron$ & 060613;080611;080915 & 4500 &  \\
19213$+$1723 (103)  & 19 23 36.9 &  +17 28 59 & H$_2$ 2.12 $\micron$  & 060513;060514 &  5400  &  0.98  \\
&  &  & H$_2$ 2.25 $\micron$ & \nodata & \nodata &  \\
&  &  &  H$_2$r 2.13 $\micron$ & 060513 & 2700 &  \\
19368$+$2239 (108)  &  19 38 58.1 &   +22 46 32 & H$_2$ 2.12 $\micron$  & 060618  & 3240 &  0.87  \\
&  &  & H$_2$ 2.25 $\micron$ & 080530 & 1680 &  \\
&  &  &  H$_2$r 2.13 $\micron$ & 060616;080530 & 3240 &  \\
19374$+$2352 (109)  & 19 39 33.2 &  +23 59 55 & H$_2$ 2.12 $\micron$  & 060606 & 2700 & 0.89  \\
&  &  & H$_2$ 2.25 $\micron$ & 080519 & 1800 &  \\
&  &  &  H$_2$r 2.13 $\micron$ & 060606;080519 & 2700 &  \\
19388$+$2357 (110) &  19 40 59.4 & +24 04 39 & H$_2$ 2.12 $\micron$  & 080611 & 2400 &  1.12  \\
&  &  & H$_2$ 2.25 $\micron$ & 080611 & 2400 &  \\
&  &  &  H$_2$r 2.13 $\micron$ & 080611 & 2400 &  \\
20050$+$2720 (114) & 20 07 06.7 &  +27 28 53 & H$_2$ 2.12 $\micron$  & 080729;080915 & 2100 &  1.00  \\
&  &  & H$_2$ 2.25 $\micron$ & 080729;080915  & 2100 &  \\
&  &  &  H$_2$r 2.13 $\micron$ & 080729;080915  & 2100  &  \\
20056$+$3350 (115) &  20 07 31.5 &  +33 59 39 & H$_2$ 2.12 $\micron$  & 080915;081109 & 2400 &  0.71 \\
&  &  & H$_2$ 2.25 $\micron$ & 080915;081109 & 2400 &  \\
&  &  &  H$_2$r 2.13 $\micron$ & 080915;081109 & 2400 &  \\
20188$+$3928 (121) &  20 20 39.3 & +39 37 52 & H$_2$ 2.12 $\micron$  & 081013 & 1740 &  1.21 \\
&  &  & H$_2$ 2.25 $\micron$ & 081013 &  1740 &  \\
&  &  &  H$_2$r 2.13 $\micron$ & 081013 & 1740  &  \\
20220$+$3728 (123)  & 20 23 55.7 & +37 38 10 & H$_2$ 2.12 $\micron$  & 110620 & 1950 &  1.34  \\
&  &  & H$_2$ 2.25 $\micron$ & 110620 & 1950  &  \\
&  &  &  H$_2$r 2.13 $\micron$ & 110620 & 1950  &  \\
20278$+$3521 (125)  &  20 29 46.9 & +35 31 39 & H$_2$ 2.12 $\micron$  & 081111;081112 & 1800 & 0.99  \\
&  &  & H$_2$ 2.25 $\micron$ & 081111;081112 &  1800 &  \\
&  &  &  H$_2$r 2.13 $\micron$ & 081111;081112  & 1800 &  \\
20286$+$4105 (126)  & 20 30 27.9 & +41 15 48 & H$_2$ 2.12 $\micron$  & 110613 & 1800 & 0.80  \\
&  &  & H$_2$ 2.25 $\micron$ & 110613 & 1800 &  \\
&  &  &  H$_2$r 2.13 $\micron$ & 110613 & 1800 &  \\
21307$+$5049 (136) & 21 32 31.5 & +51 02 22 & H$_2$ 2.12 $\micron$  & 070105 & 1800 &  1.09 \\
&  &  & H$_2$ 2.25 $\micron$ &\nodata  & \nodata &  \\
&  &  &  H$_2$r 2.13 $\micron$ & 070105 & 1800 &  \\
22172$+$5549 (143) &  22 19 09.0 &  +56 04 45 & H$_2$ 2.12 $\micron$  & 050703;051011;051016;061202 & 4740 &  0.85 \\
&  &  & H$_2$ 2.25 $\micron$ & 051016 & 3600 &  \\
&  &  &  H$_2$r 2.13 $\micron$ & 051011;051016;061202 & 4740 &  \\
22305$+$5803 (148)  &  22 32 24.3 &  +58 18 58 & H$_2$ 2.12 $\micron$  & 081109 & 1470 & 0.90  \\
&  &  & H$_2$ 2.25 $\micron$ & 081109 & 1470 &  \\
&  &  &  H$_2$r 2.13 $\micron$ & 081109 & 1470 &  \\
22506$+$5944 (151)  & 22 52 38.6 & +60 00 56 & H$_2$ 2.12 $\micron$  & 081111;081112 & 1800 & 1.00  \\
&  &  & H$_2$ 2.25 $\micron$ & 081111;081112 & 1800 &  \\
&  &  &  H$_2$r 2.13 $\micron$ & 081111;081112 & 1800 &  \\
23314$+$6033 (158) &  23 33 44.3 &  +60 50 30  & H$_2$ 2.12 $\micron$  & 071119 & 1800 &  0.93 \\
&  &  & H$_2$ 2.25 $\micron$ & 071119  & 1800 &  \\
&  &  &  H$_2$r 2.13 $\micron$ & 071119  & 1800 &  \\
23385$+$6053 (160) & 23 40 53.2 & +61 10 21  & H$_2$ 2.12 $\micron$  & 050617 & 1800 & 0.75  \\
&  &  & H$_2$ 2.25 $\micron$ & 071121 & 1200 &  \\
&  &  &  H$_2$r 2.13 $\micron$ & 071121 & 1200 &  \\
\enddata
\end{deluxetable}

\section
{SURVEY RESULTS AND DISCUSSION}

Table 2 identifies target regions (columns 1 \& 2) as {\it High} (H) or {\it Low} (L), depending upon whether the associated  {\it IRAS} source has infrared colors typical of an Ultra-Compact (UC) \ion{H}{2} region or redder, respectively (Wood \& Churchwell 1989; Palla et al. 1991). The designation (UC) indicates candidate UC \ion{H}{2} regions based on the detection of centimeter-radio continuum emission at levels $>$ 1 mJy within 40$^{\prime\prime}$ of the {\it IRAS} position (G\'omez-Ruiz et al. 2016). Columns 3 \& 4 indicate whether a targeted region contains MHOs and, if so, whether collimated outflow(s) could be identified. Columns 5 \& 6 list objects in the {\it WISE} catalog of Galactic \ion{H}{2} regions within the targeted areas and their classifications by type (Anderson et al. 2014). Columns 7 \& 8 list objects in the Red {\it Midcourse Space Experiment} ({\it MSX}: Egan et al. 2003) Source catalog (RMS: Lumsden et al. 2013) located within each targeted region and their type. In \S 3.1 we present our MHO results and in \S 3.2 we discuss their relationship to CO outflows. In \S 3.3 we discuss possible associations of MHOs with other signposts of massive star formation within surveyed regions, and in \S 3.4 we discuss identification of candidate outflow driving sources.

\begin{longrotatetable}
\begin{deluxetable}{llcclclr}
\tablecolumns{8}
\tabletypesize{\footnotesize}
\tablenum{2}
\tablewidth{0pt}
\tablecaption{Source Detections \& Associations with Massive Star Formation}
\tablehead{ \colhead{Source} &    \colhead{Type$^a$}    & \colhead{MHOs?}  & \colhead{Collimated?}  & \colhead{{\it WISE} Name$^b$} &  \colhead{\ion{H}{2} Type$^b$} &  \colhead{{\it MSX} Name}  &  {RMS Type} \\
\colhead{{\it IRAS} (Mol)}  &  \colhead{}  &  \colhead{}  &  \colhead{} &  \colhead{} &  \colhead{}  &  \colhead{}  &  \colhead{}}
\startdata
00117$+$6412 (2)   & H(UC)   & Y & Y  &   G118.964+01.893 & C & \nodata &  \nodata  \\
00420$+$5530 (3)*  & L  & Y &  Y  &   G122.002-07.084  & C & \nodata  &   \nodata   \\
05137$+$3919 (8)*  & L & Y &  Y   &  G168.066+00.819  & Q  & G168.0627+00.8221 &  YSO  \\
05168$+$3634 (9)*  &  H   &  Y & Y   & \nodata &  \nodata  & \nodata &  \nodata  \\
05274$+$3345 (10)*  &  H  &  Y &  Y  & G174.197-00.076  &  Q  & G174.1974-00.0763 &  YSO  \\
05345$+$3157 (11)*  &  L  &  Y & Y   &  G176.507+00.178  & C  & \nodata & \nodata  \\
05373$+$2349 (12)*   &  L   &  Y & Y   & G183.720-03.665 & Q  & G183.7203-03.6647 &   YSO  \\
06056$+$2131 (15)  & H(UC)  &  Y &  Y  & G189.030+00.780 & Q  & G189.0307+00.7821 &  YSO  \\
  &   & &  & G189.032+00.809  &  Q  &  G189.0323+00.8092 &  YSO  \\
06584$-$0852 (28)*    & L  &  Y &  N  & G221.955-01.993  & Q & G221.9605-01.9926  & YSO   \\
18532$+$0047 (78)   &  H(UC)  & N  &  \dots  & G034.195-00.603 & Q  & \nodata  & \nodata  \\
  &   &  & & G034.196-00.592 & K &  &    \\
    &   &  &   & G034.204-00.584 & Q &  &  \\
19213$+$1723 (103)*  &  H(UC)  &  N  &  \dots  & G052.098+01.042 & K  & G052.0986+01.0417 &  Diffuse \ion{H}{2} Region  \\
19368$+$2239 (108)   & H  &  Y  &  Y  & G058.453+00.431 & Q & G058.4670+00.4360A &  YSO  \\
  &   &  &  & G058.467+00.436 & Q  & G058.4670+00.4360B & \ion{H}{2} Region  \\
  &    &  &  & G058.477+00.427 & Q  &   &    \\
19374$+$2352 (109)*   & H(UC)  &  Y & N  & G059.602+00.911 & K &  G059.6032+00.9116A &  \ion{H}{2} Region  \\
  &    &  &  & G059.612+00.917 & G &  G059.6032+00.9116B & \ion{H}{2} Region \\
  &   &   &   &   &   & G059.6032+00.9116C & \ion{H}{2} Region \\
    &   &   &   &   &   & G059.6032+00.9116D & \ion{H}{2} Region  \\
19388$+$2357 (110)*  & H(UC)  &  Y & N   & G059.829+00.671 & Q  & G059.8329+00.6729 &  YSO  \\
20050$+$2720 (114)*  & H  &  Y & Y  & \nodata & \nodata & G065.7798-02.6121 & YSO  \\
20056$+$3350 (115)*  & H  &   N & \nodata   & G071.312+00.828 & C  &  \nodata & \nodata  \\
20188$+$3928 (121)$^c$*   &  H(UC)  &  Y & Y  &  G077.463+01.760 &  C &  G077.4622+01.7600A  & YSO   \\
 &  &  &  &   &   & G077.4622+01.7600B  & \ion{H}{2} Region  \\
20220$+$3728 (123)   &  H(UC)  &   N & \nodata & G076.180+00.064 &  Q &  G076.1877+00.0974 & \ion{H}{2} Region   \\
  &   &  &  & G076.187+00.097 &  C  &  &    \\
  &    &  &  & G076.197+00.092 & C &  &   \\
20278$+$3521 (125)    &  L   &  Y &  Y   &  G075.155-02.087 & Q & \nodata & \nodata  \\
20286$+$4105 (126)*   & H   &   Y & N  & G079.869+01.180 & C &  G079.8749+01.1821 &  \ion{H}{2} Region   \\
21307$+$5049 (136)*   & L(UC)  &  Y & N  &  G094.263-00.414 &  C &  G094.2615-00.4116  & YSO  \\
22172$+$5549 (143)*  & L  &  Y  & Y  & \nodata  & \nodata  & G102.8051-00.7184A &  YSO  \\
  &  &  &  &  &  & G102.8051-00.7184B &  YSO  \\
   &  &  &  &  &  & G102.8051-00.7184C & YSO   \\
22305$+$5803 (148)*   & H  &  Y & Y   & G105.509+00.230 & Q  &  G105.5072+00.2294  &  YSO   \\
22506$+$5944 (151)    &  H &  Y & Y  & G108.603+00.494 & Q   & G108.5955+00.4935A  &  YSO   \\
  &  &  &  &  &  & G108.5955+00.4935B & YSO   \\
   &  &  &  &  &  & G108.5955+00.4935C & YSO   \\
23314$+$6033 (158)  &  L  &  Y & N   & G113.614-00.615  &  G  & G113.6041-00.6161  &   \ion{H}{2} Region  \\
23385$+$6053 (160)$^d$   & L   &  Y &  N  &  G114.526-00.543 &  C & \nodata &  \nodata \\
\enddata
\tablenotetext{a}{Refers to objects that have {\it IRAS} colors similar to ultracompact (UC) \ion{H}{2} regions (H) and redder colors (L) (Wood \& Churchwell 1989; Palla et al. 1991). The designation (UC) indicates a candidate UC \ion{H}{2} region based on the detection of centimeter-radio continuum emission at levels $>$ 1 mJy within 40$^{\prime\prime}$ of the {\it IRAS} position (G\'omez-Ruiz et al. 2016; Molinari et al. 1998; Palau et al. 2010; Kurtz et al. 1994).}
\tablenotetext{b}{Entries are from the {\it WISE} catalog of Galactic \ion{H}{2} regions (Anderson et al. 2014). Classifications indicate known (K) , grouped (G), and candidate (C) \ion{H}{2} regions, and radio quiet (Q) objects.}
\tablenotetext{c}{G077.4622+01.7600A is coincident with core C and anomalous H$_2$ emission reported by Wolf-Chase et al. (2013) and a deeply embedded source reported by Yao et al. (2000).}
\tablenotetext{d}{Bipolar, but very compact, MHOs about Mol 160 A (Wolf-Chase et al. 2012).}
\tablecomments{Targets marked with an asterisk were included in a near-infrared imaging survey conducted by Varricatt et al. (2010). }
\end{deluxetable}
\end{longrotatetable}

\subsection{H$_2$ Emission}

To choose our final list of MHOs, we excluded emission where the continuum-subtracted H$_2$ 2.12-$\micron$/H$_2$ 2.25-$\micron$ flux ratio is $<$ 3.  Such emission is typically associated with fluorescence from photo-dissociation regions (PDRs) and is generally diffuse and extended. Some previously reported MHOs do not meet our flux ratio criterion, but are included in Table 3 for the sake of completeness. We used morphology as an additional criterion to exclude PDRs, but in no case did we find regions exhibiting PDR morphology (diffuse, extended arcs) where the H$_2$ 2.12-$\micron$/H$_2$ 2.25-$\micron$ flux ratio is $>$ 4. Since the vast majority of MHOs have flux ratios significantly greater, we are confident that this method has successfully identified the MHOs in our targets. 

For each target region containing MHOs, Table 3 lists its {\it IRAS} source designation and Molinari number  (column 1), identified MHOs (column 2),  positions of the peak emission (columns 3 \& 4), areas used in the aperture photometry (column 5), H$_2$ 2.12-$\micron$ line fluxes (column 6), H$_2$ 2.25-$\micron$ line fluxes (column 7), H$_2$ 2.12-$\micron$/H$_2$ 2.25-$\micron$ flux ratios (column 8), and comments relevant to certain catalog entries (column 9).  Where the H$_2$ 2.25-$\micron$ line flux is below 3 times the noise (3$\sigma$) within an aperture, the H$_2$ 2.25-$\micron$ line is regarded as undetected and no value is given in column 7. In these cases, lower limits to the H$_2$ 2.12-$\micron$/H$_2$ 2.25-$\micron$ flux ratios are calculated using the 3$\sigma$ value for that aperture as an upper limit for the H$_2$ 2.25-$\micron$ line flux.

MHOs identified in Table 3 follow the numbering scheme used by the on-line catalog of these objects, which is currently hosted by the University of Kent\footnote{\url{http://astro.kent.ac.uk/~df/MHCat/}} and was initially published by Davis et al. (2010). In general, MHOs that appear to be associated with linear features are assigned the same number and given additional letter designations to identify individual knots. This is also the convention we use for MHOs that appear in groups. MHO numbers that are followed by ``\_\#'' indicate knots we have identified which we associate with previously identified MHO groups or outflows. This does not necessarily imply that all MHOs with the same given number are part of the same outflow. We discuss these associations on a case-by-case basis in the Appendix. 

For the high excitation temperatures associated with J-shocks, one expects H$_2$  2.12-$\micron$/2.25-$\micron$ $\sim$ 3 $-$ 4, and for the lower-excitation C-shocks, $\sim$ 6 $-$ 20. Above 20, the excitation temperature is $<$ 1000 K, and observable emission is not expected (Smith 1994, 1995; Smith et al. 2003). For the most part, our computed flux ratios are consistent with C-type shocks. A few MHOs with flux ratios that are more consistent with J-type shocks lie in the direction of PDRs and may include fluorescent emission. It is interesting to note that three of our surveyed regions, Mol 109, Mol 121 (WAS13), and Mol 126, exhibit one or more spots where the H$_2$ 2.12-$\micron$/2.25-$\micron$ flux ratio $<$ 1, which cannot be explained by either fluorescence or shocks. 

{\bf
\begin{deluxetable}{rrrrrrrrr}
\tablecolumns{9}
\tabletypesize{\footnotesize}
\tablenum{3}
\tablewidth{0pt}
\tablecaption{MHOs and Fluxes.\label{fluxtable}}
\tablehead{ \colhead{Name} &
\colhead{MHO} &  \colhead{$\alpha$(J2000)}  &  \colhead{$\delta$(J2000)}  & \colhead{Area}   & \colhead{$F_{2.12}$} & \colhead{$F_{2.25}$} & \colhead{$F_{2.12}$/$F_{2.25}$} & \colhead{Comments} \\
\colhead{ IRAS (Mol)}  &  \colhead{} &  \colhead{h m s}  &  \colhead{$^{\circ}$ $^{\prime}$ $^{\prime\prime}$}  &  \colhead{10$^{-10}$ sr} &  \colhead{$10^{-18}~\mathrm{W~m^{-2}}$} &  \colhead{$10^{-18}~\mathrm{W~m^{-2}}$} &  \colhead{}  &  \colhead{}}
\startdata
00117$+$6412 (2)  & 2967A  &  00 14 33.6 & +64 28 39 & 3.341 & $6.83 \pm 0.52$ & $0.45 \pm 0.11$ & $15.1 \pm 3.7$  & \\
& 2967B  & 00 14 32.4 & +64 28 37 & 4.969 & $9.09 \pm 0.69$ & $0.57 \pm 0.14$ & $16.1 \pm 4.0$  & \\
& 2967C  &  00 14 30.9 & +64 28 35  & 1.165 & $1.42 \pm 0.12$ & $0.20 \pm 0.06$ & $7.1 \pm 2.0$ & \\
& 2967D  &  00 14 30.4 &  +64 28 39 &  1.594 & $2.15 \pm 0.18$ & $0.22 \pm 0.07$ & $9.9 \pm 3.4$ & \\
& 2967E  & 00 14 28.8 & +64 28 34 &  1.405 & $3.73 \pm 0.29$ &  \nodata  & $ >19.5 $  & \\
&  2967F  &  00 14 28.8 &  +64 28 27 &  4.609 & $6.84 \pm 0.53$ & $1.03 \pm 0.15$ & $6.6 \pm 1.1$ & \\
& 2967G  & 00 14 27.8 & +64 28 26 &  1.045 & $1.66 \pm 0.13$ &  \nodata  & $ >11.1 $ & \\
& 2967H  & 00 14 23.6 & +64 28 21 &  0.685 & $0.78 \pm 0.07$ &  \nodata  & $ >6.6 $  & \\
& 2968A  & 00 14 26.4  & +64 29 00 & 3.495 & $9.05 \pm 0.69$ & $0.99 \pm 0.14$ & $9.1 \pm 1.5$ & \\
&  2968B & 00 14 26.5 & +64 29 05 &  1.422 & $2.38 \pm 0.19$ & $0.36 \pm 0.07$ & $6.7 \pm 1.4$ & \\
\enddata
\tablecomments{Refs. - Chen et al. (1999, 2003, 2005); Khanzadyan et al. (2011); Varricatt et al. (2010); Wolf-Chase et al. (2012); Wolf-Chase et al. (2013)
(This table is available in its entirety in machine-readable form.)} 
\end{deluxetable}
}

\subsection{Relationship of MHOs to Massive CO Outflows}

The targets for our survey were all chosen from sources associated with massive CO outflows listed in Table 2 of ZHB05, with the exception of Mol 10, which was mapped in CO by Zhang et al. (2007). We were able to identify MHOs in 22 of the 26 regions we observed (Table 3). Across these 22 regions,  MHO detections fall into 3 morphological categories: (1) co-linear MHOs that define the position angle of one dominant outflow in the region; (2) multiple outflows along different position angles; and (3) isolated or scattered MHOs where no clear outflows or position angles can be identified. Table 4 lists candidate outflows (column 1) identified from MHO arrangements (column 2), possible driving sources (column 3), outflow position angles (column 4), and distinguishing MHO features (column 5). 

In spite of the low resolution of the CO observations, there are several regions where chains of MHO detections follow the large-scale morphology of CO outflows presented in Figure 1 of ZHB05, including Mol 8, 9, 11, 12, 15, 109, 114 \& 143.  In some cases, both CO and H$_2$ emission appear to trace PDRs or other complex structures (e.g., Mol 2, 28, 126, \& 136). This is not surprising, as CO emission has been associated with shells produced by expanding \ion{H}{2} region bubbles identified from mid-infrared images (Dewangan et al. 2016; Churchwell et al. 2006). In a few cases, there is no clear relationship between the CO and H$_2$ emission (e.g., Mol 110, 125, 148, 151). There are several possible explanations for this: (1) While CO emission traces swept-up ambient material entrained in outflows, MHOs trace the current location of shocks, where the underlying wind is interacting with the ambient medium. It is possible that outflow axes change in time due to the precession of jets. (2) MHOs and CO emission may trace outflows from different sources. (3) Some differences may be explained by the size of the outflow and resolution of the observations. For example, Mol 151 displays a very compact ($<$ 1$^{\prime}$ in length) dramatic bipolar H$_2$ jet along an east-west direction, while the high-velocity CO emission shows little evidence of bipolarity except for a small east-west separation of $\sim$ 30$^{\prime\prime}$ between the peaks of the blue- and redshifted lobes, comparable to the resolution of the CO observations (ZHB05). (4) In some cases, extinction effects and proximity to the edges of molecular clouds may explain observed differences. 

\subsection{Association with \ion{H}{2} Regions and YSOs}

Table 2 lists all sources in the {\it WISE} Catalog of Galactic \ion{H}{2} Regions (Anderson et al. 2014) that are found within our targeted regions (Column 5). Column 6 indicates their classifications as known (K), grouped (G), candidate (C) or radio quiet (Q). K sources have measured Radio Recombination Lines (RRL) or H$\alpha$ emission that confirms their \ion{H}{2} region status; C sources are spatially coincident with radio continuum emission, but do not have RRL or H$\alpha$ observations; G sources represent K \& C sources that are associated positionally, typically within the radius of a PDR; and Q sources lack detected radio continuum emission and are presumably associated with pre-UC \ion{H}{2} region objects or lower-luminosity intermediate-mass star-forming regions. Anderson et al. (2014) estimate that $\sim$95\% of C sources are bona fide \ion{H}{2} regions. 

Ten of our target regions contain only Q sources (Table 2 - Mol 8, 10, 12, 15, 28, 108, 110, 125, 148, \& 151), although two of these (Mol 15 \& Mol 110) are identified as candidate UC \ion{H}{2} regions by G\'omez-Ruiz et al. (2016). All of these regions contain MHOs, and all but Mol 28 \& 110 contain collimated outflows. Three of our targets with collimated MHOs have no entries in the {\it WISE} Catalog of Galactic \ion{H}{2} Regions (Mol 9, 114, \& 143). Of the four regions that harbor K or G sources (Mol 78, 103, 109, 158), two lack MHO detections entirely (Mol 78 \& 103) and the other two (Mol 109 \& 158) lack collimated flows. Although the three targets containing K sources all have {\it IRAS} colors associated with `High' (H) objects, there is no obvious relationship between \ion{H}{2} region class and {\it IRAS} color. The regions containing only Q sources represent a mix of H \& L objects. 

Table 2 also lists RMS sources found within our targeted regions (Column 7) and their type (Column 8). The RMS catalog is the largest statistically selected catalog of massive protostars and \ion{H}{2} regions to date (Lumsden et al. 2013). It is thought to be complete for the detection of a B0 V star at the distance of the Galactic center, although inclusion in the catalog depends upon source detection by {\it MSX} and specific color criteria that may have excluded very young, compact objects containing a high fraction of ionized polycyclic aromatic hydrocarbons (PAHs) in thick PDRs (Kerton et al. 2015). 

We additionally used infrared color criteria established by  Koenig \& Leisawitz (2014) and Fischer et al. (2016) to identify candidate Class 0 and Class I YSOs from the {\it ALLWISE} database, which combines data from the {\it WISE} cryogenic and {\it NEOWISE} post-cryogenic surveys (Wright et al. 2010; Mainzer et al. 2011). Class 0 and Class I YSOs are associated, respectively, with protostars in the main and late accretion phases of pre-main sequence evolution (Lada 1987; Andr\'e et al. 1993). 

Candidate Class I objects were required to satisfy all of the following criteria:

\begin{eqnarray}
W2-W3>2.0 \\
W2-W3<4.5 \\
W1-W2>0.46\times(W2-W3)-0.9 \\ 
W1-W2>-0.42\times(W2-W3)+2.2
\end{eqnarray}

Class 0 candidates were required to satisfy both of these criteria:

\begin{eqnarray}
W2-W3<1.8\times(W3-W4)-6.5 \\
W1-W2>1
\end{eqnarray}

W1, W2, W3, \& W4 refer to the four {\it WISE} bands at 3.4 $\micron$, 4.6 $\micron$, 12 $\micron$, \& 22 $\micron$, respectively.

\subsection{Candidate Driving Sources}

Assuming collimated outflows are generated by circumstellar disks, one would expect them to shut off once massive YSOs begin to ionize their surroundings. Since entries in the  {\it WISE} Catalog of Galactic \ion{H}{2} Regions were first identified by their characteristic mid-infrared ``bubble'' morphology, most of them are very large and clearly not good candidates for driving the collimated outflows identified in these regions. Four Q sources, which may be associated with younger or less luminous objects, do lie near the centers of outflows associated with Mol 8, 12, 15, \& 148, but all are clustered with other candidate driving sources. 

RMS YSOs can be identified as candidate driving sources for collimated MHO outflows in Mol 8, 12, 15, 143, 148, \& 151. RMS YSOs may also drive MHO flows in Mol 10, 28, 110, 114, \& 136, but the relationship is less clear cut, either because MHOs are either isolated detections or form linear arrangements offset from the YSO. (See Appendix A for further discussion.) We can also identify {\it ALLWISE} sources that fit Class 0/I YSO color criteria as candidate driving sources for outflows in 50\% of the regions containing MHOs (11/22), with the caveat that candidates may contain multiple objects, since the resolution of {\it WISE} at 22 $\micron$ is 12$^{\prime\prime}$. Additionally, we note that some {\it ALLWISE} sources that satisfy Class 0/I color criteria are coincident with bright MHOs, which suggests that in some instances, these sources may be tracing shocked gas rather than YSOs. In any case, infrared colors alone cannot distinguish definitively between different classes of objects and further observations will be required to confirm the natures of these sources. 

Table 4 presents 36 candidate outflows and driving sources, including cross-references to driving sources proposed in other publications.  We identify candidate driving sources that fit {\it WISE} Class 0/I YSO color criteria (Koenig \& Leisawitz 2014; Fischer et al. 2016) and/or are listed as YSOs in the RMS catalog for 26 of these outflows. We note that only 3 of the 36 ($\sim$ 8\%) identified outflows (Mol 15 Flow 1, this paper and Mol 121 Flows 1 \& 3 , WAS13) are associated with driving sources that are candidate UC \ion{H}{2} regions; the remaining driving sources appear to be younger objects. This suggests that the jet phase shuts off shortly after the YSO has begun to ionize it surroundings; however, future high-resolution observations of the candidate UC \ion{H}{2} regions, particularly radio recombination line data, would be particularly useful in establishing the evolutionary status of these objects. Seven of the candidate driving sources of outflows listed in Table 4 are both RMS YSOs and YSOs we identified from {\it ALLWISE} sources. Only in two cases (Mol 10 Flow 4, Mol 151 Flow 1) is the proposed driving source a RMS YSO lacking a counterpart identified from {\it ALLWISE}. 
Although evidence suggests that most of the collimated outflows identified in this study are driven by massive YSOs, much work remains to be done to establish source and outflow properties. NIR spectroscopy is needed to determine extinctions towards MHOs, which is essential for estimating MHO luminosities, and higher resolution observations are necessary to distinguish individual source contributions to bolometric luminosities. 

\begin{longrotatetable}
\begin{deluxetable}{lclrr}
\tablecolumns{6}
\tabletypesize{\footnotesize}
\tablenum{4}
\tablewidth{0pt}
\tablecaption{Candidate Outflows \& Driving Sources}
\tablehead{ \colhead{Designation} &  \colhead{MHOs}  &  \colhead{Candidate Source(s)}  & \colhead{P.A. MHOs}   &  \colhead{Notes} \\
\colhead{ Mol \# Flow \#}  &  \colhead{}  &  \colhead{}  &  \colhead{degrees} &  \colhead{}}
\startdata
Mol 2 Flow 1  &  2967A-H  & MM2$^a$,   &  75  & bright knots, complex morphology  \\
 &   &  J001427.00+642819.3,  &  &  \\
 &  &   J001425.05+642822.4,  &   &   \\
 &    & J001423.45+642824.9, &    &   \\
 &    &  J001422.48+642836.8 &   &   \\
Mol 2 Flow 2   & 2968A-B   & MM1$^a$? &  10 & bright knots, complex morphology  \\
Mol 3 Flow 1 & 2903  &  J004457.30+554718.1 (D)* &   58 &  bow-shaped knot, faint bridge of emission     \\
Mol 3 Flow 2 & 2971-2974, 2976  & J004458.06+554656.7 (B)* &  -80  &  curved chain of knots     \\
Mol 3 Flow 3 & 2969, 2970, 2900, 2975 &  J004457.30+554718.1 (D)*?  & 25 & colinear knots \\
Mol 8 Flow 1 &  1002,1003-1003\_2 &  J051713,74+392219.6 (A,B)* & 15-20   &  bright bow-shaped knots, precessing jet?    \\
 &  & (G168.0627+00.8221) &   &   RMS counterpart \\
Mol 8 Flow 2 &  1004\_2, 1004\_3 & J051713,74+392219.6 (A,B)* & -20  &  bright connected chain    \\
Mol 9 Flow 1 & 1055G & J052022.03+363756.5  & 60  &  bright bow-shaped feature   \\
Mol 9 Flow2?  & 1055A,C,E  &  J052022.03+363756.5  & 28 & faint knots  \\
Mol 10 Flow 1 &  1005  &  MM-1/MM-2 (B)* & 50 & bright jet \\
Mol 10 Flow 2 & 1011, 1009, 1059?  & MM-1/MM-2  & 5 & MHOs along ``middle jet''$^b$ \\
Mol 10 Flow 3 & 1057, 1007, 1060?  & \dots  & -60 & MHOs along ``long jet''$^c$  \\
Mol 10 Flow 4 & 1006  & G174.1974-00.0763 (A)* & -90 & bright bow-shaped feature  \\
Mol 11 Flow 1 & 1015, 1015\_2, 1015\_3 & J053752.03+320003.9  &  64 &   bright bow-shaped features  \\
Mol 11 Flow 2 & 1061A-F & J053752.03+320003.9? & -72 &  linear chain of emission knots  \\
Mol 11 Flow 3$^c$ & 1018-1018\_3, 1016-1016\_3 & (A)* &  -49 &  bright, complex morphology  \\
Mol 11 Flow 4 & 1062A & J053752.99+315934.8 & -50 &   very bright jet feature  \\
Mol 12 Flow 1 & 738 group, \& 742A? & J054024.23+235054.6 (A)* & 31 & faint SW-NE chain  \\
  &  &  (G183.7203-03.6647) &  & RMS counterpart \\
Mol 12 Flow 2 & 744A, 734 \& 740 groups, 751 & J054024.79+235048.0 & -10 &  S-shaped chain of knots   \\
Mol 12 Flow 3 & 739, 744B & J054024.23+235054.6 (A)* & 63 &    \\
  &  &  (G183.7203-03.6647) &  & RMS counterpart \\
Mol 12 Flow 4 & 742B, 742C, 737? & J054024.95+235220.6, & 95 & bright knots \\
 &    & J054026.46+235222.0 &  &  \\
Mol 15 Flow 1 &  1216A-F &  J060840.45+213102.0  &  -35 &  chain of knots   \\
  &  &  (G189.0307+00.7821)  &  & RMS counterpart \\
Mol 108 Flow 1 & 2623A-D  &  J193859.37+224656.0 & 65  &  chain of knots    \\
Mol 114 Flow 1 & 2608 \& 2609 complexes & (C,F)*  & $\sim$ -87,-72 & curved chain of knots, bright bow  \\
Mol 121 Flow 1$^d$ &  864-865, 863, 891-892, 937 & Associated with Core D & 47 &  bipolar MHOs   \\
Mol 121 Flow 2$^d$ & 867,938-939  & Associated with Core A & 90 & 867 bright bow feature    \\
Mol 125 Flow 1  & 949A-D  & \dots & 68  &  faint chain of knots  \\
Mol 126 Flow 1  &  961 & J203029+411558.6 (C)* & -57  &  ``goldfish'' bow, bright knot  \\
Mol 143 Flow 1 & 2765-2765\_2, 2766-2766\_2 & J221908.42+560501.2  (A)* &  10 &     \\
  &  &  (G102.8051-007184A) &  &  RMS counterpart \\
Mol 143 Flow 2 & 2772A-B &  J221908.42+560501.2 (A)* &  40 &     \\
Mol 143 Flow 3 & 2773, 2776A-B &  J221908.42+560501.2 (A)*  &  -52 &  linear arrangement of knots   \\
&  &  (G102.8051-007184B)  &  & RMS counterpart \\
Mol 148 Flow 1 & 2777D-G  & J223223.87+581859.7 (A)*  & 22  &    \\
  &  &  (G105.5072+00.2294)  &  & RMS counterpart  \\
Mol 148 Flow 2 & 2777A-C  &  J223219.41+581750.7 & 73 &  linear chain of 3 faint knots  \\
Mol 151 Flow 1 & 2778A-K & G108.5955+00.4935B & 85 & bright bow features A \& B  \\
Mol 151 Flow 2 &  2779A-D & J225242.60+600041.5 & 10 &  curved chain of knots  \\
Mol 160 Flow 1$^e$ &  2921-2922 &  A$^f$ &  & compact `bipolar' knots   \\
\enddata
\tablenotetext{a}{Palau et al. (2010)}
\tablenotetext{b}{These outflows correspond roughly to jets identified by Chen et al. (2005) and Zhang et al. (2007). See Mol 10 section for details.}
\tablenotetext{c}{First identified by VDR10.}
\tablenotetext{d}{Wolf-Chase et al. 2013}
\tablenotetext{e}{Wolf-Chase et al. 2012}
\tablenotetext{f}{24 $\micron$ point source (Molinari et al. 2008)}
\tablecomments{Sources marked with an asterisk were included in a near-infrared imaging survey conducted by Varricatt et al. (2010). }
\end{deluxetable}
\end{longrotatetable}

\section
{SUMMARY AND CONCLUSIONS}

\begin{enumerate}

\item Using NICFPS on the ARC 3.5-m telescope at the APO, we have detected a total of 236 MHOs, 156 of which are new detections, in 22 out of 26 regions associated with massive CO outflows. 

\item We find an excellent agreement between predicted H$_2$ 2.12-$\micron$/2.25-$\micron$ flux ratios for shocks (H$_2$ 2.12-$\micron$/2.25-$\micron$ $>$ 3) and fluorescent emission (1 $<$ H$_2$ 2.12-$\micron$/2.25-$\micron$  $<$ 3) and morphology: MHOs are typically ``knots'', bow-shaped or other compact features, while fluorescent emission has the morphology of bubbles (e.g., as seen in regions Mol 2, 10, 11, 109, 125, 126, 151), filaments or arcs (e.g., Mol 3, 15, 136, 158), or pillars (e.g., Mol 143). Based on low H$_2$ 2.12-$\micron$/2.25-$\micron$ flux ratios, we note that some previously reported MHOs may have been misidentified as such. 

\item For MHOs with detected H$_2$ 2.25-$\micron$ emission, H$_2$ 2.12-$\micron$/2.25-$\micron$ flux ratios are typically between $\sim$ 6 and 20, consistent with C-type shocks. Where H$_2$ 2.12-$\micron$/2.25-$\micron$ $< \sim$ 5, the MHO is either faint or lies in the direction of a PDR. MHOs 1018 and 1018$\_3$ in Mol 11 are good examples of the latter. A few MHOs have H$_2$ 2.12-$\micron$/2.25-$\micron$ $>$ 20 (e.g., the bright bow features MHO 1015$\_2$ in Mol 11 and MHOs 2778 A\&B in Mol 151), but in no case is H$_2$ 2.12-$\micron$/2.25-$\micron$ $>$ 30. Three of our survey targets, Mol 109, Mol 121 (WAS13), and Mol 126, exhibit one or more spots where the H$_2$ 2.12-$\micron$/2.25-$\micron$ flux ratio $<$ 1, which cannot be explained by either fluorescence or shocks. 

\item MHO arrangements fall into three morphological categories: (1) co-linear MHOs that define the P.A. of one dominant outflow in the region; (2) multiple outflows along distinct P.A.s; and (3) isolated or clustered MHOs where no clear P.A. can be identified. 

\item In over half the regions with MHO detections, MHO arrangements and fluorescent H$_2$ structures trace features present in low-resolution high-velocity CO maps presented in ZHB05, suggesting the CO emission traces a combination of complex dynamical effects that may include expanding \ion{H}{2} regions as well as protostellar outflows. 

\item All but three (Mol 9, 114, \& 143) of our 26 target regions contain entries in the {\it WISE} Catalog of Galactic \ion{H}{2} Regions (Anderson et al. 2014); however, the majority of these are radio quiet (Q) sources thought to be either massive objects in a pre-UC \ion{H}{2} region phase, or intermediate-mass star-forming regions. In only four cases does a source lie near the center of a collimated MHO outflow, and in each of these cases, the source is clustered with other possible driving sources.

\item We identify candidate driving sources that fit {\it WISE} Class 0/I YSO criteria (Koenig \& Leisawitz 2014; Fischer et al. 2016) and/or are listed as YSOs in the RMS catalog for 26 out of the 36 outflows we distinguish across our target regions.

\item Only 3 of the 36 outflows ($\sim$ 8\%) identified in Table 4 (Mol 15 Flow 1, this paper and Mol 121 Flows 1 \& 3 , WAS13) are associated with driving sources that are candidate UC \ion{H}{2} regions; the remaining driving sources appear to be younger objects. This suggests that the jet phase shuts off shortly after the YSO has begun to ionize it surroundings. Future high-resolution observations of the candidate UC \ion{H}{2} regions, particularly radio recombination line data, would be particularly useful in establishing the evolutionary status of these objects.

 \end{enumerate}

\acknowledgments

This research is based on observations obtained with the Apache Point Observatory 3.5-meter telescope, which is owned and operated by the Astrophysical Research Consortium. We particularly thank the 3.5-meter Observing Specialists and Al Harper for assistance in acquiring the data, as well as Michael Medford, who performed the original NICFPS data reduction. This research has made use of SAOImage DS9, developed by the Smithsonian Astrophysical Observatory; data products from the Wide-field Infrared Survey Explorer, which is a joint project of the University of California, Los Angeles, and the Jet Propulsion Laboratory/California Institute of Technology, funded by the National Aeronautics and Space Administration; and data products from the Midcourse Space Experiment. Processing of Midcourse Space Experiment data was funded by the Ballistic Missile Defense Organization with additional support from NASA Office of Space Science. This research has also made use of the NASA/ IPAC Infrared Science Archive, which is operated by the Jet Propulsion Laboratory, California Institute of Technology, under contract with the National Aeronautics and Space Administration. GW-C was funded in part through NASA's Illinois Space Grant Consortium, and the authors gratefully acknowledge support from the Brinson Foundation grant in aid of astrophysics research at the Adler Planetarium. KA thanks Jamie Riggs for help in developing the near-infrared continuum subtraction procedure, {\bf and GW-C thanks Geza Gyuk for assistance converting Table 3 to machine-readable format.}

\facilities{ARC, WISE, MSX}

\software{IRAF, SAOImage DS9, Aladin}

\appendix
\section{INDIVIDUAL REGIONS: RESULTS AND DISCUSSION}

Figures 1 - 20 present images for all targets with MHO detections, with the exception of Mol 160 and Mol 121, which were included in previous papers (WSS12; WAS13). For each region, we present (a) a 3-color rgb image that combines continuum-subtracted H$_2$ 2.12-$\micron$ (blue), continuum-subtracted H$_2$ 2.25-$\micron$ (red), and an average of the continuum-subtracted H$_2$ 2.12-$\micron$ and H$_2$ 2.25-$\micron$ images (green); and (b) a continuum-subtracted H$_2$ 2.12-$\micron$ greyscale image with MHO numbers and regions (magenta) indicating the apertures used to compute fluxes listed in Table 3. The rgb images include positions of {\it WISE} Class 0/I candidates (labelled yellow circles indicating the resolution of {\it WISE} at 22 $\micron$), which were identified using the criteria presented in \S 3.3; RMS sources (orange +: YSO, diamond: \ion{H}{2} region); and entries in {\it WISE} Catalog of Galactic \ion{H}{2} Regions (purple boxes labelled according to type: C, G, K, or Q).  {\it IRAS} source error ellipses (green) are included on the greyscale images. Maps of the CO outflows for all of the regions discussed below, except Mol 10, were presented in Figure 1 of ZHB05. CO maps of Mol 10 are presented in Zhang et al. (2007). We refer the reader to VDR10 for further discussion of regions marked with an asterisk below and in Tables 2 \& 4.

\subsection{IRAS 00117+6412 (Mol 2)}

The brightest H$_2$ emission towards this target is due to fluorescence from a PDR  centered on the {\it IRAS} source position and C-type \ion{H}{2} region, G118.964+01.893 (Figures 1a \& b). The CO outflow of ZHB05 shows N-S bipolarity. The blueshifted emission peaks at the position of the {\it IRAS} source and the redshifted emission peaks $\sim$ 30$^{\prime\prime}$ to the south. There is a steep drop-off in emission to the north and east, consistent with the location of G118.964+01.893 at the eastern border of the molecular cloud. 

Palau et al. (2010; hereafter, PSB10) identified three intermediate-mass YSOs in this region. One can be associated with G118.964+01.893 (which they identify as a UC \ion{H}{2} region). Two others are associated with millimeter sources, MM1 \& MM2 (MM1 has multiple components). MM1 is located just west of the {\it IRAS} error ellipse ($\alpha$(2000) = 00$^h$14$^m$26.05$^s$, $\delta$(2000) = 64$^{\circ}$28$^{\prime}$43.7$^{\prime\prime}$) and MM2 is located $\sim$ 16$^{\prime\prime}$ to the south ($\alpha$(2000) = 00$^h$14$^m$26.31$^s$, $\delta$(2000) = 64$^{\circ}$28$^{\prime}$27.8$^{\prime\prime}$). PSB10 estimate the star powering the UC \ion{H}{2} region to be $\sim$ 1 Myr and $\sim$ 6 M$_{\odot}$. MM1 is reported to be a Class 0/I source embedded in a $\sim$ 3 M$_{\odot}$ dust core with estimated luminosity of 400-600 L$_{\odot}$.  PSB10 mapped a CO outflow along a NE-SW direction, centered on MM1, using CO J=2$\rightarrow$1 observations acquired with the Submillimeter Array (SMA: Ho et al. 2004). The blueshifted emission lies to the NE.  MM2 is a deeply embedded object associated with H$_2$O maser emission, but not centimeter or near-infrared emission. PSB10 noted that MM2 appears to be a $\sim$ 1.6 M$_{\odot}$ Class 0 object, but curiously lacked CO outflow emission. 

It seems likely that MM1 is the driving source of MHOs 2968A \& B to the NE of this source, particularly since this is the direction of the CO outflow emission. We suggest that MM2 may drive the H$_2$ outflow defined by the string of MHO 2967 knots along a P.A. $\sim$ 75$^{\circ}$ to the south of the \ion{H}{2} region; however, we identify additional Class I candidates (J001427.00+642819.3, J001425.05+642822.4, J001423.45+642824.9) and one Class 0 candidate (J001422.48+642836.8) that can't be ruled out as driving sources, given the dispersion and morphology of the MHO knots. It is not clear whether these MHOs trace a single outflow and position angle.  It is possible that PSB10 did not detect an outflow associated with MM2 because of the E-W orientation of the outflow traced by the MHOs and a lack of molecular material eastward of the UC \ion{H}{2} region.

\begin{figure}[htb!]
\figurenum{1} 
\includegraphics[angle=0,scale=0.65]{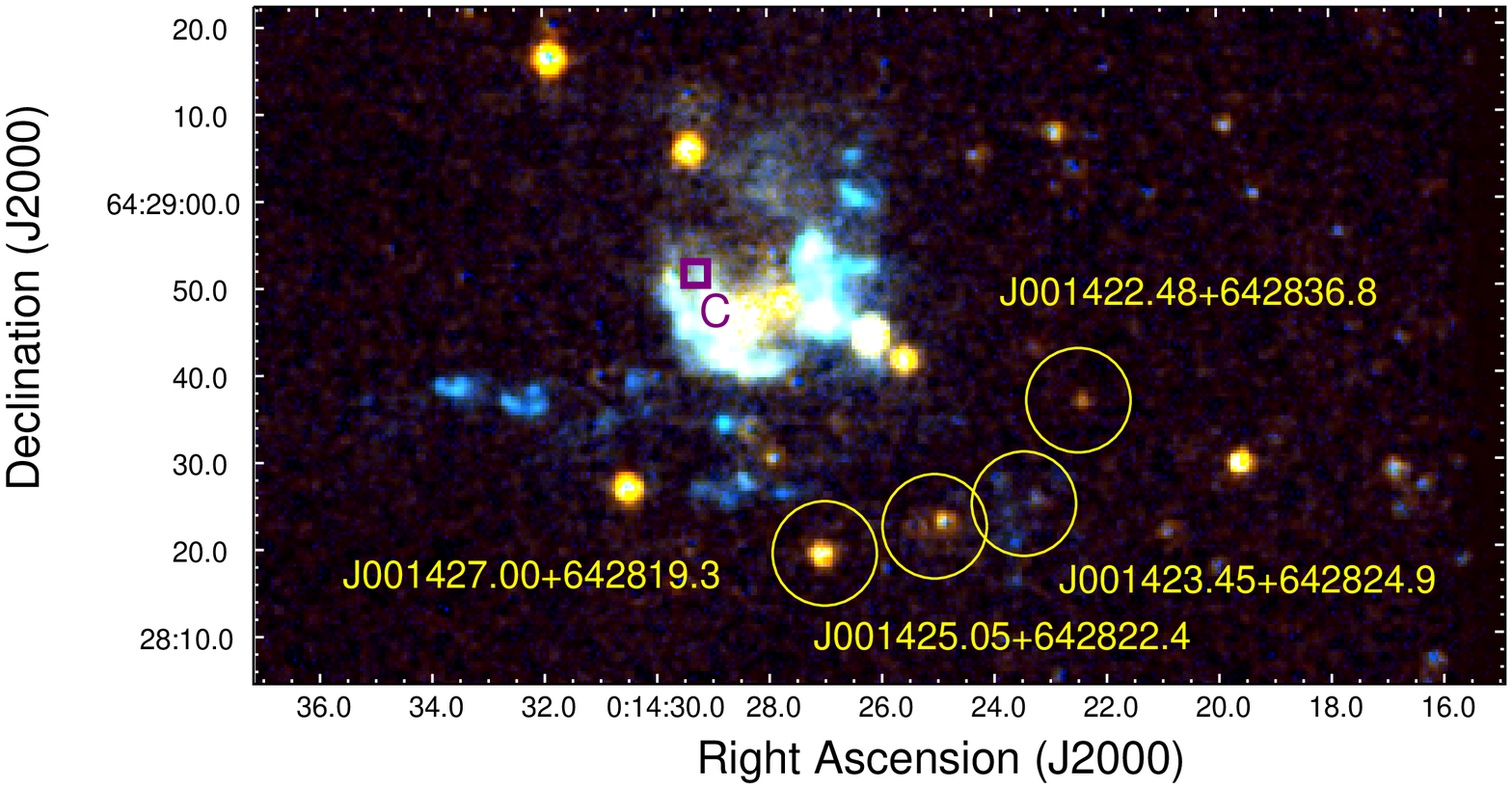}
\includegraphics[angle=0,scale=0.65]{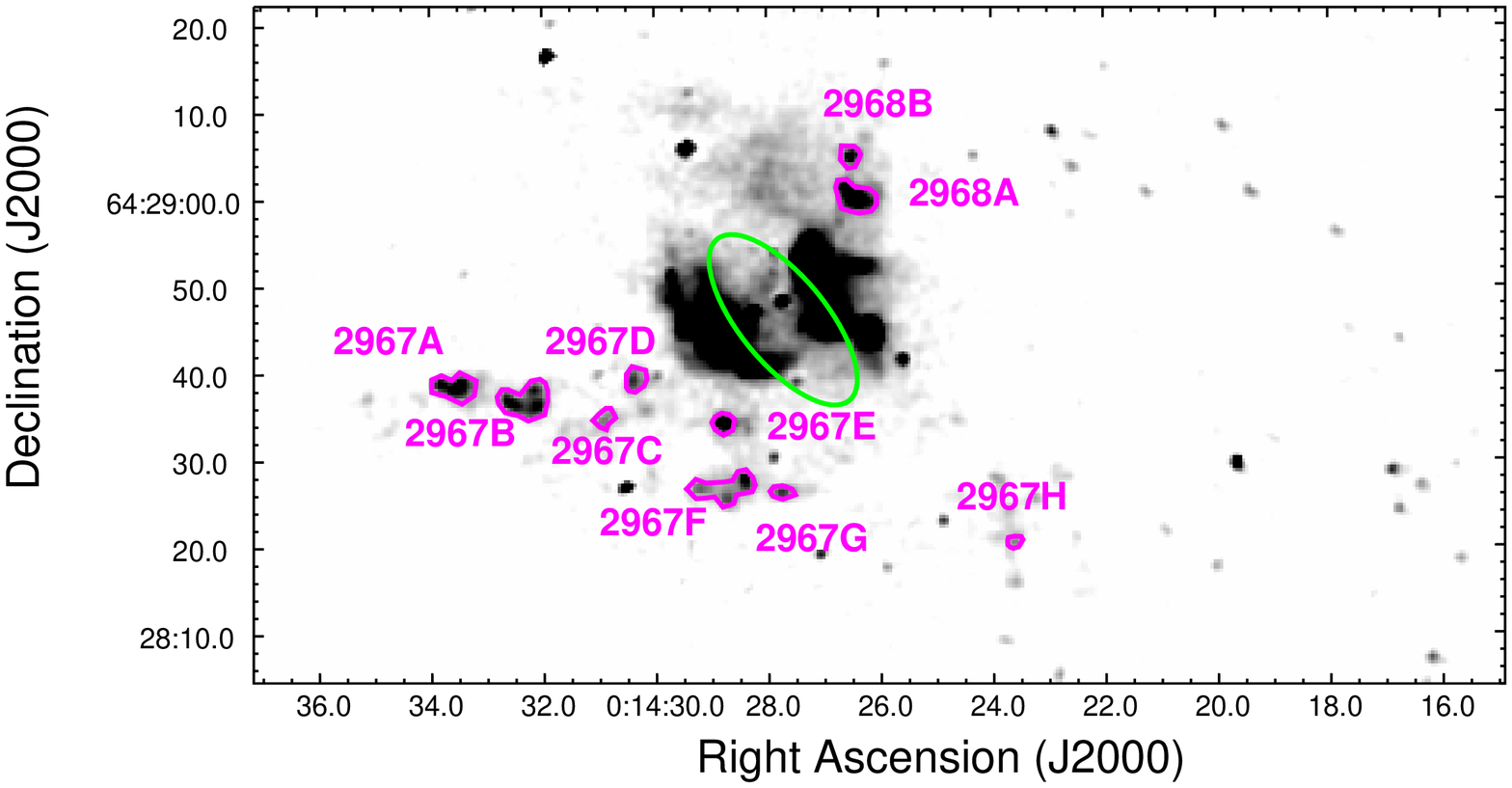}
\caption{Mol 2 (a) rgb image and (b) greyscale image, with labels as described in \S A.}
\end{figure}

\subsection{IRAS 00420+5530 (Mol 3)$^*$}

We identify three Class I candidates (J004458.06+554656.7, J004457.30+554718.1, J004451.03+554639.9) and one Class 0 candidate (J004500.80+554627.0) in this region (Figures 2a \& b). Four of the MHOs (2900-2903) were previously discovered by VDR10. One of these (MHO 2902) lies along a PDR that is probably associated with the {\it IRAS} source, which is coincident with J004457.30+554718. MHO 2902 has a very low H$_2$ 2.12-$\micron$/2.25-$\micron$ flux ratio of 2.9, so, at the very least, the PDR is likely to contribute to the flux caught by the aperture. A candidate \ion{H}{2} region (G122.002-07.084) lies near a ridge of fluorescent emission in the southwest quadrant of the field. The CO high-velocity gas mapped by ZHB05 has a complex morphology with multiple peaks. 

MHOs appear to lie along at least three distinct position angles in this region. The bow-shaped knot MHO 2903 appears to connect to J004457.30+554718.1 via a faint bridge of H$_2$ emission along a P.A. of $\sim$ 58$^{\circ}$. J004457.30+554718.1 coincides with VDR10 source ``D'', H$_2$O masers, and a VLA source (Palla et al. 1991; Molinari et al. 2002, hereafter MTR02). The faint chain of knots (MHOs 2971-2974, 2976) along a P.A. of $\sim$ -80$^{\circ}$ appears to align with Class I candidate  J004458.06+554656.7 (VDR10 source ``B''), which is also associated with MM 2 (MTR02) and a mid-infrared source at $\alpha(2000) = 00^h44^m57.24^s$, $\delta(2000) = 55^{\circ}46^{\prime}50.2^{\prime\prime}$ that was identified using the MIRLIN camera (Ressler et al. 1994) on NASA's 3-m Infrared Telescope Facility (IRTF) on Mauna Kea (J. O'Linger-Luscusk, private communication). H$_2$O masers are located between MHOs 2971 \& 2973 (Harju et al. 1998). Although the bright knot (MHO 2901) appears to be associated with J004458.06+554656.7, it is not colinear with the fainter knots and its relationship to a specific outflow is unclear. Contours associated with overlapping red- and blue-shifted CO emission peaks extend in a general east-west direction roughly 15-30$^{\prime\prime}$ south of the {\it IRAS} source (ZHB05) consistent with the orientation of the MHO chain. MHOs 2969, 2970, 2900, and 2975 are approximately colinear with J004457.30+554718.1 along a P.A. of $\sim$ 25$^{\circ}$.  

\begin{figure}[htb!]
\figurenum{2} 
\includegraphics[angle=0,scale=0.44]{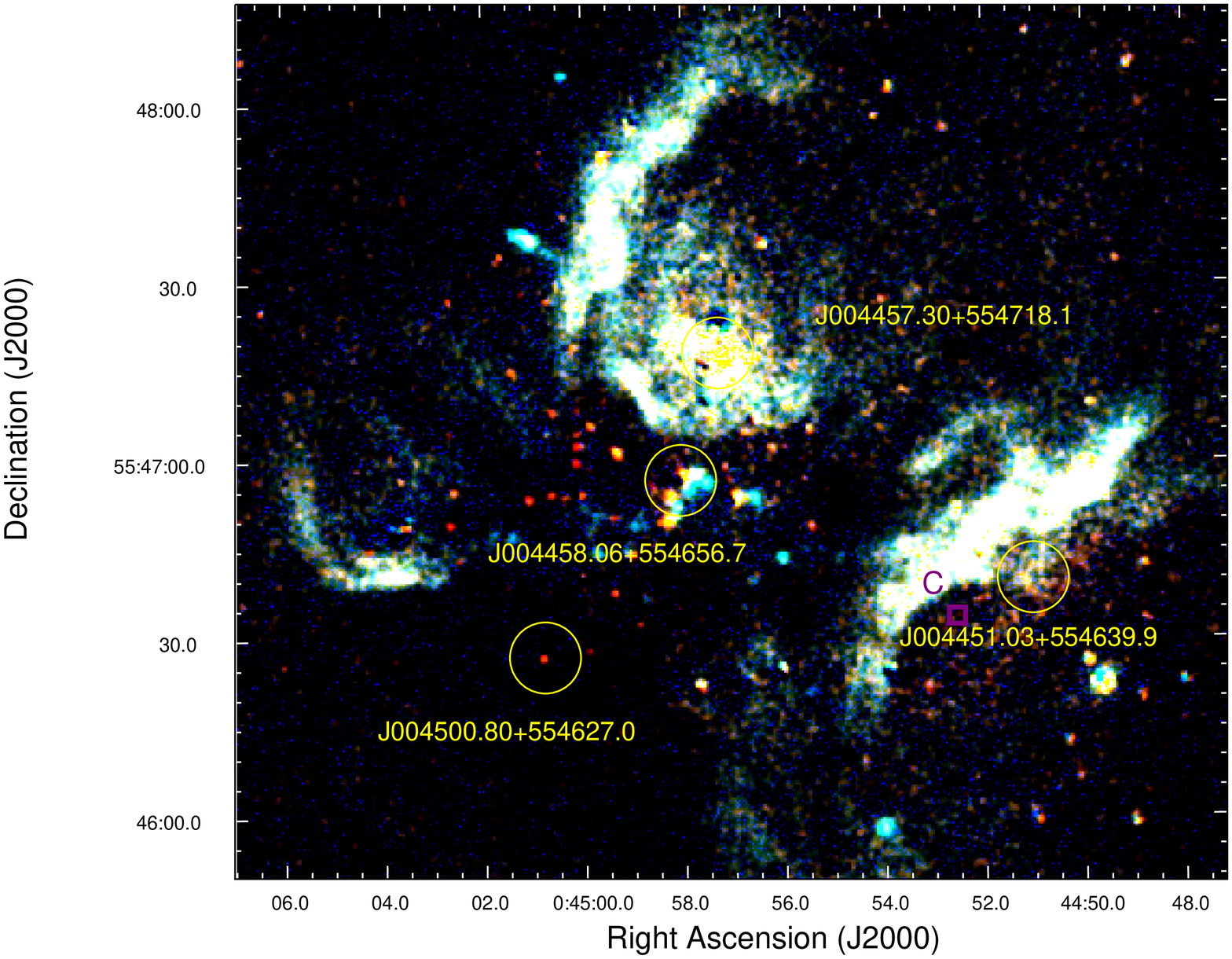}
\includegraphics[angle=0,scale=0.44]{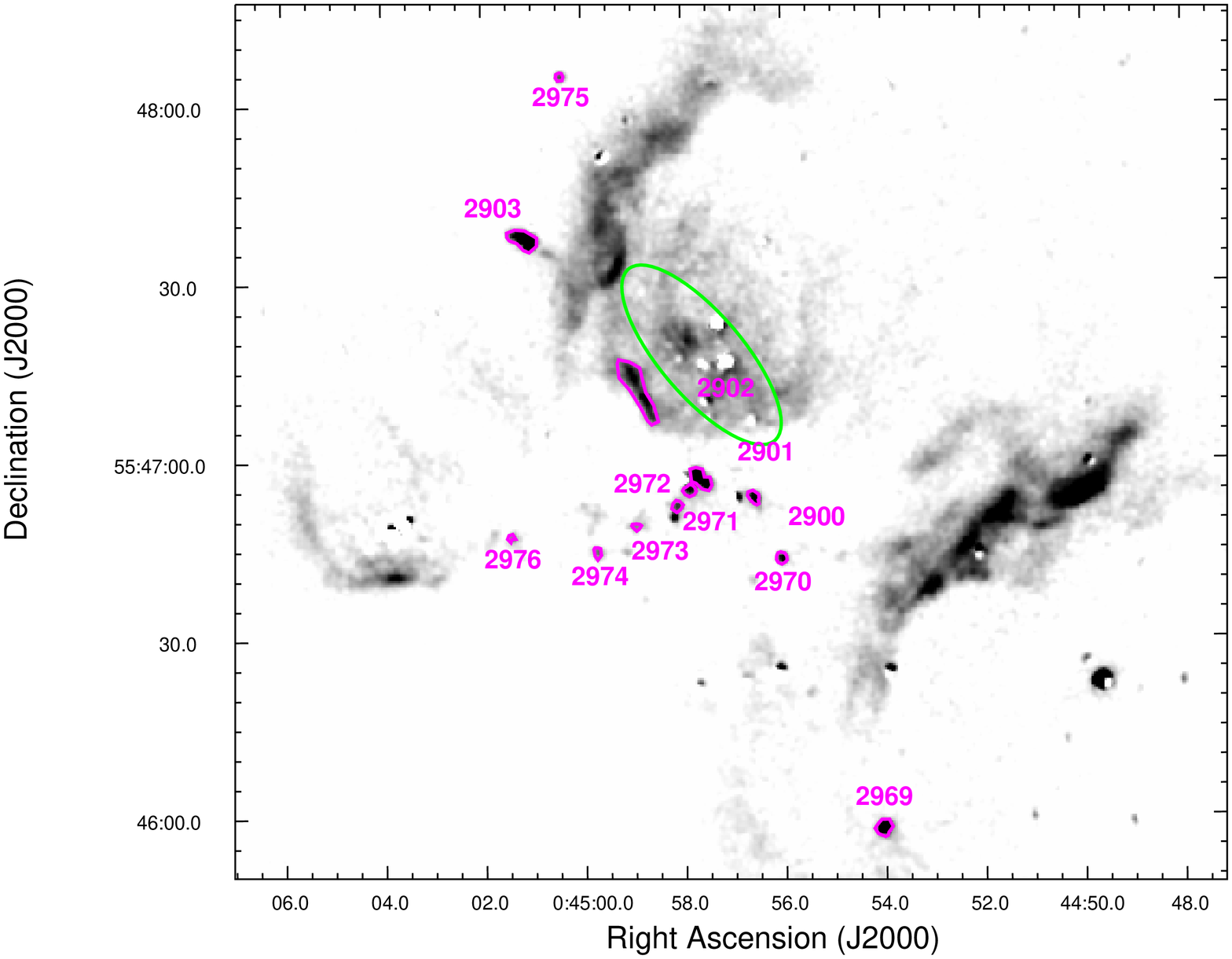}
\caption{Mol 3 (a) rgb image and (b) greyscale image, with labels as described in \S A.}
\end{figure}

\subsection{IRAS 05137+3919 (Mol 8)$^*$}

We identify substructure in two outflows previously reported by VDR10 (Figures 3a \& b). The outflow along a P.A. of $\sim$15-20$^{\circ}$ is defined by the bright, complex bow-shaped features 1002 \& 1003, which are approximately equidistant from RMS YSO G168.0627+00.8221. As suggested by VDR10, our 2.12 $\micron$ image confirms that the jet driving this outflow may be precessing. Although the component of MHO 1003 furthest from the RMS YSO (1003\_2) lies along a P.A. of $\sim$20$^{\circ}$, the brighter, closer, component, MHO 1003, suggests a P.A. of $\sim$15$^{\circ}$. MHOs 1002 \& 1003 are the brightest MHOs in this region, and follow the position angle of the CO outflow identified by ZHB05. Another outflow along a P.A. of $\sim$ -20$^{\circ}$ is defined by the MHO 1004 substructures 1004\_2 \& 1004\_3.  The morphology of the remaining MHO 1004 components is complex and it is not clear whether they (and MHO 1020) are associated with either outflow. 

The two outflows intersect at the position of the RMS YSO / Class I candidate J051713.74+392219.6, which is also coincident with MM 1 (MTR02), the K sources VDR10 denote as ``A'' and ``B'', and a mid-infrared source detected with MIRLIN (J. O'Linger-Luscusk, private communication, unpublished data). The fluxes derived from the MIRLIN observations (F(12.5 $\micron$) = 4.8 Jy; F(20.8 $\micron$) = 11.5 Jy; F(24.5 $\micron$) = 27.6 Jy)  rise steeply in the mid-infrared  and are comparable to {\it WISE} (F(12 $\micron$) = 4.4 Jy; F(22 $\micron$) = 18.6 Jy)and {\it MSX} (F(12 $\micron$) = 4.6 Jy; F(21 $\micron$) = 13.7 Jy) fluxes at similar wavelengths. Using the distance (11.5 kpc) and luminosity (2.25$\times 10^5$ L$_{\odot}$) derived from modeling the SED of this source (Molinari et al. 2008; VDR10), the total H$_2$ (2.12 $\micron$) luminosity, (L$_{2.12}$), calculated from summing the MHO fluxes in Table 3 and assuming no extinction, is 2.99 L$_{\odot}$. Following WAS13, and assuming the 2.12 $\micron$ luminosity is 5-10\% of the total rovibrational H$_2$ emission, L$_{H_2}$ $\sim$ 29.9-59.8 L$_{\odot}$. Since the assumption of no extinction yields a lower limit to the true H$_2$ luminosity, this places the Mol 8 outflows at or above the linear fit to Log(L$_{H_2}$) vs. Log(L$_{bol}$) for high-mass YSOs presented in Caratti o Garatti (2015, Fig. 9). 

\begin{figure}[htb!]
\figurenum{3} 
\includegraphics[angle=0,scale=0.65]{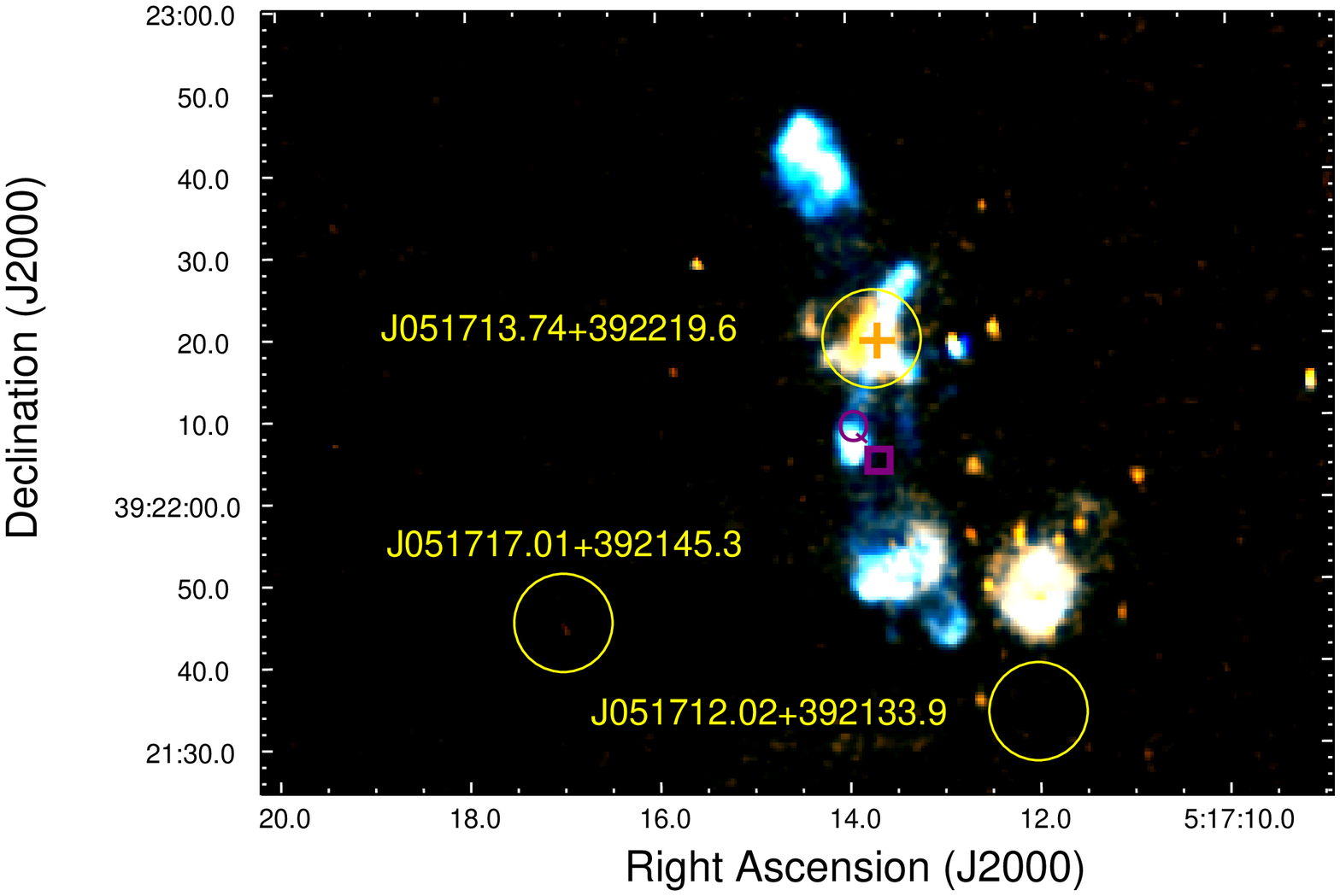}
\includegraphics[angle=0,scale=0.65]{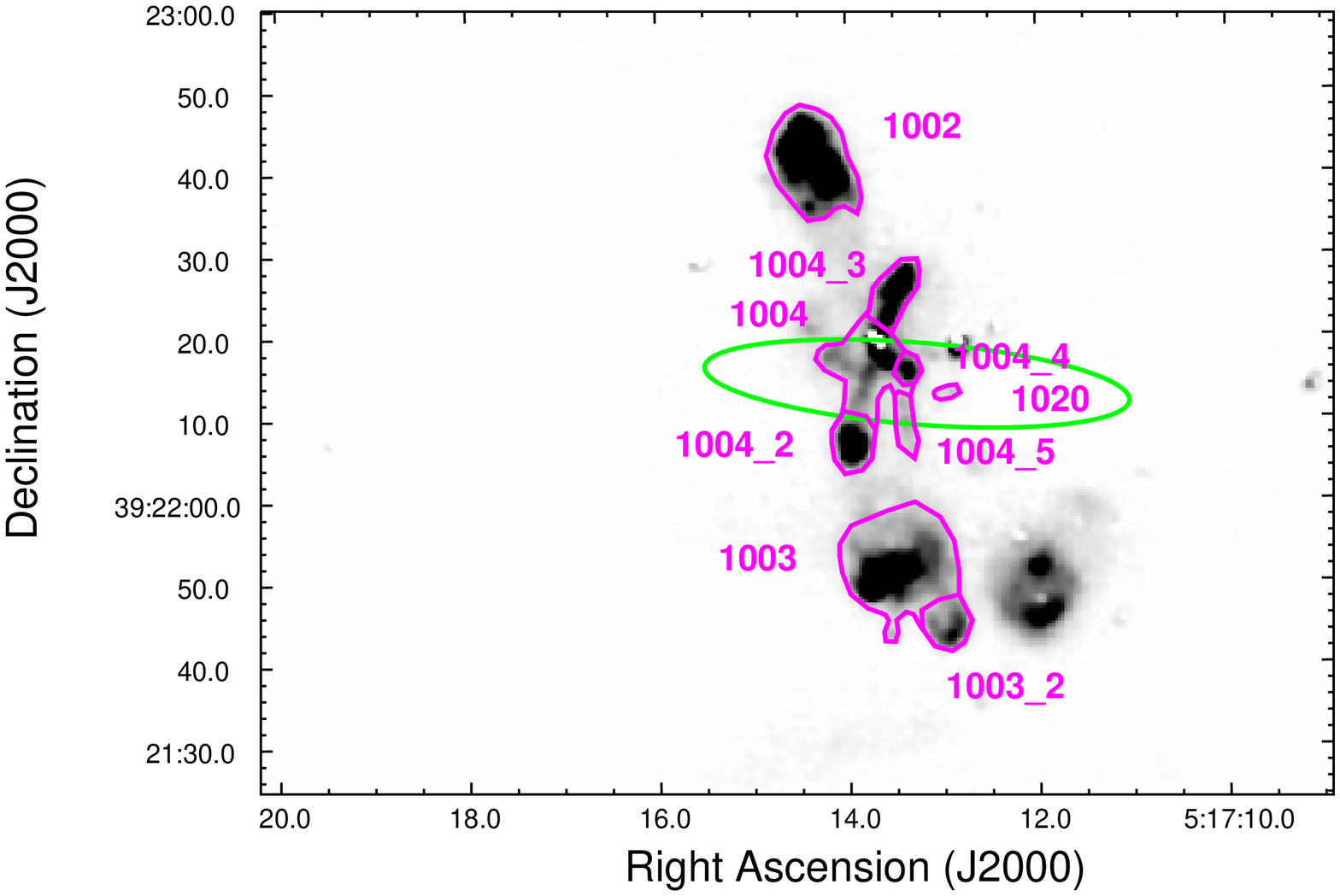}
\caption{Mol 8 (a) rgb image and (b) greyscale image, with labels as described in \S A.}
\end{figure}

\subsection{IRAS 05168+3634 (Mol 9)$^*$}

We identify nine Class I candidates, one Class 0 candidate (J052023.40+363745.9), and four sources that satisfy both Class 0 \& Class I  color criteria (J052023.04+363813.0, J052022.03+363756.5, J052010.31+363743.9, \& J052010.76+363635.0) in this region (Figures 4a \& b). VDR10 found no MHOs near the {\it IRAS} source position, but we note that the low-resolution CO outflow mapped by ZHB05 is centered near J052022.03+363756.5 and a dense core (BGPSv2\_G170.661-00.249) identified by the Bolocam Galactic Plane Survey (BGPS v2.1: Ginsburg et al. 2013), not on the {\it IRAS} source. Whereas the CO outflow of ZHB05 lies along a P.A. of $\sim$40 $^{\circ}$, the bright bow-shaped MHO 1055G lies along a P.A. of $\sim$ 60 $^{\circ}$ from J052022.03+363756.5. MHOs 1055 A, C, \& E lie along a P.A. of $\sim$ 28$^{\circ}$. It is possible that J052022.03+363756.5 is the source of two outflows that produce the large-scale CO morphology, but this is unclear. Three faint MHOs (1055B, D, \& F) lie to the southwest of the {\it IRAS} source and one to northwest (MHO 1056), but their associations are also unclear.  

\begin{figure}[htb!]
\figurenum{4}
\includegraphics[angle=0,scale=0.45]{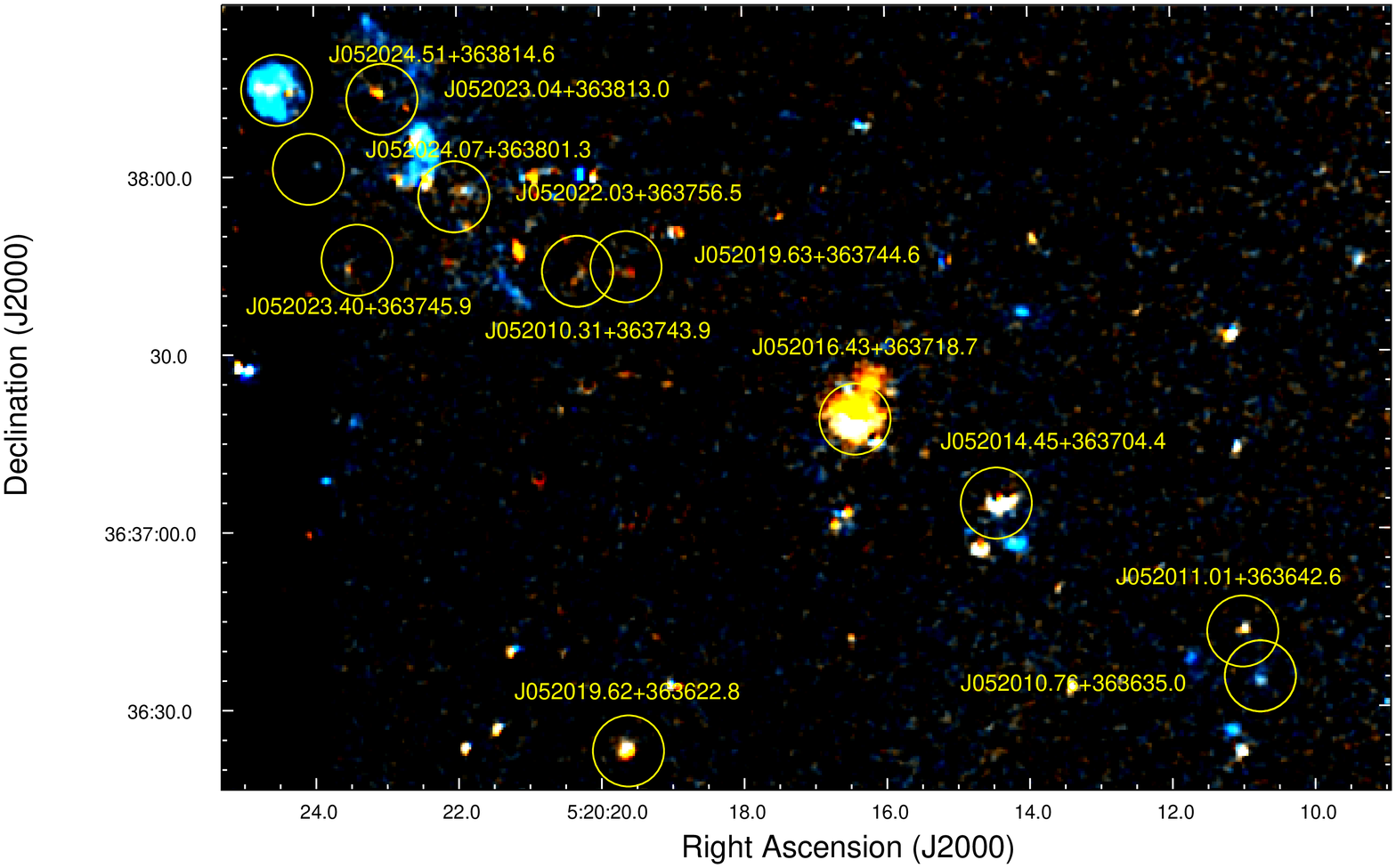}
\includegraphics[angle=0,scale=0.45]{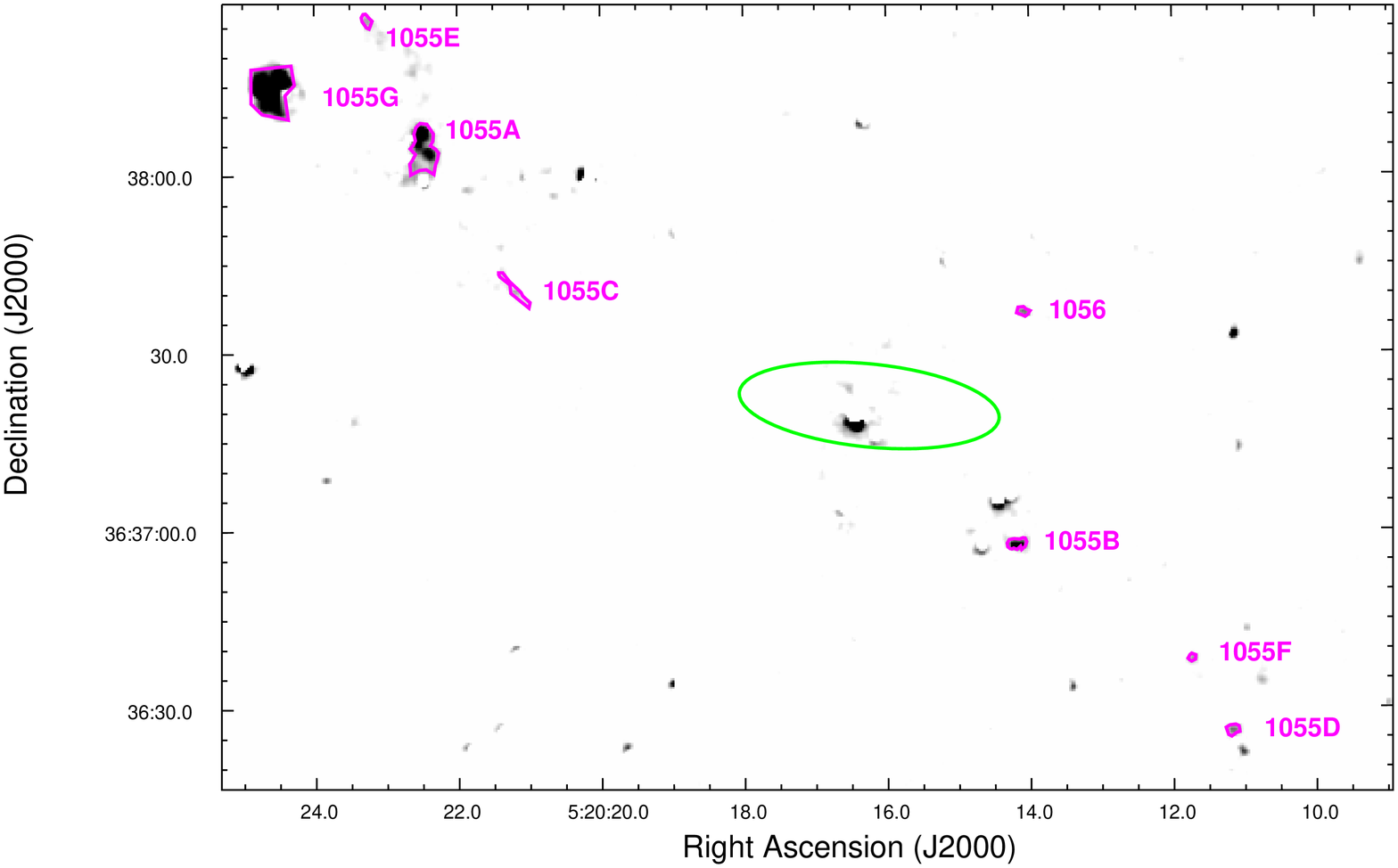}
\caption{Mol 9 (a) rgb image and (b) greyscale image, with labels as described in \S A.}
\end{figure}

\subsection{IRAS 05274+3345 (Mol 10)$^*$}

VDR10 present a detailed background of this region, which contains multiple signs of star formation. We identify five {\it ALLWISE} sources in this region that satisfy Class I color criteria; two of these also satisfy Class 0 color criteria (J053049.21+334818.4, J053049.01+334746.7). The RMS YSO candidate G174.1974-00.0763 (VDR10 source `A') lies close to the center of the {\it IRAS} error ellipse and is the brightest infrared source in this field (Figures 5a \& b). At least three distinct outflows were identified from near-infrared observations (Chen et al. 2005) and via multiple spectral lines using the Submillimeter Array (Zhang et al. 2007). Chen et al. (2005) identified 28 MHOs in this region, which were not entered in the on-line MHO catalog (Davis et al. 2010). Several of these objects lie along three axes they identified as the ``short'', ``middle'', and ``long'' jets.  The three outflows A, B, and C, identified by Zhang et al. (2007) along  P.A.s=5$^{\circ}$, 35$^{\circ}$ \& -60$^{\circ}$, respectively, intersect in the vicinity of the the hot cores they refer to as MM-1 and MM-2. These cores lie approximately at the center of a triangle defined by MHOs 1008, 1009, \& 1010 (Figure 5b). G\'omez-Ruiz et al. (2016) identified several methanol masers at 44 GHz in this region, which are clustered towards MHOs 1007 \& 1008, as well as between MHOs 1008, 1009, \& 1010.

Based on our 2.12-$\mu$m image (Figure 5b), we can make the following associations: MHO 1005 corresponds to the feature VDR10 identify as `1' and Chen et al. (2005) associate with a feature they refer to as the `short jet' (see Table 4, Mol 10 Flow 1). Similarly, MHOs 1006, 1007, 1008, 1009, 1010, \& 1011, correspond to VDR10 features 2, 3, 4, 5, 6, \& 7, respectively. Zhang et al. (2007) outflow `A' (Chen et al. 2005 `middle jet') appears as a faint bridge of emission connecting MHOs 1009 and 1011, and possibly also MHO 1059 (Figure 5b and Table 4, Mol 10 Flow 2). The `long jet' identified by Chen et al. (2005), along a P.A. of $\sim$-60$^{\circ}$, connects MHOs 1057, 1007, and possibly 1060 (Figure 5b) along a line slightly south of MM-1 and MM-2 (see Table 4, Mol 10 Flow 3). Although it is tempting to associate Zhang et al. (2007) outflow `B' with the `short jet' detected by Chen et al. (2005), VDR10, and the present study, we note that outflow `B' lies along a P.A. $\sim$ 35$^{\circ}$, while the jet  associated with MHO 1005 lies along a P.A. of $\sim$ 50$^{\circ}$, so these features may not be tracing the same outflow. Outflow `C' lies along the same P.A. as the long jet. Although this outflow may be associated with MM-1/MM-2, there are several near-infrared point sources (Figure 5a) between these sources and Class I candidate J053046.98+334748.3 that are viable driving source candidates, as is the RMS source, which is the probable driving source of the large-scale CO outflow mapped by Hunter et al. (1995). It is not clear whether MHO 1060 is associated with this outflow, but it lies along the same P.A. of -60$^{\circ}$.

The bow-shaped feature MHO 1006 lies directly west of the RMS source, along a P.A. of -90$^{\circ}$, and is likely driven by this source (Mol 10 Flow 4). MHO 1006 has the lowest 2.12/2.25 flux ratio of the MHOs we identified in this region and may be dominated by fluorescent emission from a compact PDR associated with with the RMS YSO/ Q-type object G174.1974-00.0763, which is in the {\it WISE} Catalog of Galactic \ion{H}{2} Regions (Anderson et al. 2014). The associations of MHOs 1008, 1010 \& 1058 with outflows in this region is unclear.

\begin{figure}[htb!]
\figurenum{5}
\includegraphics[angle=0,scale=0.52]{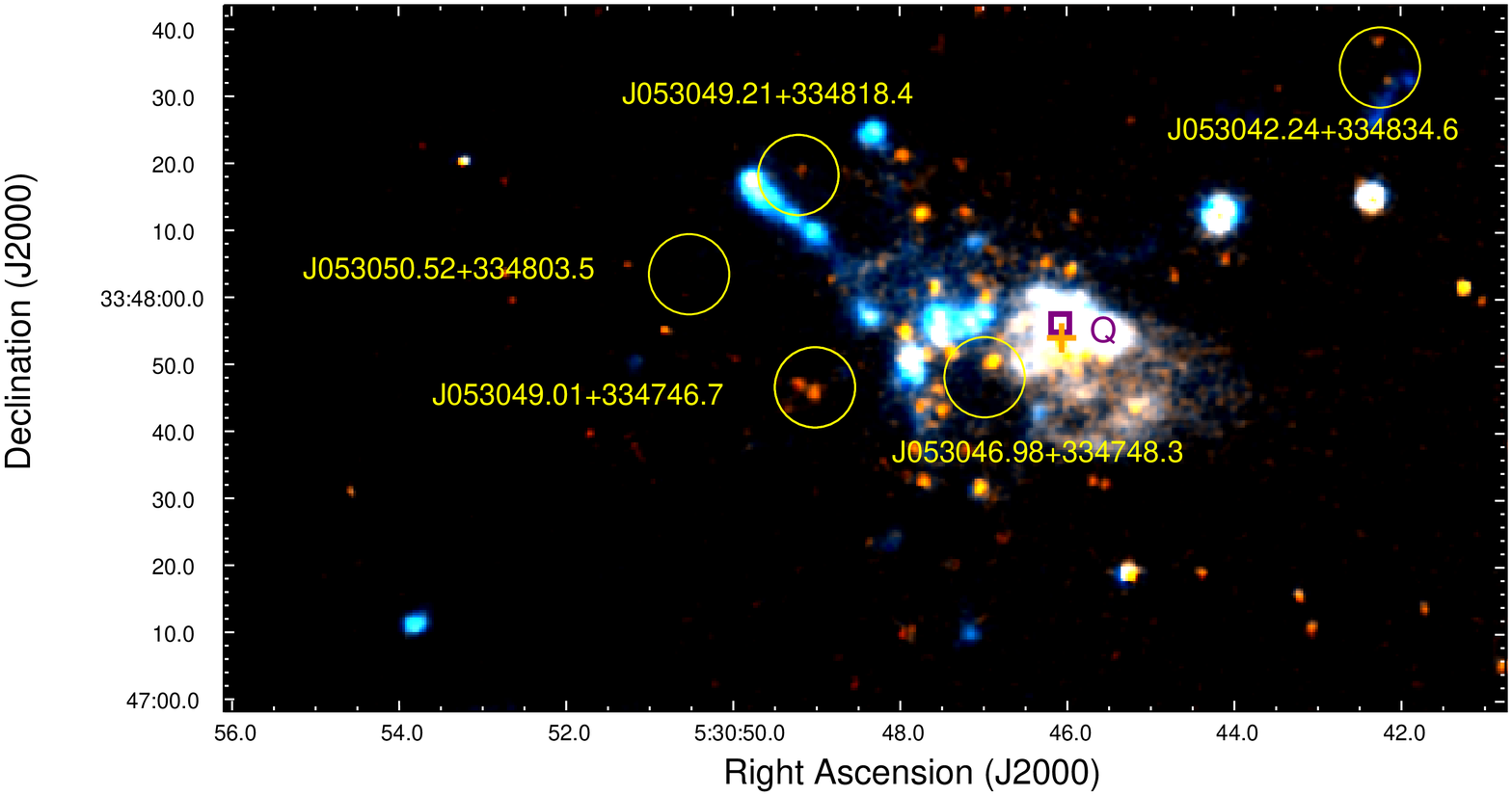}
\includegraphics[angle=0,scale=0.52]{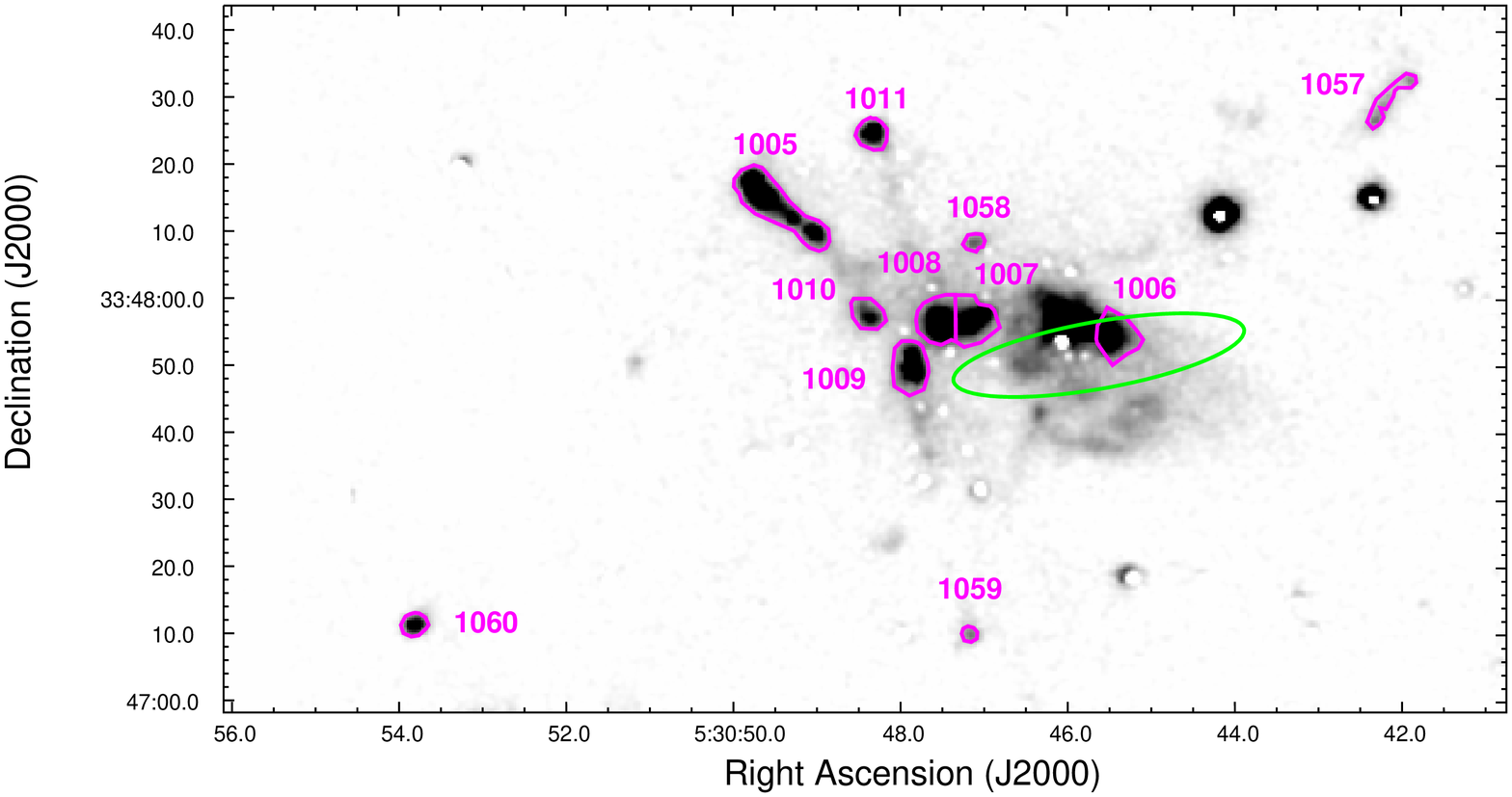}
\caption{Mol 10 (a) rgb image and (b) greyscale image, with labels as described in \S A.}
\end{figure}

\subsection{IRAS 05345+3157 (Mol 11)$^*$}

The complex arrangements of MHOs clearly indicate multiple outflows in this region (Chen et al. 1999; Chen et al. 2003; VDR10). We identify ten {\it ALLWISE} sources that fit Class I color criteria, four of which also fit Class 0 criteria (J053753.90+320015.7, J053752.99+315934.8, J053752.11+320020.0, J053751.45+320021.5). There are at least two distinct epochs of star-formation (Figures 6a \& b): (1) an infrared cluster that is centered on the {\it IRAS} position and is enveloped in a PDR (with no Class 0/I associations); and (2) younger objects to the northeast and northwest of the PDR, as indicated by the MHOs, Class 0/I candidates, and seven dense cores (Lee et al. 2011). Lee et al. (2011) suggested that two of the dense cores (Cores 1 \& 3) are forming massive protostars. The low-resolution CO map of ZHB05 shows the molecular outflow to be centered at least 30$^{\prime\prime}$ to the northeast of the {\it IRAS} source, and elongated in a general ESE - WNW direction, with blueshifted outflow gas in the direction of the MHO 1061 knots.

The Class I candidate, J053752.03+320003.9, is coincident with Core 3. It is equidistant from the bright bow-shaped features MHO 1015 and 1015\_2, which define an outflow along a position angle of 64$^{\circ}$ (Table 4, Mol 11 Flow 1). VDR10 identified a source they refer to as `B' at the southwest end of MHO 1015 as the possible driving source of this outflow, but our observations suggest J053752.03+320003.9 is the better candidate. The linear chain of knots MHO 1061 A-F (Table 4, Mol 11 Flow 2) also intersect Core 3 and may connect with the MHO 1063 knots to the southeast at a position angle of -72$^{\circ}$, but this is unclear. There are several {\it WISE} Class I candidates in the vicinity of the MHO 1061 knots that are also viable candidates. VDR10 suggested an outflow along a position angle of -49$^{\circ}$ (Table 4, Mol 11 Flow 3) connecting features we identify with the MHO 1018 and 1016 complexes. Their candidate driving source (`A ') coincides with a bright `orange' point source on our rgb image (Figure 6A), which lies at the center of this outflow. The Class 0/I candidate, J053752.99+315934.8 (coincident with Core 1), lies at the northwestern end of MHO 1062A (Table 4, Mol 11 Flow 4). MHO 1062A is the second brightest MHO in this region. Although it aligns with the faint knots MHO 1062 B \& C to the northwest, it is unclear whether these are part of this outflow. The remaining MHOs cannot be clearly linked to distinct outflows. MHO 1019 lies to the southwest of the infrared cluster and may be excited by a source at the edge of the PDR.

\begin{figure}[htb!]
\figurenum{6}
\includegraphics[angle=0,scale=0.38]{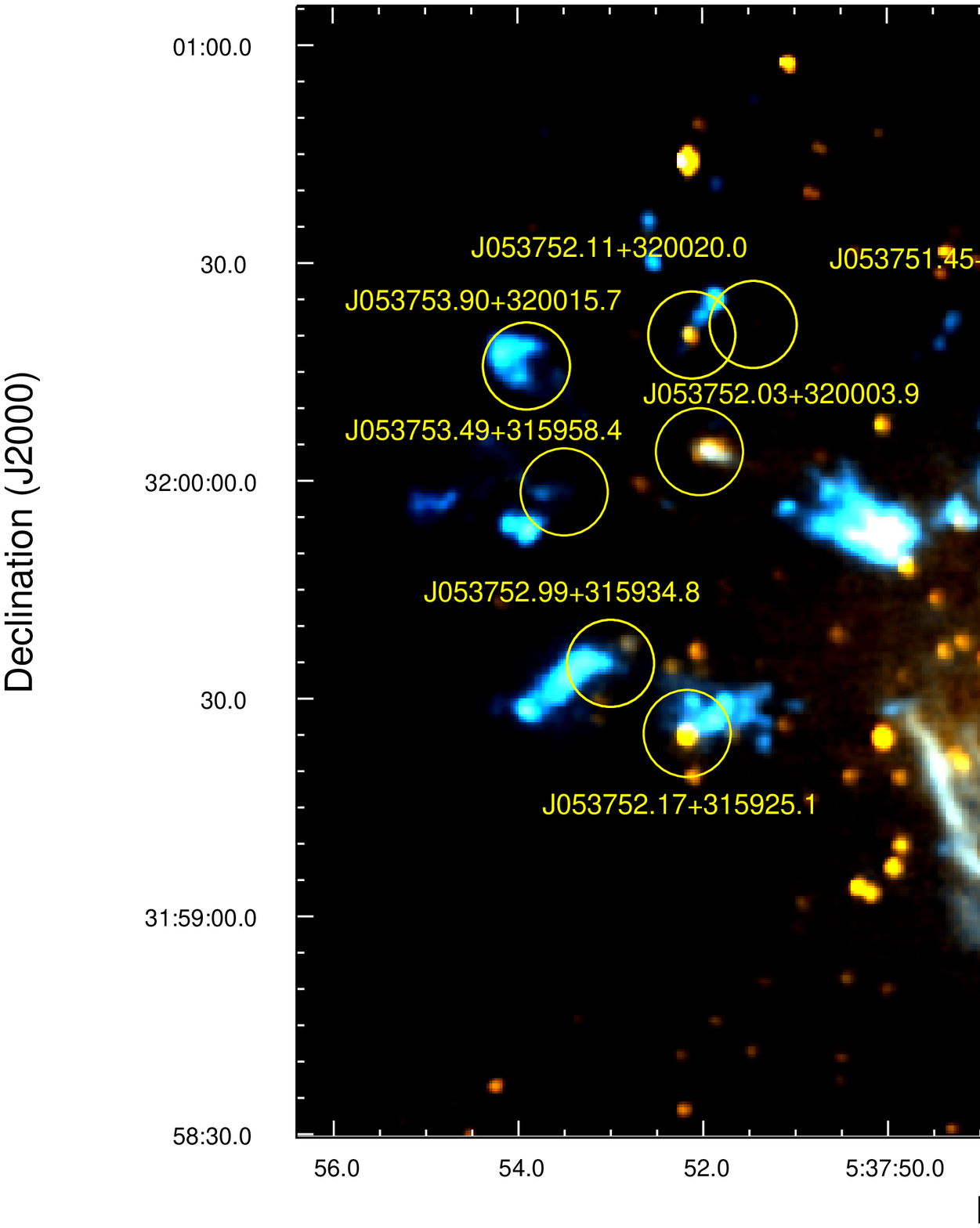}
\includegraphics[angle=0,scale=0.38]{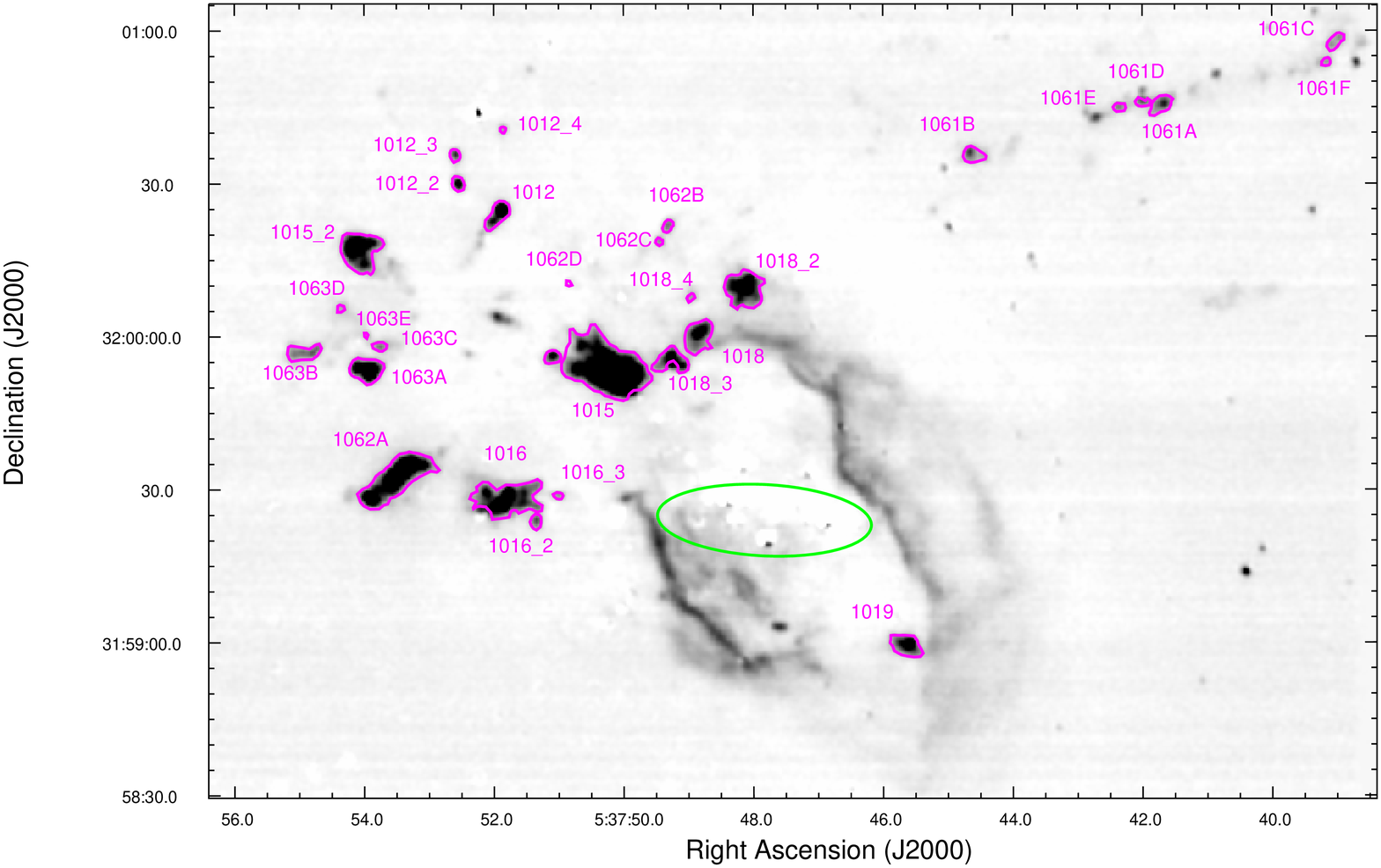}
\caption{Mol 11 (a) rgb image and (b) greyscale image, with labels as described in \S A.}
\end{figure}

\subsection{IRAS 05373+2349 (Mol 12)$^*$}

Except for MHO 751, all of the MHOs in this region were previously detected (VDR10; Khanzadyan et al. 2011). We identify six {\it ALLWISE} sources that satisfy Class I color criteria and two that satisfy Class 0 (J054020.76+235031.0, J054026.46+235222.0) color criteria. The CO outflow mapped by ZHB05 is slightly elongated along a position angle of $\sim$ 31$^{\circ}$ about the {\it IRAS} source position. The Class I candidate J054024.79+235048.0, RMS YSO candidate G183.7203-03.6647, and Q-type \ion{H}{2} region G183.720-03.665 are positionally coincident near the center of the {\it IRAS} error ellipse (Figures 7a \& b).

Our observations suggest the presence of at least four outflows, as well as isolated knots not clearly connected to any outflows in this region. MHOs defining two of these outflows (Table 4, Mol 12 Flow 1 \& Mol 12 Flow 3) intersect at the position of the {\it IRAS} source.  Mol 12 Flow 1 is defined by the faint MHO 738 knots along a P.A. of $\sim$ 31$^{\circ}$, similar to the large-scale CO outflow. The faint knot in the northeast, MHO 742A, lies along this position angle and may be part of this outflow, but the brighter knots, MHOs 742B \& C, lie along a distinctly different position angle of $\sim$ 95$^{\circ}$. Potential driving sources for this outflow (Table 4, Mol 12 Flow 4) include the Class I candidate J054024.95+235220.6 or the nearby Class 0 candidate J054026.46+235222.0.  The faint emission knot to the west, MHO 737, may be part of this outflow. Mol 12 Flow 3 joins MHO 739 and MHO 744B along a P.A. of $\sim$ 63$^{\circ}$. 

Mol 12 Flow 2 is defined by the S-shaped series of MHOs along a position angle of $\sim$ -10$^{\circ}$, which stretches from MHO 744A to MHO 740A, and includes the MHO 734 knots and MHO 751. Based on the arrangements of these MHOs, we suggest that Class I candidate J054024.79+235048.0 drives this outflow. This source is coincident with the brighter of two mid-infrared sources identified by MIRLIN (J. O'Linger-Luscusk, private communication). The differences in mid-infrared fluxes derived from the high-resolution MIRLIN observations (F(12.5 $\micron$) = 4.19, 1.37 Jy; F(20.8 $\micron$) = 7.48, 4.79 Jy; F(24.5 $\micron$) = 11.38, 8.98 Jy) , compared with {\it MSX} (F(12 $\micron$) = 4.65 Jy; F(21 $\micron$) = 12.27 Jy) and {\it WISE} (F(12 $\micron$) = 6.0 Jy; F(22 $\micron$) = 20.7 Jy) fluxes, suggest that multiple sources contribute to the mid- and far-infrared luminosity in this region. The remote MHO 735 nearly coincides with Class I candidate, J054019.59+235202.5, but since this is an isolated knot, it is not possible to identify an outflow.  

\begin{figure}[htb!]
\figurenum{7}
\includegraphics[angle=0,scale=0.48]{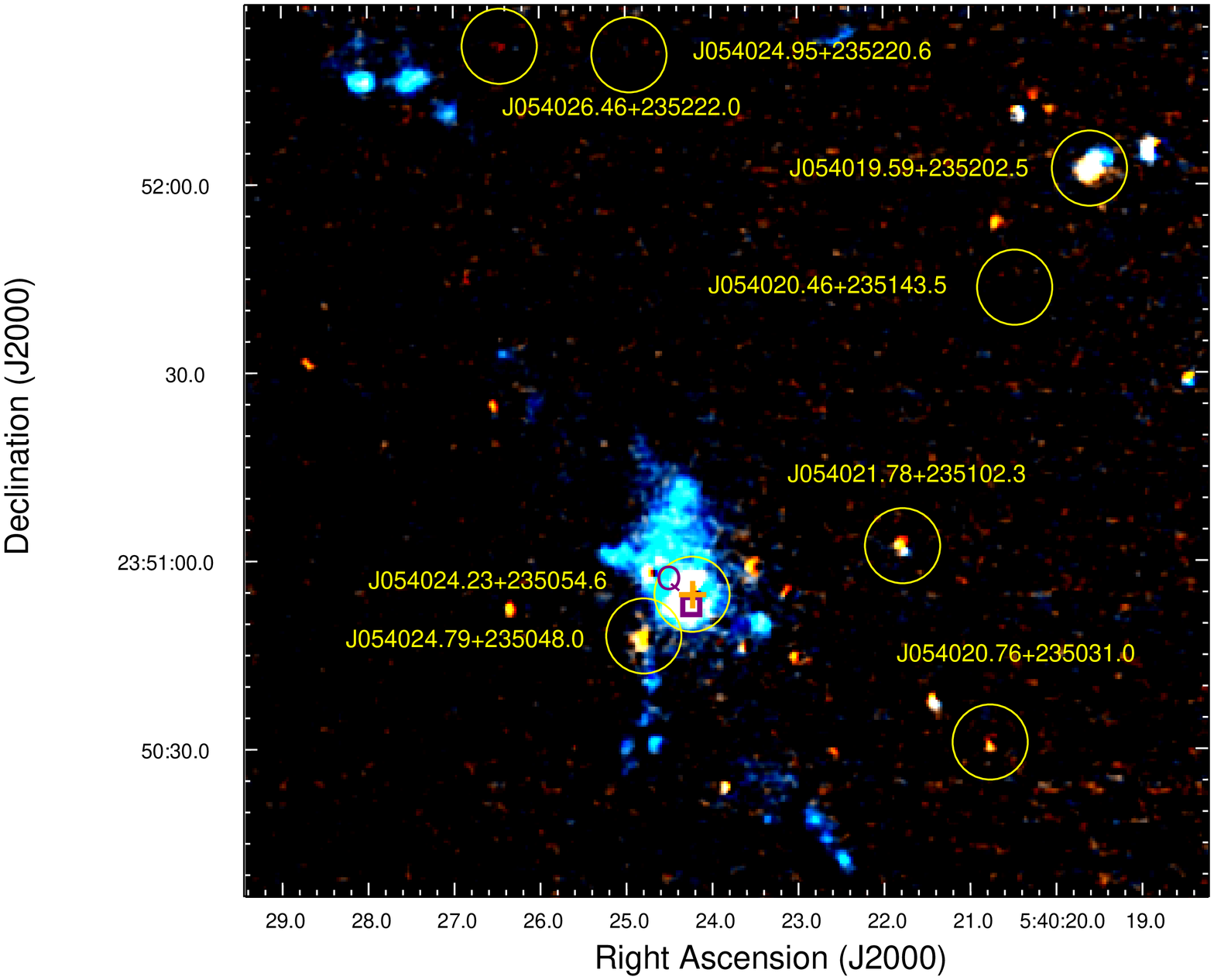}
\includegraphics[angle=0,scale=0.48]{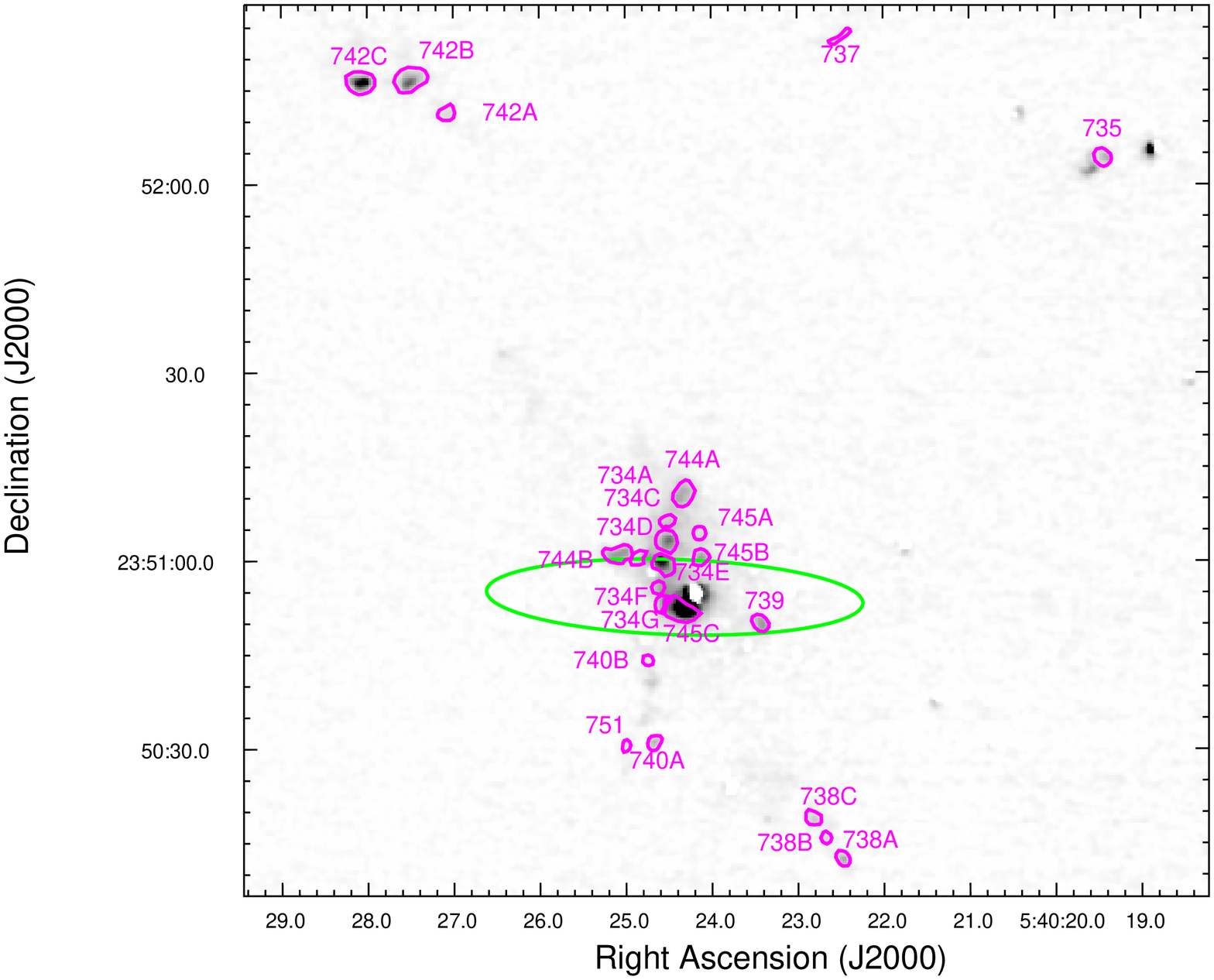}
\caption{Mol 12 (a) rgb image and (b) greyscale image, with labels as described in \S A.}
\end{figure}

\subsection{IRAS 06056+2131 (Mol 15)}

The arrangement of MHOs in this region closely mirrors the complex morphology of the high-velocity CO emission mapped by ZHB05, which in turn clearly indicates there are multiple outflows (Figures 8a \& b). We identify nine Class I candidates, two Class 0/I candidates (J060838.48+213043.8, J060842.88+213118.7), and one Class 0 candidate (J060836.97+213048.8). The western peak of the blueshifted CO emission of ZHB05 is centered on the {\it IRAS} source position, which is coincident with Class I candidate J060840.45+213102.0, RMS YSO G189.0307+00.7821, and Q-type \ion{H}{2} region G189.030+00.780. The eastern blueshifted CO peak is centered $\sim$15$^{\prime\prime}$ south of Class I candidate, J060846.72+213144.0, which is coincident with RMS YSO G189.0323+00.8092 and Q-type \ion{H}{2} region G189.032+00.809. The redshifted CO emission of ZHB05 displays a U-shaped morphology that roughly traces the MHO 1216, 1217, and 1218 complexes.  Both \ion{H}{2} regions lie near dense cores identified by BGPS. Mol 15 Flow 1 lies along a position angle of $\sim$ -35$^{\circ}$, including the MHOs in the 1216 group. This outflow is likely driven by RMS YSO G189.0307+00.7821/ Class I candidate J060840.45+213102.0. MHOs 1217A-C lie along a position angle of $\sim$ 88$^{\circ}$ degrees centered north of G189.0307+00.7821/ J060840.45+213102.0, and the MHO 1218 complex may be associated with an outflow produced by RMS YSO G189.0323+00.8092/ Class I candidate J060846.72+213144.0, but the complex arrangement of these MHOs makes it impossible to link these to specific outflows. Two other Class I candidates (J060843.63+213150.7, J060843.08+213147.2) are also potential driving sources.

\begin{figure}[htb!]
\figurenum{8}
\includegraphics[angle=0,scale=0.45]{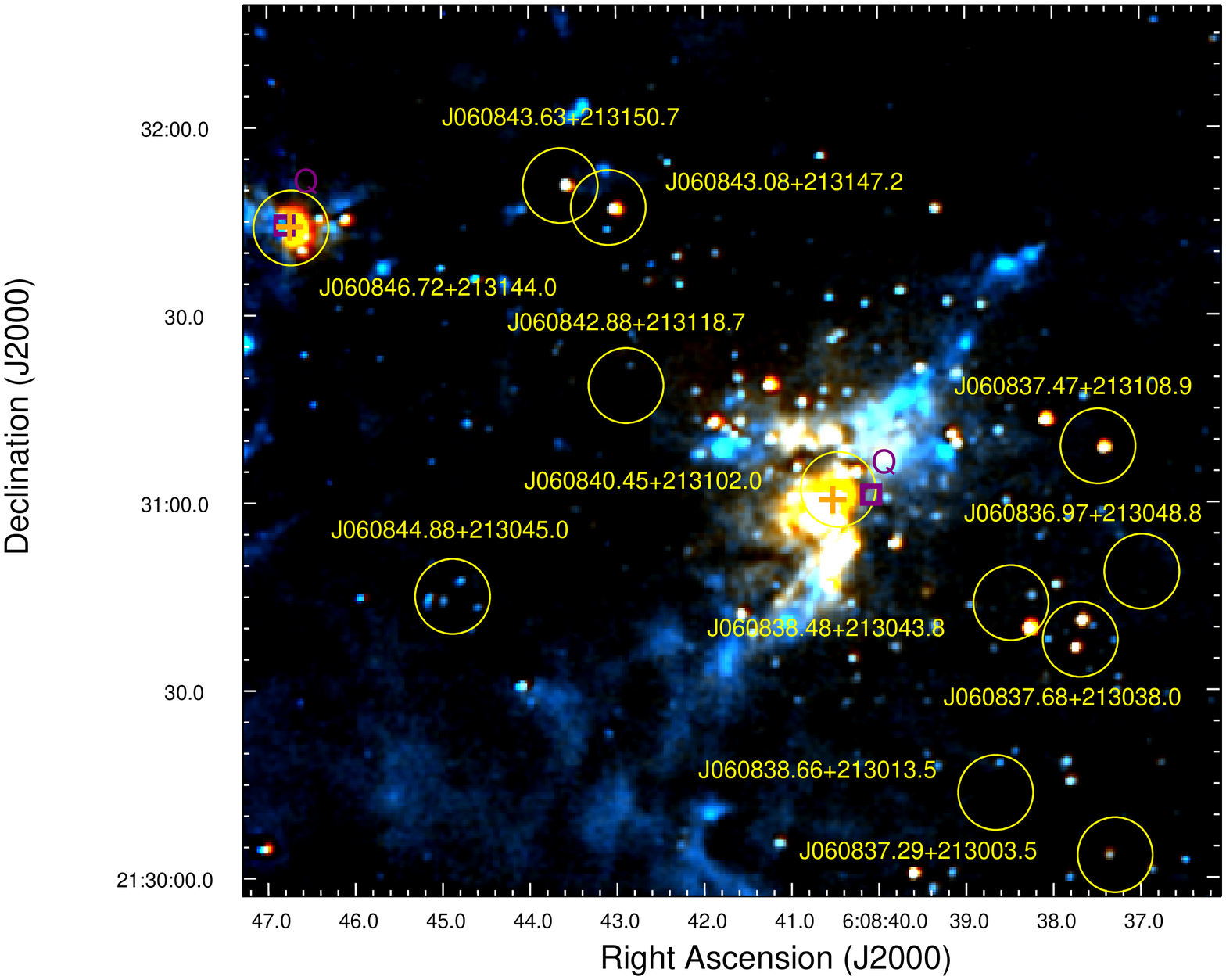}
\includegraphics[angle=0,scale=0.45]{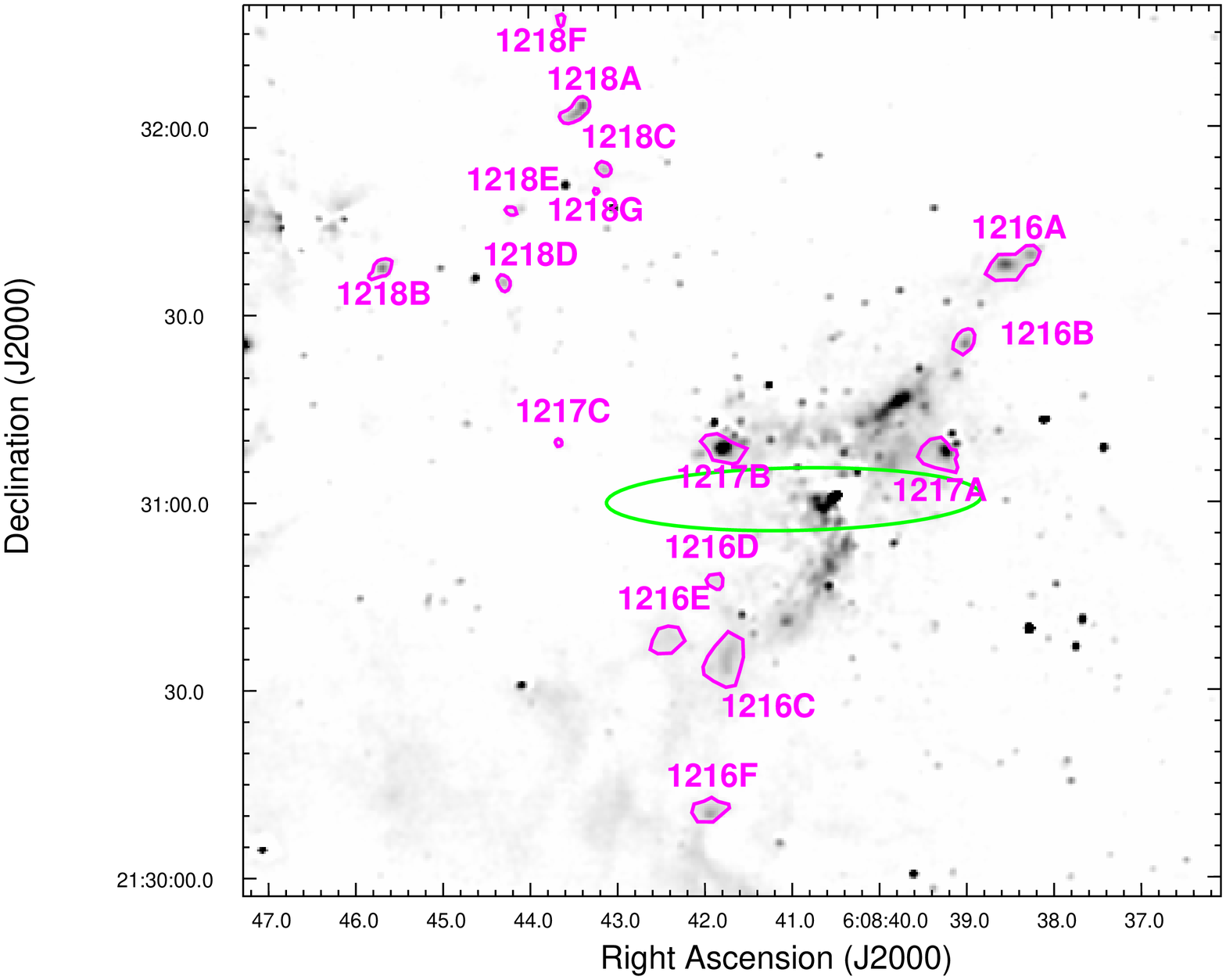}
\caption{Mol 15 (a) rgb image and (b) greyscale image, with labels as described in \S A.}
\end{figure}

\subsection{IRAS 06584-0852 (Mol 28)$^*$}

Most of the H$_2$ emission in this region is due to fluorescence, some of which may be associated with Q-type \ion{H}{2} region G221.955-01.993 (Figures 9a \& b). Interestingly, the morphology of the fluorescent emission closely parallels the complex CO high-velocity emission identified by ZHB05. VDR10 detected no MHOs in this region; however, we detect three isolated knots. MHO 3139 lies directly to the northwest of fluorescent H$_2$ that coincides with overlapping blue- and redshifted CO emission detected by ZHB05. MHO 3140 is coincident with Class I candidate, J070051.00-085629.8 (RMS YSO G221.9605-01.9926), and a double mid-infrared source identified by MIRLIN (O'Linger-Luscusk, private communication). The MIRLIN emission (F(12.5 $\micron$) = 5.27, 1.16 Jy; F(17.9$\micron$) = 9.98, 1.10 Jy; F(20.8 $\micron$) = 10.46, 1.38 Jy; F(24.5 $\micron$) = 11.32, $<$0.72 Jy) appears to trace the two stars VDR10 identify as reddened YSOs `A' and `B'. Mol 3141 is a faint emission knot with no evident association.

\begin{figure}[htb!]
\figurenum{9}
\includegraphics[angle=0,scale=0.58]{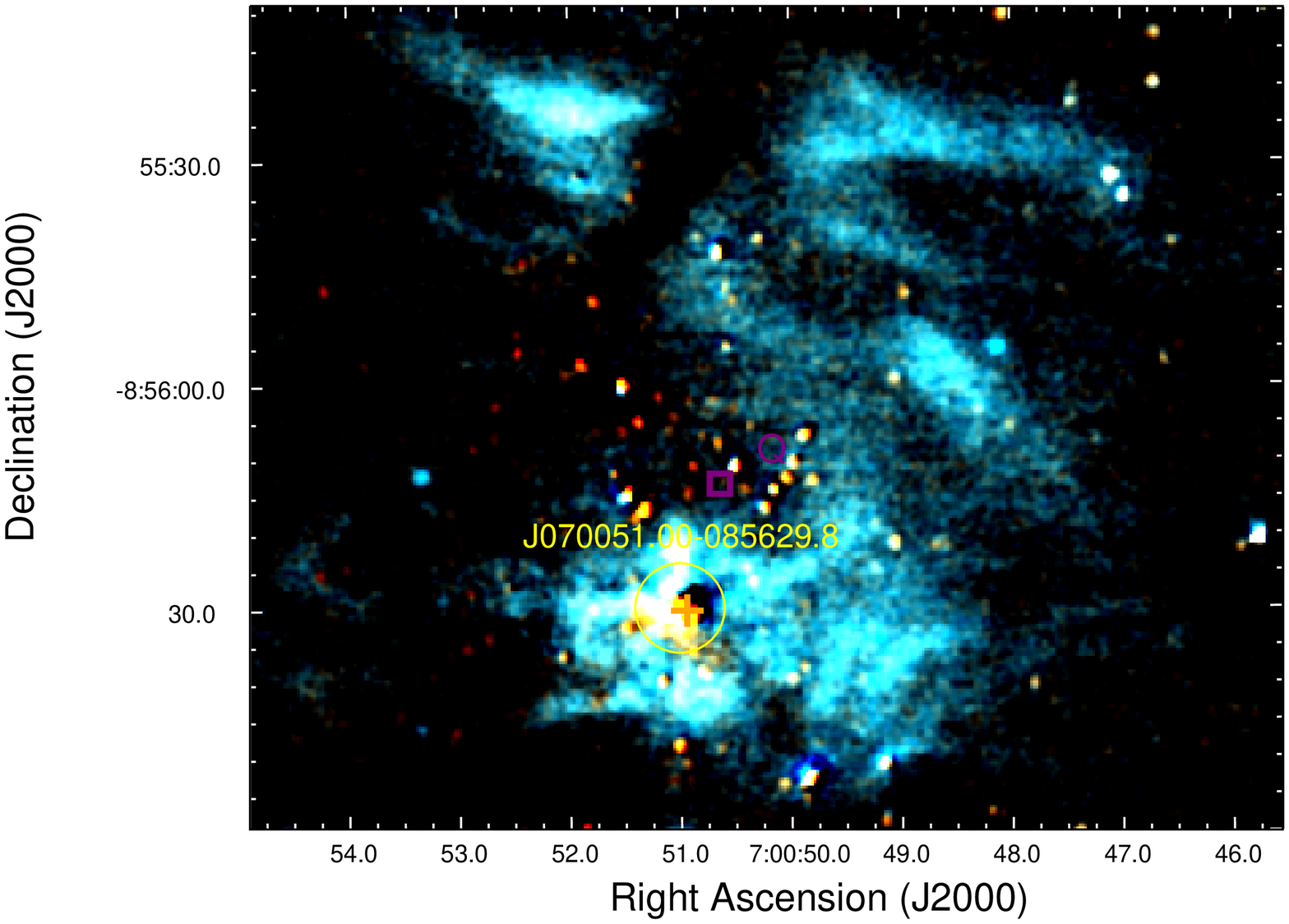}
\includegraphics[angle=0,scale=0.58]{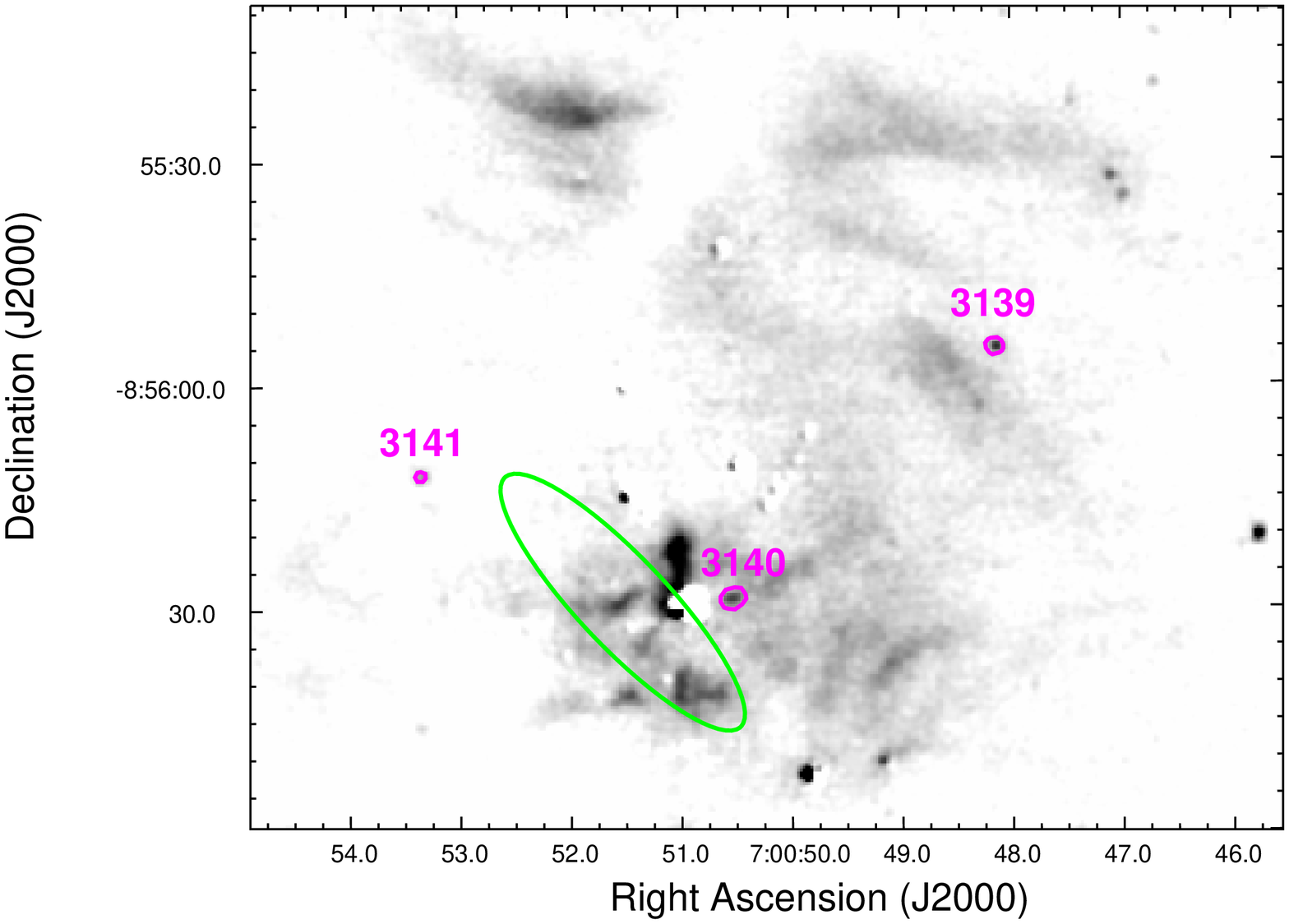}
\caption{Mol 28 (a) rgb image and (b) greyscale image, with labels as described in \S A.}
\end{figure}

\subsection{IRAS 19368+2239 (Mol 108)}

This region contains multiple signposts of star formation (Figures 10a \& b), including three Q-type \ion{H}{2} regions, a RMS source identified as an \ion{H}{2} region (G058.4670+00.4360B), a RMS YSO (G058.4670+004360A), and twelve 44-GHz masers (G\'omez-Ruiz et al. 2016). The CO outflow mapped by ZHB05 is complex and multi-lobed, suggesting multiple outflows in this region, with most of the high-velocity emission located north of the {\it IRAS} position (and RMS sources). We identified eight Class I candidates, two of which are positionally coincident with the RMS \ion{H}{2} region (J193857.17+224624.5) and RMS YSO (J193856.89+224632.2). We detected a chain of four MHOs (Mol 108 Flow 1), along a position angle of $\sim$ 65$^{\circ}$, located north of the RMS YSO, but colinear with Class I candidate J193859.37+224656.0, which clearly shows multiple near-infrared components (see Figure 10a). 

\begin{figure}[htb!]
\figurenum{10}
\includegraphics[angle=0,scale=0.5]{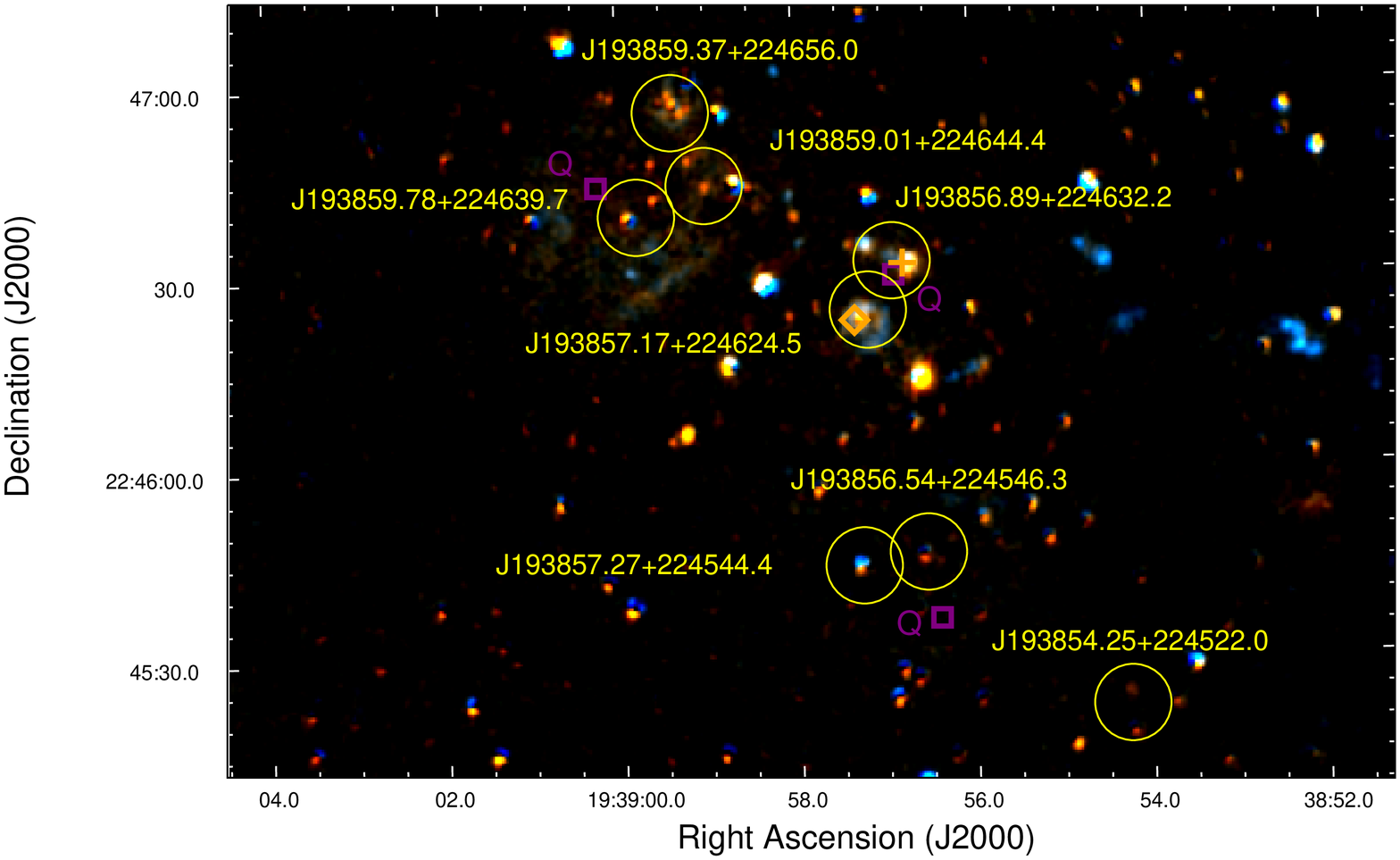}
\includegraphics[angle=0,scale=0.5]{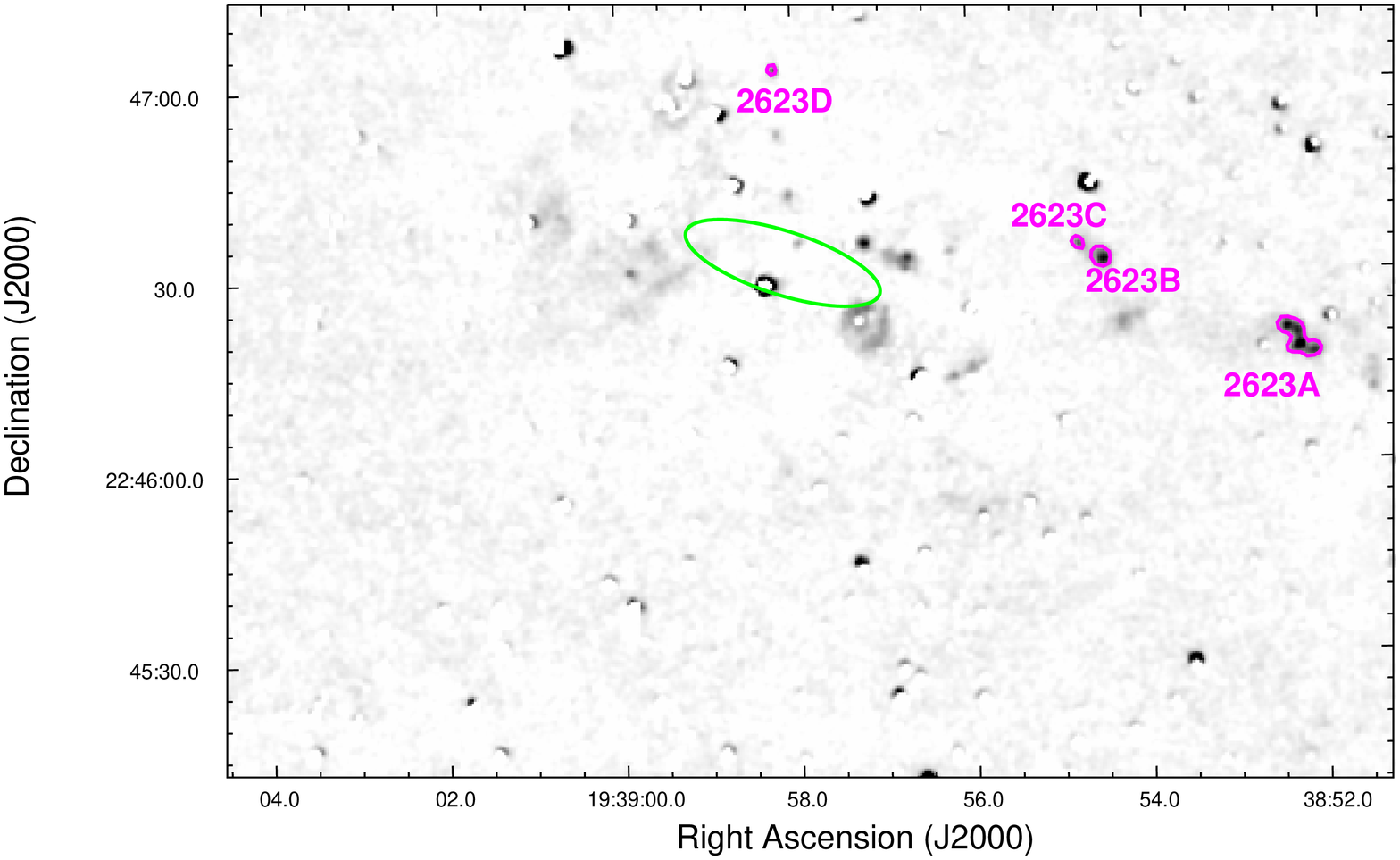}
\caption{Mol 108 (a) rgb image and (b) greyscale image, with labels as described in \S A.}
\end{figure}

\subsection{IRAS 19374+2352 (Mol 109)$^*$}

VDR10 detected three MHOs in this region (2600 - 2602). Two of these (2601 \& 2602) are located along the periphery of the {\it IRAS} error ellipse, just south of a PDR associated with G059.612$+$00.917, a grouped (G-type) \ion{H}{2} region (Figures 11a \& b). Known  (K-type) \ion{H}{2} region G059.602$+$00.911,  four RMS objects catalogued as \ion{H}{2} regions, and a dense core detected by the APEX Telescope Large Area Survey of the Galaxy (ATLASGAL: Schuller et al. 2009) are clustered near the center of the CO outflow mapped by ZHB05, which lies to the southeast of the {\it IRAS} position. G\'omez-Ruiz et al. (2016) detected two 44-GHz methanol masers just northeast of MHO 2600. 

We identified three Class I candidates and three additional faint MHOs toward this target. Class I candidate J193934.64+235948.3 is positionally coincident with the ATLASGAL core, as well as an \ion{H}{2} region catalogued in both the RMS survey and {\it WISE} catalog of galactic \ion{H}{2} regions (Lumsden et al. 2013; Anderson et al. 2014). This source is particularly interesting since it exhibits ``anomalous'' H$_2$ emission, where the 2.12-$\micron$/2.25-$\micron$ ratio is less than one. We note that WAS13 detected such emission toward a deeply embedded source (DES: Yao et al. 2000) in Mol 121. The DES is thought to be a massive YSO driving a very young outflow. It is tempting to associate J193934.64+235948.3 with an outflow connecting MHOs 2600 -2602 along a P.A. of $\sim$ -40$^{\circ}$, but the faint chain of MHOs that extend to the southwest of MHO 2600 along a P.A. of $\sim$ 30$^{\circ}$ calls into question this interpretation. The CO outflow of ZHB05 shows multiple peaks, suggesting more than one outflow in this region. Faint MHO 2628 may be associated with Class I candidate J193933.43+235836.2, at the southern end of the field.

\begin{figure}[htb!]
\figurenum{11}
\includegraphics[angle=0,scale=0.56]{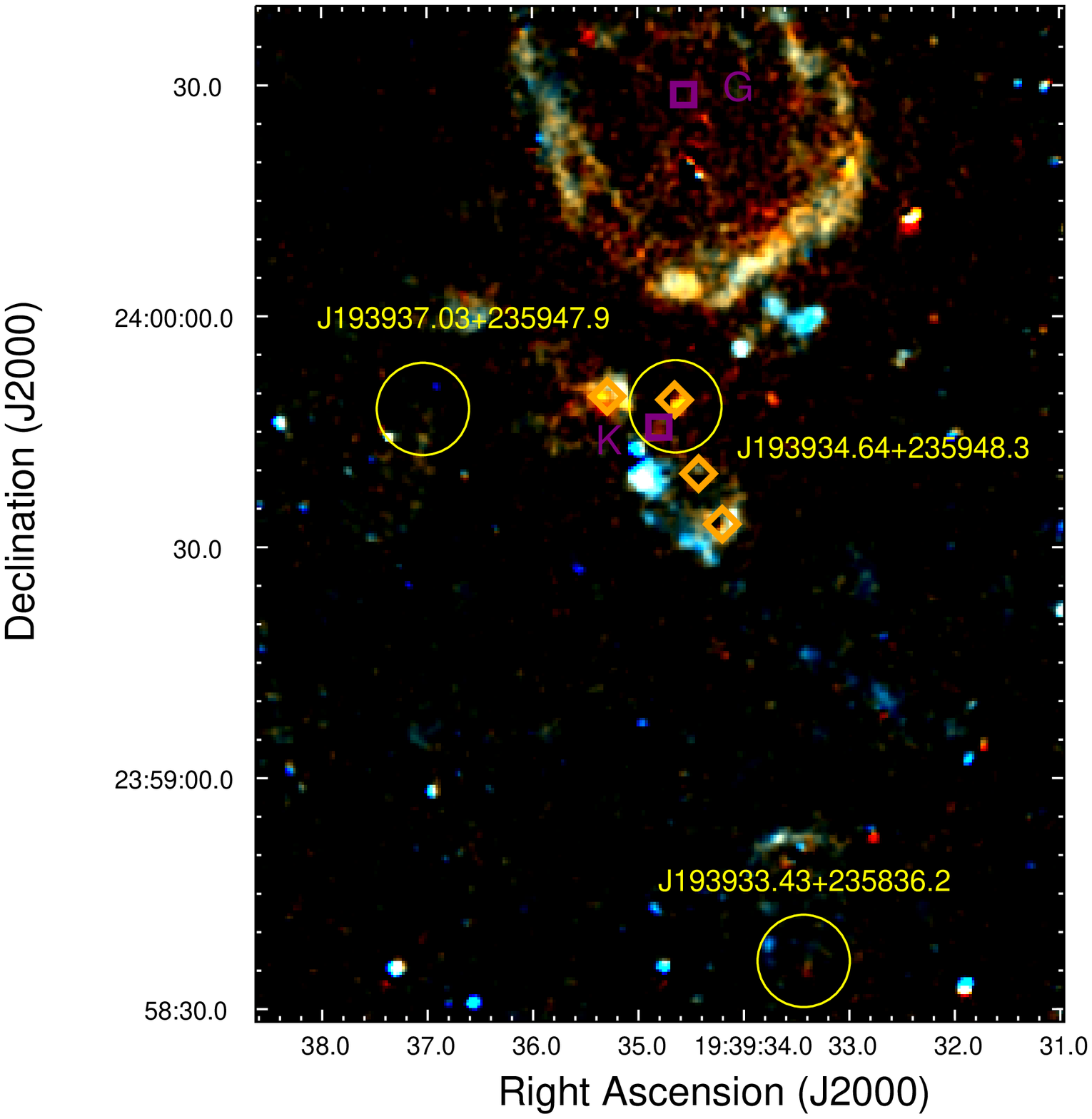}
\includegraphics[angle=0,scale=0.56]{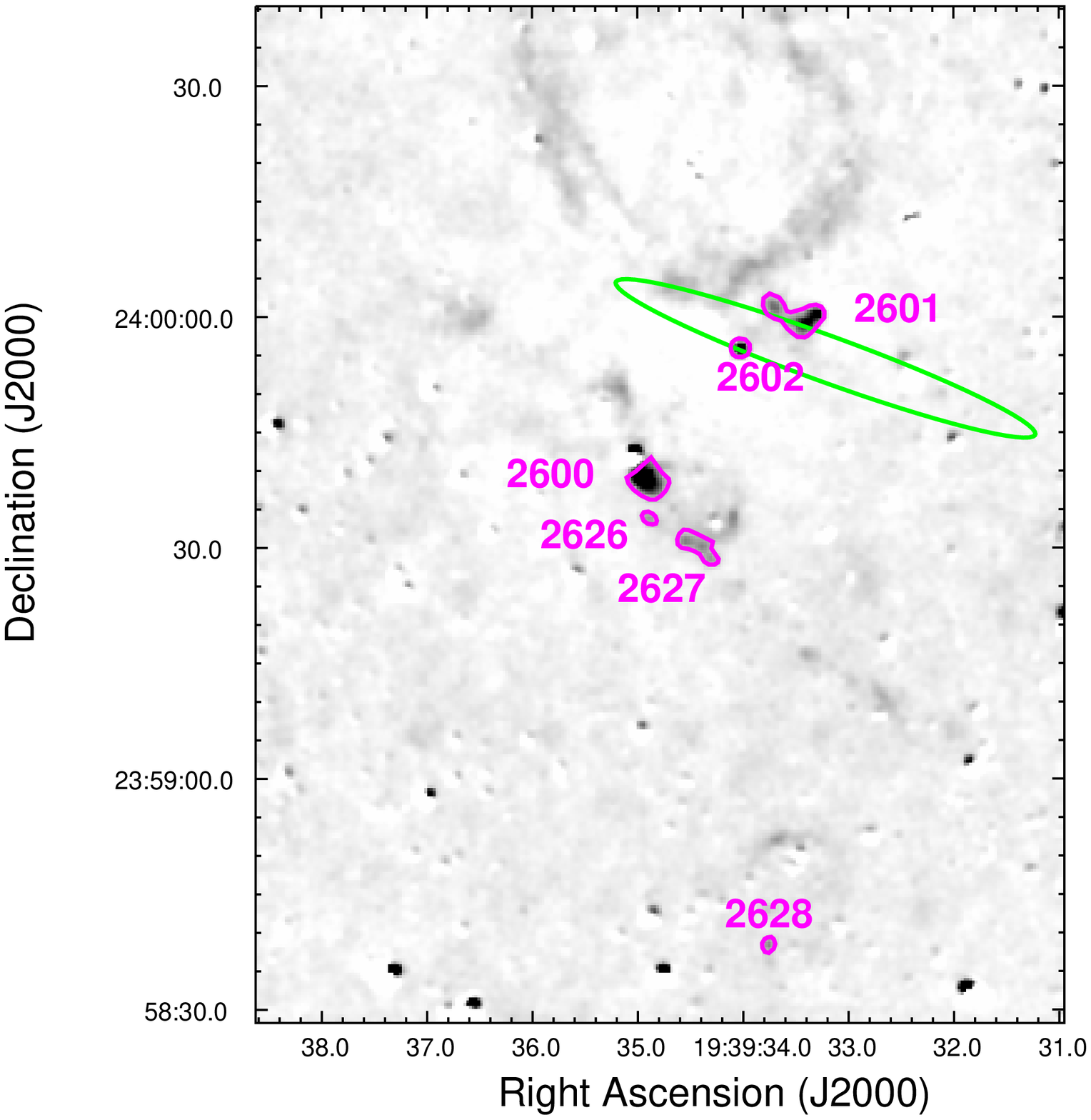}
\caption{Mol 109 (a) rgb image and (b) greyscale image, with labels as described in \S A.}
\end{figure}

\subsection{IRAS 19388+2357 (Mol 110)$^*$}

All three of the MHOs in this region (2613 - 2615) were previously identified by VDR10. The MHOs, RMS YSO G059.8329+00.6729, and two 44-GHz methanol masers (G\'omez-Ruiz et al. 2016) are coincident with Class I candidate J194059.39+240443.9, to within the resolution of {\it WISE} at 22 $\micron$ (Figures 12a \& b). MHOs 2614 \& 2615 have low 2.12-$\micron$/2.25-$\micron$ flux ratios, suggesting these features are dominated by fluorescent emission; however, MHO 2613 has a flux ratio of 5.4, lower than typical values associated with C-type shocked emission, but higher than typical values for either fluorescent or J-type shocked emission.  Although MHO 2613 is  roughly elongated in the direction of the RMS YSO, the complex morphology of the H$_2$ emission in this region makes association with an outflow tenuous. Furthermore, the peak of the high-velocity CO emission mapped by ZHB05, which shows no hint of bipolarity, lies to the south of J194059.39+240443.9, closer to the center of the  Q-type \ion{H}{2} region G059.829+00.671.  The CO emission may trace an outflow driven by a yet-to-be-identified source in this region, or it may trace the dynamics of an expanding \ion{H}{2} region/PDR.

\begin{figure}[htb!]
\figurenum{12}
\includegraphics[angle=0,scale=0.5]{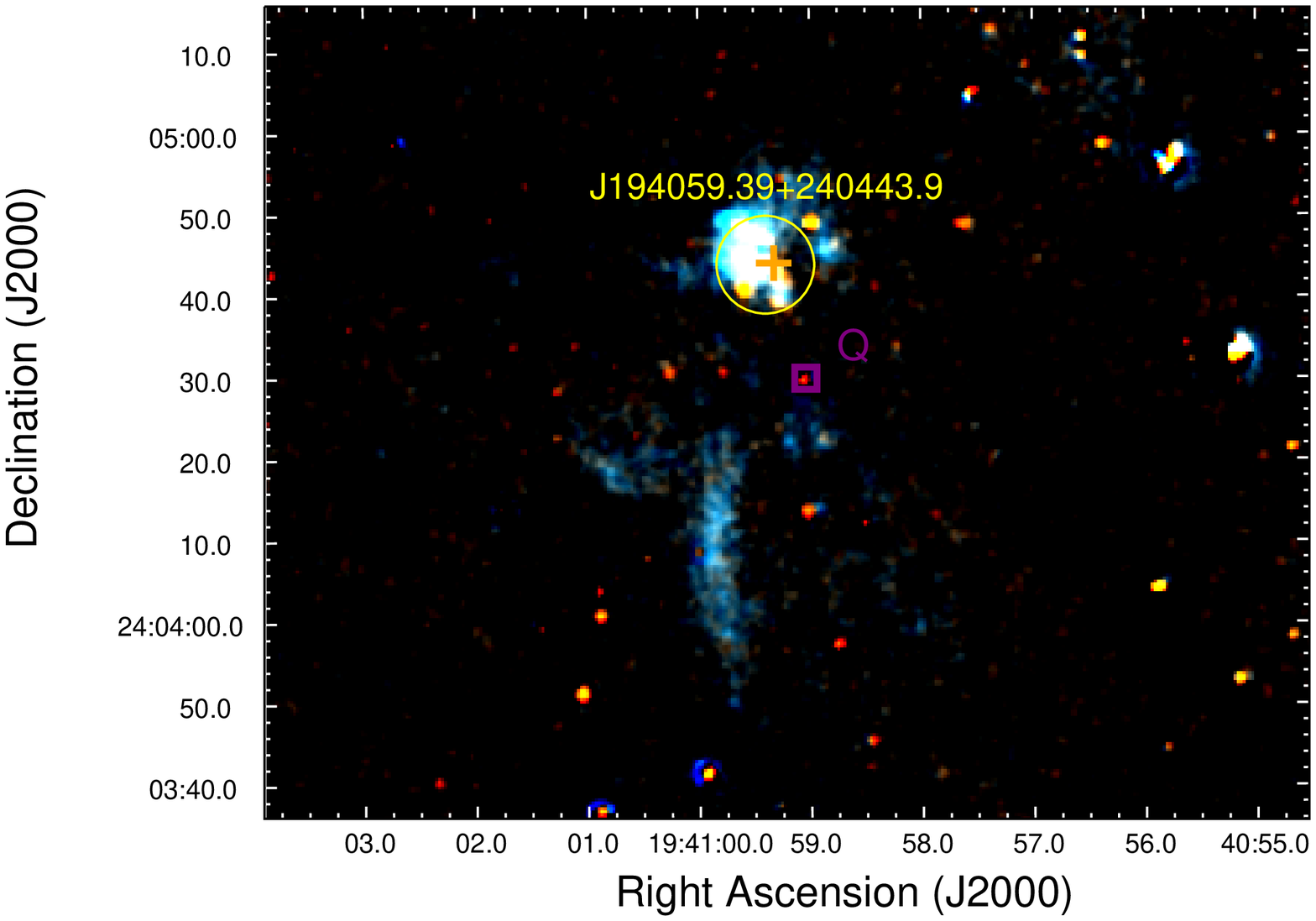}
\includegraphics[angle=0,scale=0.5]{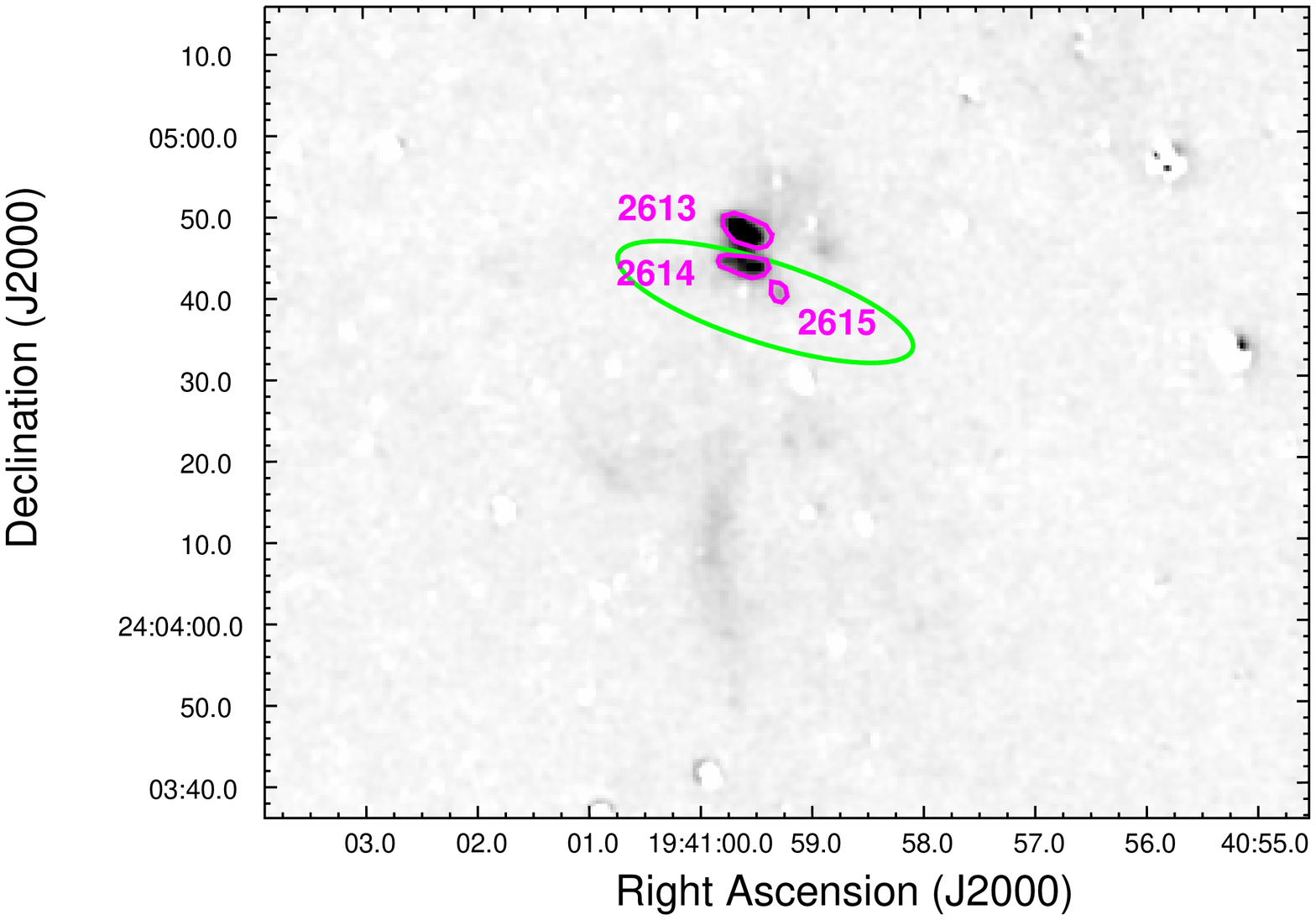}
\caption{Mol 110 (a) rgb image and (b) greyscale image, with labels as described in \S A.}
\end{figure}

\subsection{IRAS 20050+2720 (Mol 114)$^*$}

Bachiller et al. (1995) discovered an extremely high-velocity multipolar CO outflow in this region. The CO outflow map of ZHB05  likewise is oriented roughly E-W and is centered to the north of the {\it IRAS} position. VDR10 identified several MHOs, including a jet-like feature extending to the southeast of their candidate driving sources, `C' and `F'. Our images (Figures 13a \& b) reveal more components to the jet, which appears curved (Mol 114 Flow 1). VDR10 sources `C' and `F' lie close to a 44-GH maser (G\'omez-Ruiz et al. 2016), which is located between MHOs 2608 \& 2608\_6.  Although the jet defined by the MHO 2608 group to the southeast of the candidate driving sources lies along a P.A. of $\sim$ -87$^{\circ}$, the direction of the jet to the northwest, defined by the MHO 2609 group, lies along a P.A. of $\sim$ -72$^{\circ}$.  This curved jet is likely associated with the  CO outflow along a P.A. of $\sim$-77$^{\circ}$, denoted ``A'' by Bachiller et al. (1995: Figure 3). 

Although the bright {\it WISE} Class I candidate, J200706.50+272850.3 (coincident with VDR10 source `A' and RMS YSO G065.7798-02.6121) appears to be the main contributor the the {\it IRAS} flux, this source(s) lies $\sim$ 10$^{\prime\prime}$ to the south of the MHOs that define Flow 1, making it an unlikely driving source candidate. We agree that VDR10 source `C' or `F' is more likely to drive the jet, and we note that the energetic parameters calculated by ZHB05 for the high-velocity CO emission (also located north of the {\it IRAS} position) do not suggest a particularly massive outflow. The bolometric luminosity of this outflow (3.88$\times10^2$ L$_{\odot}$) is the lowest of all the outflows in the ZHB05 study. There are also several isolated groups of MHOs in this region, which cannot be clearly linked to driving sources, and many YSO candidates that may be contributing to different outflows. 

\begin{figure}[htb!]
\figurenum{13}
\includegraphics[angle=0,scale=0.45]{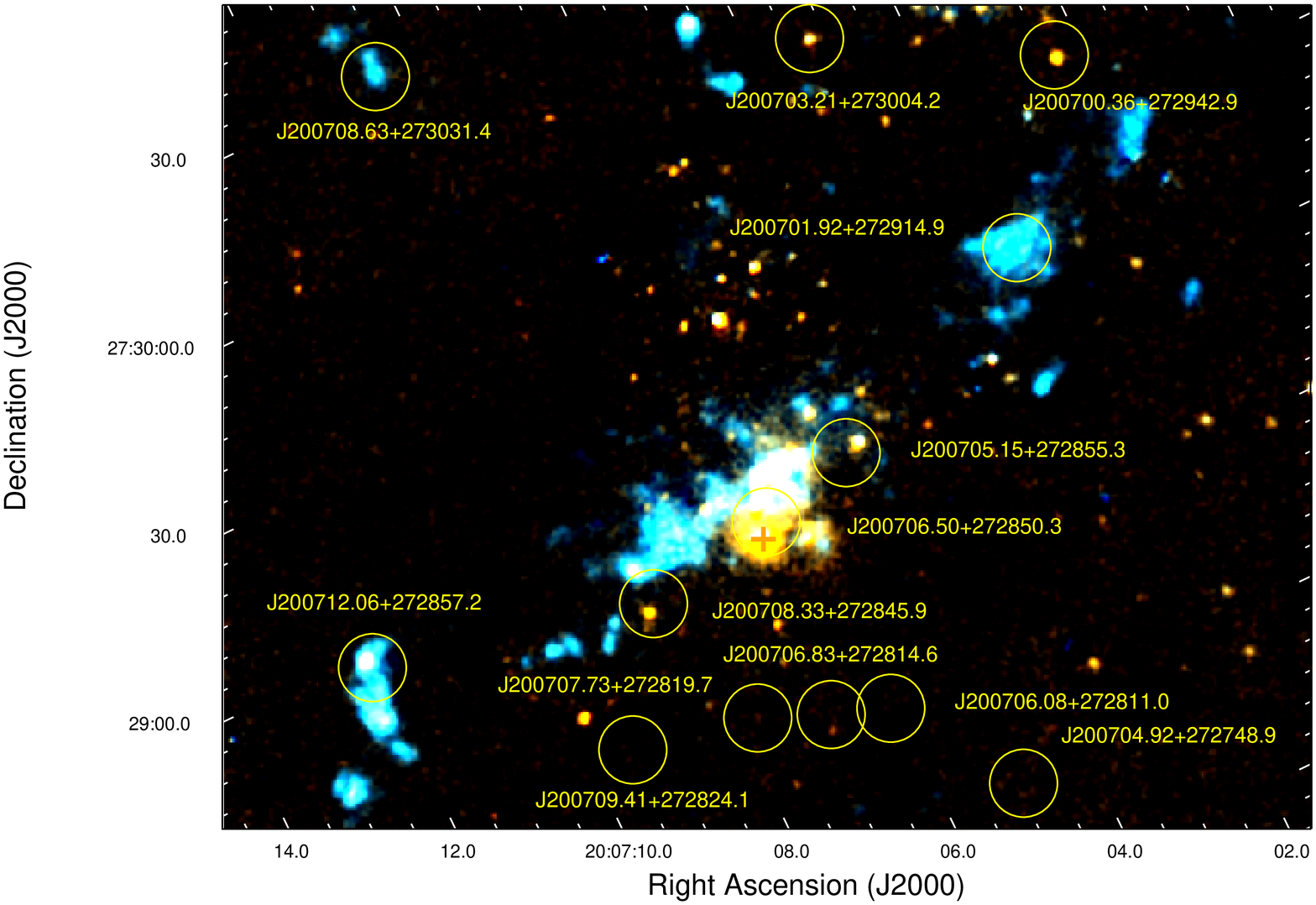}
\includegraphics[angle=0,scale=0.45]{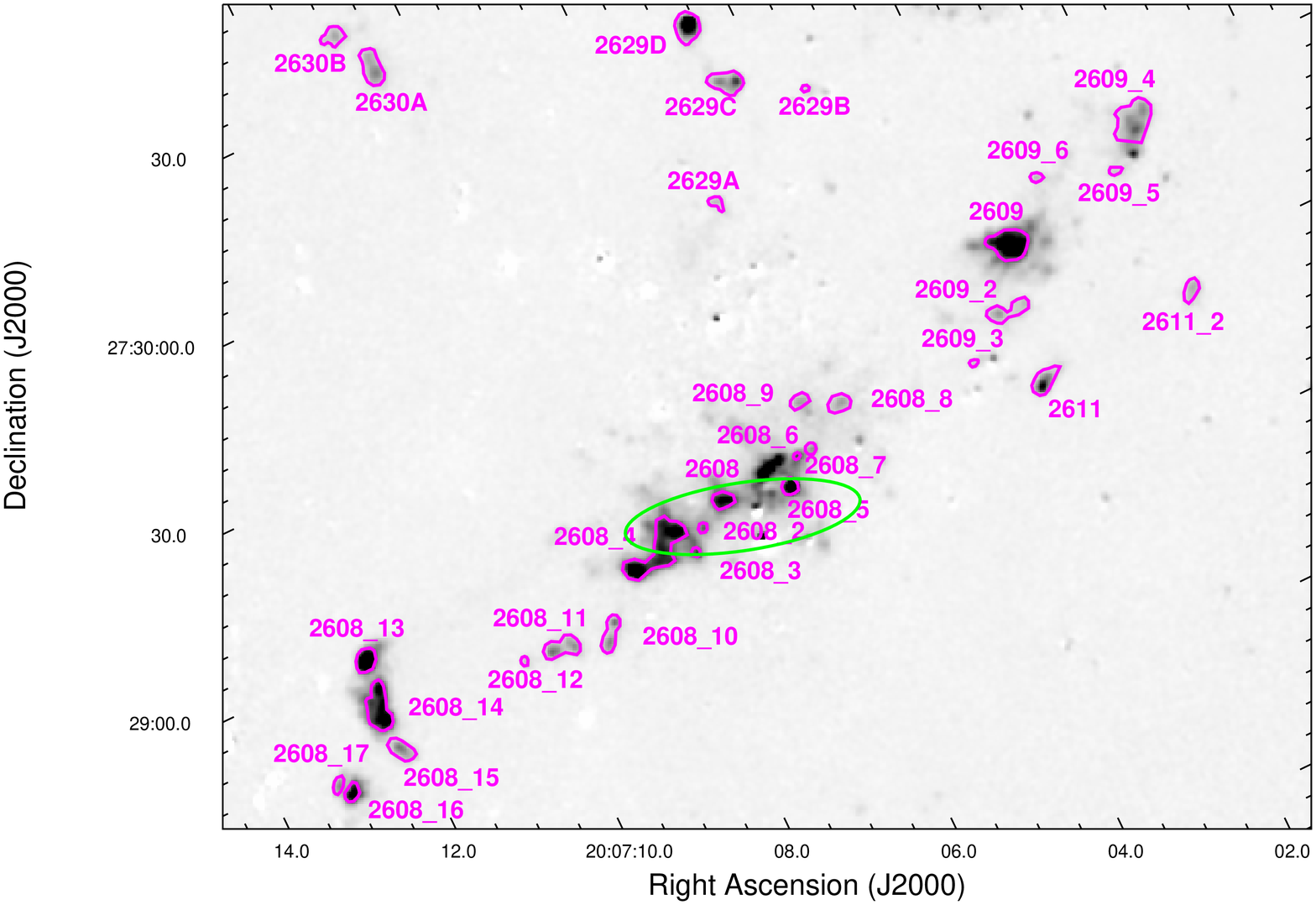}
\caption{Mol 114 (a) rgb image and (b) greyscale image, with labels as described in \S A.}
\end{figure}

\subsection{IRAS 20278+3521 (Mol 125)}

Most of the H$_2$ emission in this region is due to a PDR associated with the Q-type {\it WISE} source, G075.155-02.087, which is positionally coincident with the {\it IRAS} source position (Figures 14a \& b). A linear arrangement of four faint MHOs along a position angle of $\sim$ 68$^{\circ}$ defines Mol 125 Flow 1, but the axis is offset 10$^{\prime\prime}$ to the south of the {\it IRAS} source and no known objects within the mapped field can be identified as potential driving sources. We note that the center of the CO outflow identified by ZHB05 also lies about 10$^{\prime\prime}$ southwest of the {\it IRAS} source; however, the positional discrepancy is well within the $\sim$ 30$^{\prime\prime}$ resolution of the CO observations, and the CO emission shows no evidence of bipolarity, while the MHOs form a linear chain in the plane of the sky.

\begin{figure}[htb!]
\figurenum{14}
\includegraphics[angle=0,scale=0.55]{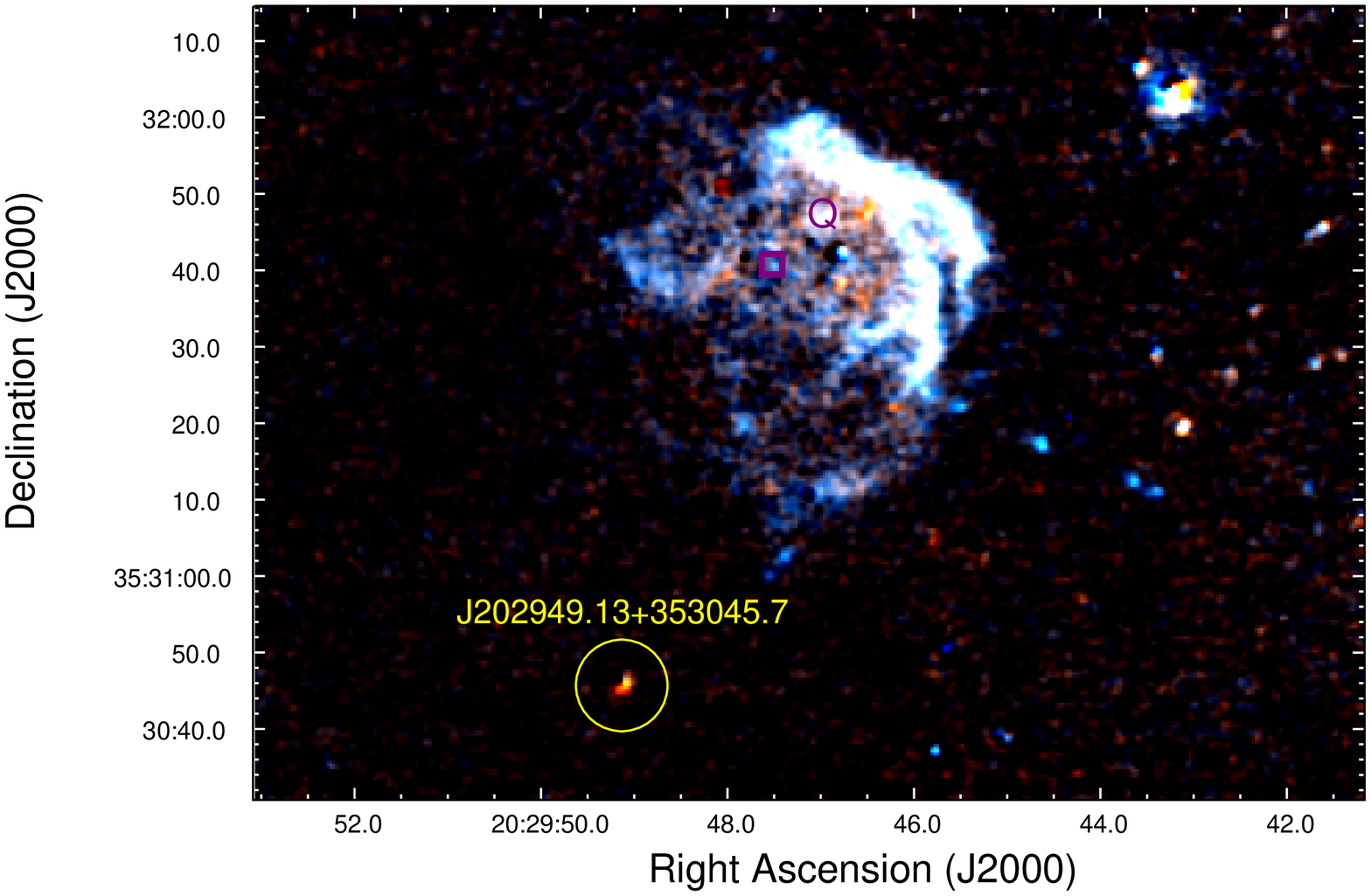}
\includegraphics[angle=0,scale=0.55]{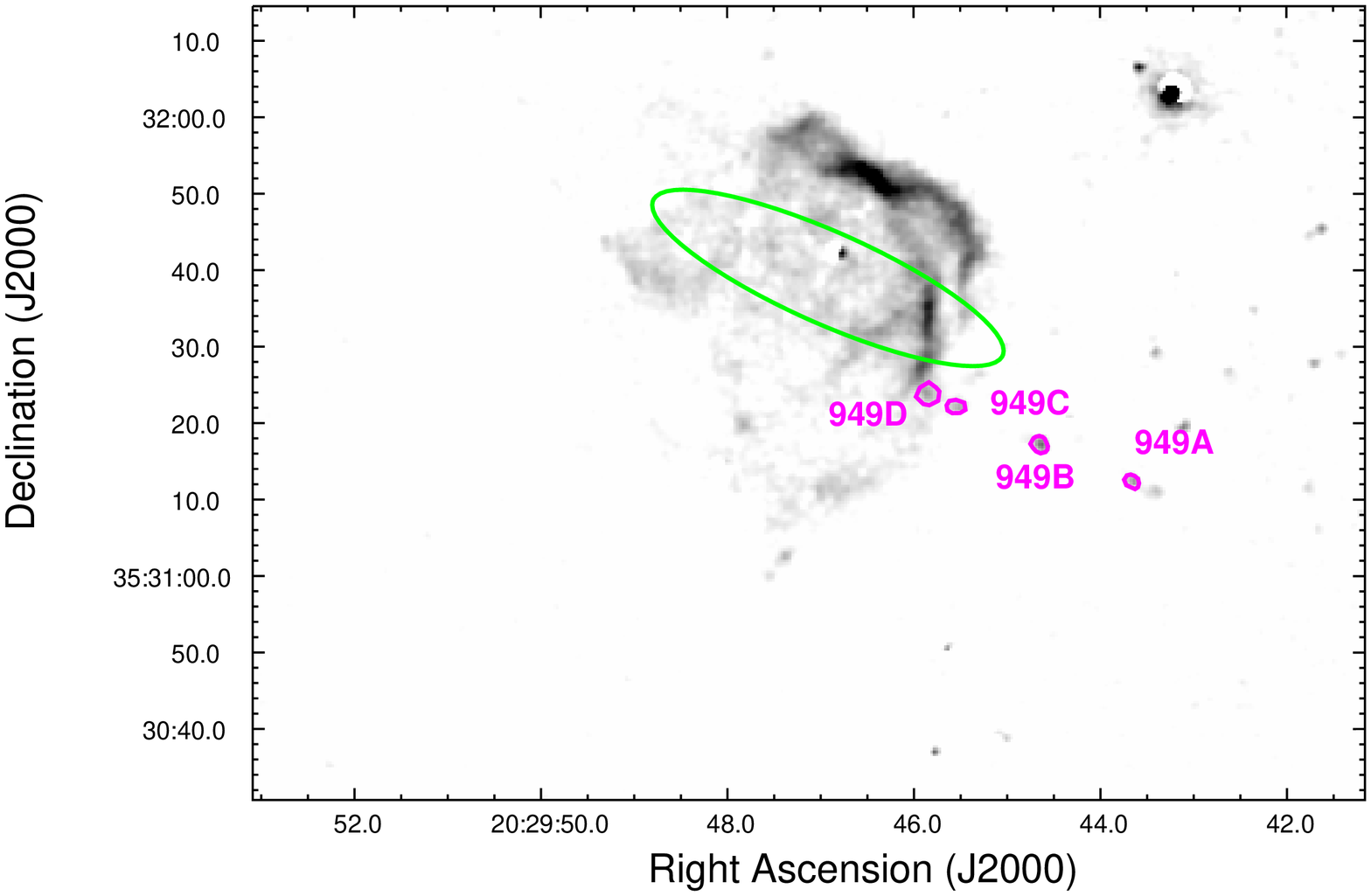}
\caption{Mol 125 (a) rgb image and (b) greyscale image, with labels as described in \S A.}
\end{figure}

\subsection{IRAS 20286+4105 (Mol 126)$^*$}

Most of the H$_2$ emission in this region traces fluorescence whose morphology suggests multiple PDRs (Figures 15a \& b). The CO outflow mapped by ZHB05 is elongated to the west (and slightly north) of the {\it IRAS} source position, in the direction of extended fluorescent emission. VDR10 identified four YSOs in this region, which they labeled A-D. Source `A' coincides with  the position of G079.8749+01.1821, listed as a \ion{H}{2} region in the RMS catalog ; source `C'  coincides with the {\it WISE} Class I candidate, J203029.51+411558.6; and source `D' lies at the center of an arc that defines the southernmost PDR in our images. J203029.51+411558.6, and two 44-GHz masers $\sim$ 10$^{\prime\prime}$ to the southwest (G\'omez-Ruiz et al. 2016), lie along the rim of the PDR associated with G079.8749+01.1821. It is of interest to note that there are three spots of anomalous H$_2$ emission in this region. One of these spots coincides with J203029.51+411558.6 (`C'), which VDR10 noted was embedded in strong nebulosity. J203029.51+411558.6 also lies at the center of the ZHB05 CO outflow.

VDR10 suggested the bright emission feature they identify as `2' (MHO 961) is driven by source `B', which lies directly between G079.8749+01.1821 and MHO 961. In our images, MHO 961 has a curious ``goldfish''-shaped morphology that points back toward J203029.51+411558.6 (`C'), suggesting J203029.51+411558.6 is the likely source of MHO 961 as well as the CO outflow (Mol 126 Flow 1). J203029.51+411558.6 may also be the source of MHO 960, which appears to connect to a curving bridge of emission that points back toward this source; however, this is unclear due to the preponderance of fluorescent emission. MHOs 962 \& 963 cannot be linked clearly to sources in this region.  

\begin{figure}[htb!]
\figurenum{15}
\includegraphics[angle=0,scale=0.55]{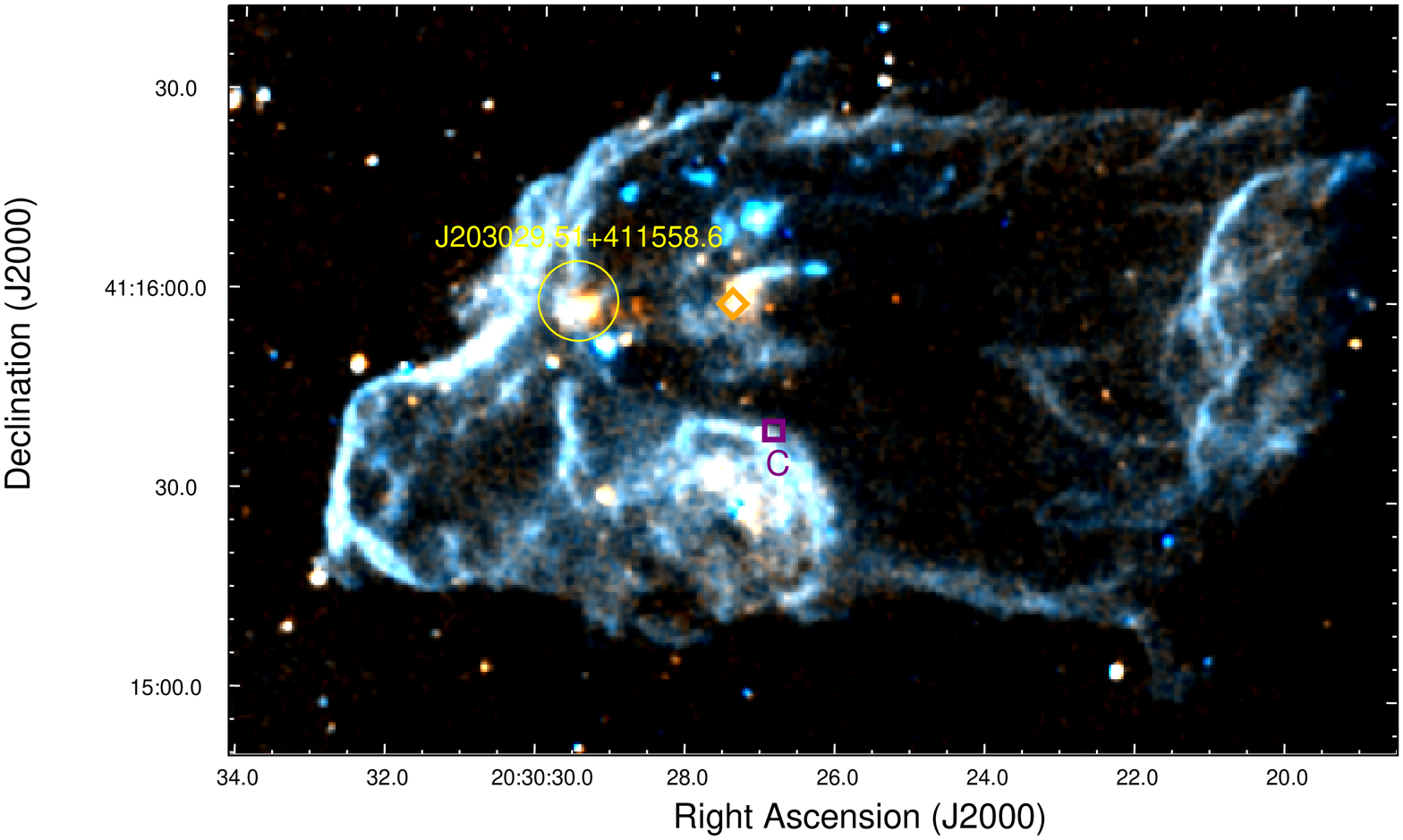}
\includegraphics[angle=0,scale=0.55]{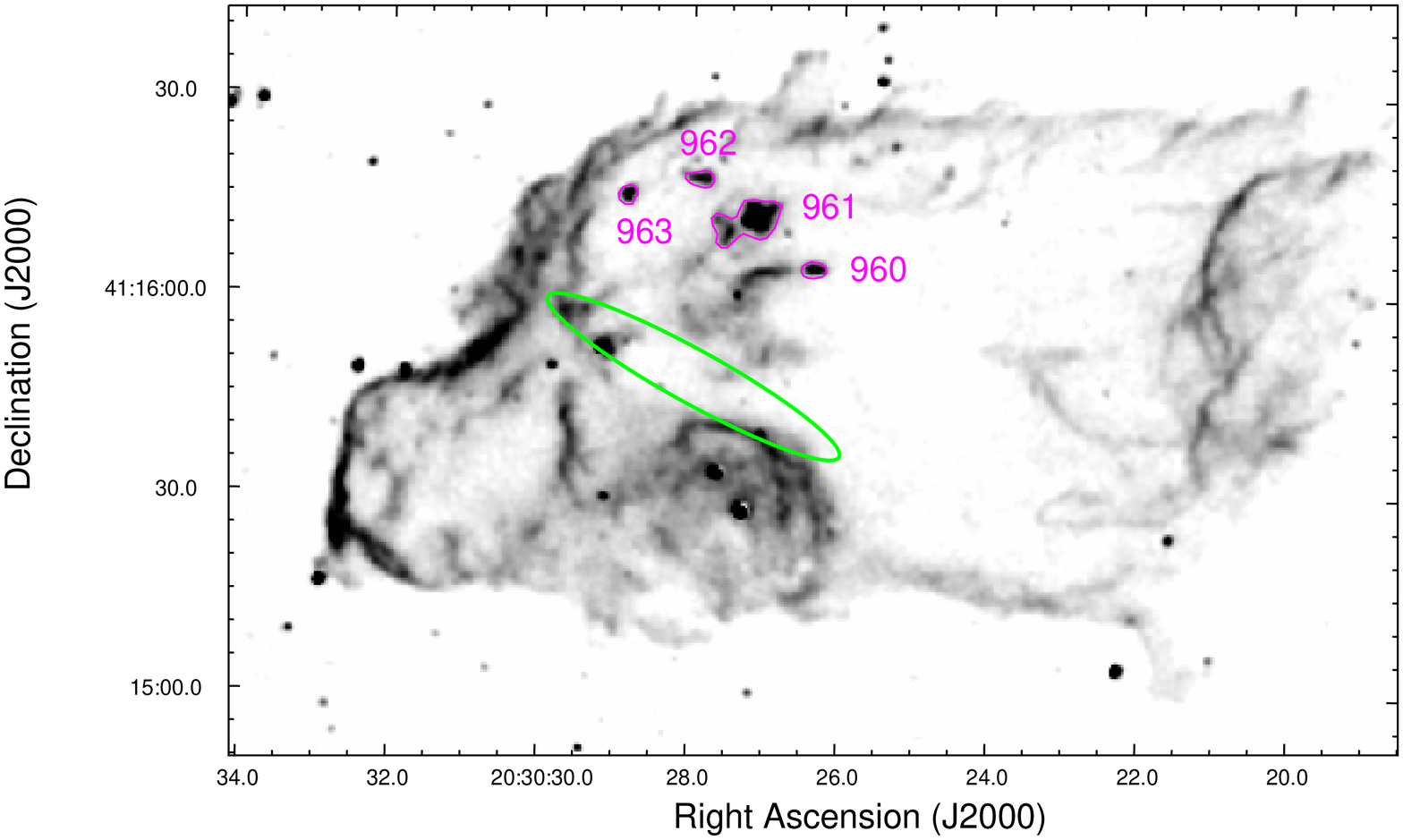}
\caption{Mol 126 (a) rgb image and (b) greyscale image, with labels as described in \S A.}
\end{figure}

\subsection{IRAS 21307+5049 (Mol 136)$^*$}

MHO 882 (detected by VDR10) has a very low 2.12 $\micron$/ 2.25 $\micron$ flux ratio of 1.1, calling into question the nature of this object. VDR10 noted it lies just $\sim$ 3.7$^{\prime\prime}$ of a highly-reddened YSO they identify as `A', which is coincident with RMS YSO G094.2615-00.4116 (Class I candidate J213230.61+510216.1). VDR10 identified `A' with a cometary nebula that opens to the northwest. High-resolution ($\sim$ 6$^{\prime\prime}$) CO observations, obtained by Fontani et al. (2004) at the Owens Valley Radio Observatory (OVRO), suggest the RMS YSO (VDR10 `A') is the source of a compact CO outflow. Our continuum-subtracted H$_2$ 2.12 $\micron$ images indicate a bipolar nebula with an hourglass shape centered roughly 16$^{\prime\prime}$ to the northeast of the YSO, close to the {\it IRAS} position and coincident, within positional accuracy, with the {\it WISE} C-type \ion{H}{2} region, G094.263-00.414 (Figures 16a \& b). The multi-peaked CO outflow mapped by ZHB05 is similar in direction (NW-SE) and extent to the bipolar nebula, which suggests the fluorescent H$_2$ emission may trace the periphery of a bipolar cavity produced by the CO outflow.

\begin{figure}[htb!]
\figurenum{16}
\includegraphics[angle=0,scale=0.52]{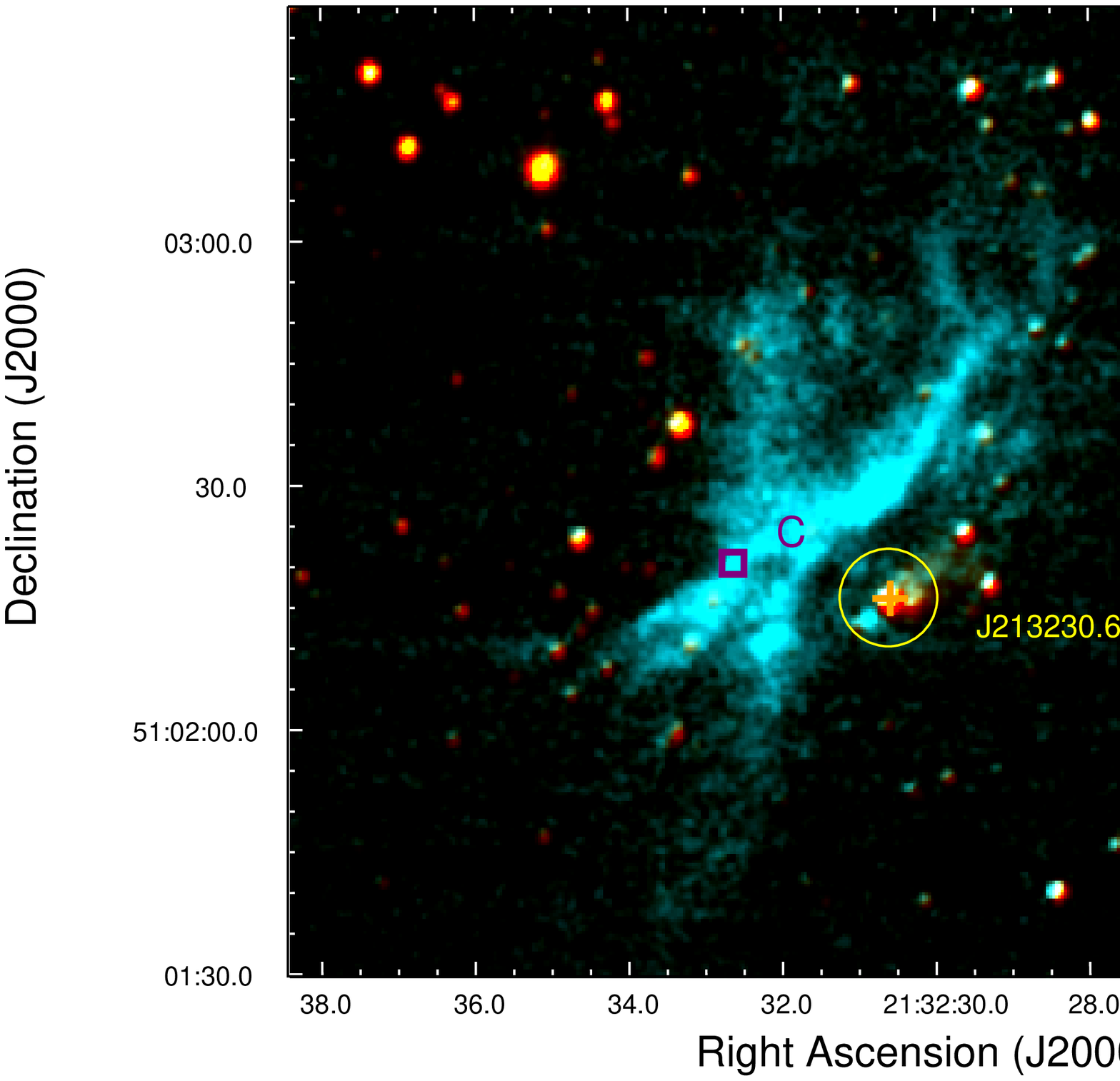}
\includegraphics[angle=0,scale=0.52]{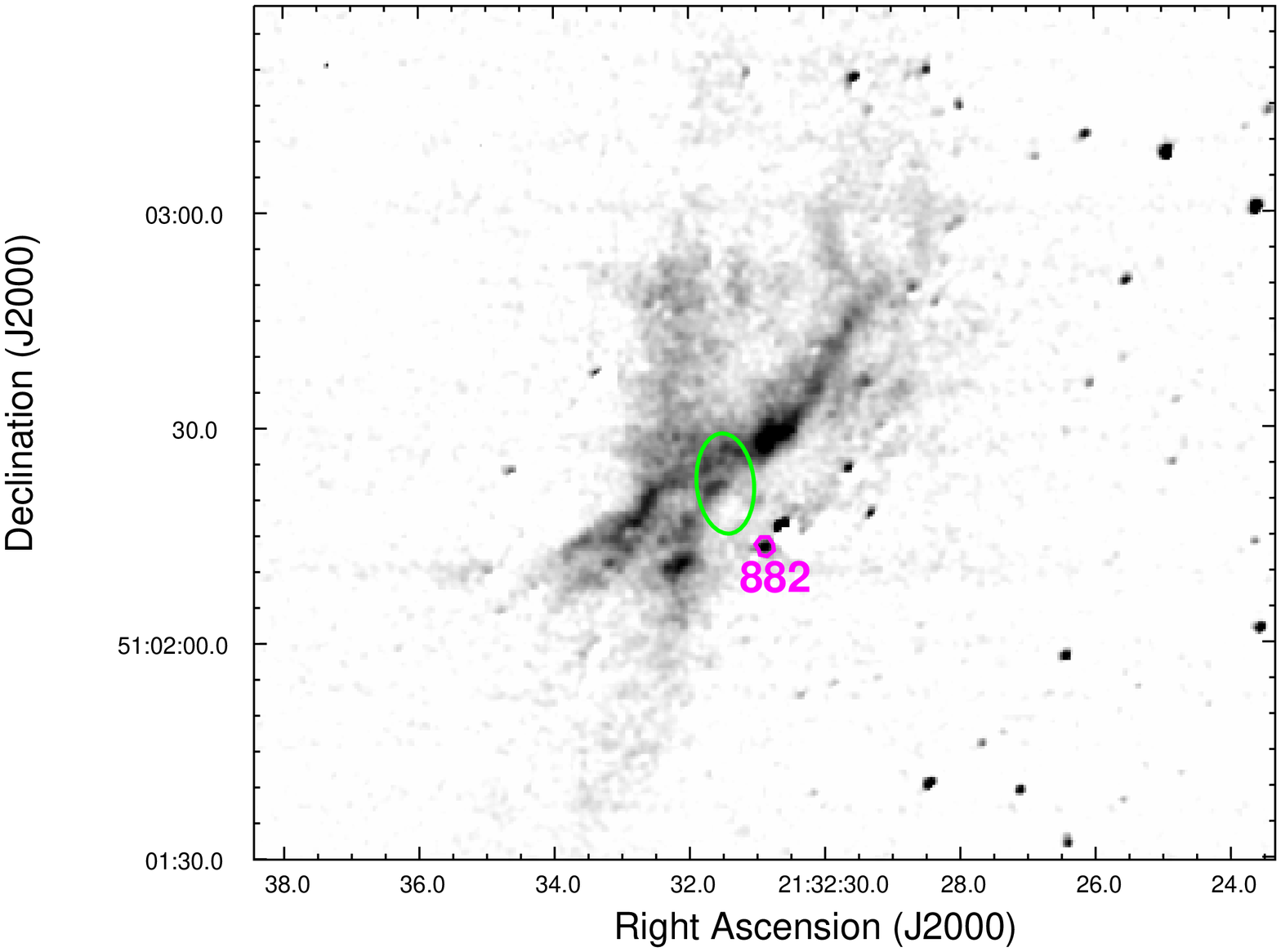}
\caption{Mol 136 (a) rgb image and (b) greyscale image, with labels as described in \S A.}
\end{figure}

\subsection{IRAS 22172+5549 (Mol 143)$^*$}

VDR10 noted that this is a known \ion{H}{2} region; however, Molinari et al. (1998) did not find radio emission associated with the {\it IRAS} source and it does not appear in the {\it WISE} Catalog of Galactic \ion{H}{2} regions (Anderson et al. 2014). Molinari et al. (2002) suggested that the O star HD211883, situated $\sim$ 2$^{\prime}$ north of a complex ridge of 3.6 cm emission, is the source of the ionization front.  Our images (Figures 17a \& b) indicate a PDR with `pillar' morphology, suggesting the source of the \ion{H}{2} region lies to the northwest of our imaged region, consistent with this interpretation. Three RMS YSOs (G102.8051-00.7184A, B, \& C) are located near the apex of the pillar, roughly 20$^{\prime\prime}$ north of the {\it IRAS} position. MTR02 identified a dense core, via 1.3 and 3 mm continuum observations, that peaks close to the position of the southernmost RMS YSO, G102.8051-00.7184B. 

ZHB05 detected multi-lobed CO emission in this region, suggesting more than one outflow. This is also evident in the $\sim$ 29$^{\prime\prime}$ CO map shown in Figure 14a of Fontani et al. (2004), whose higher-resolution ($\sim$ 6$^{\prime\prime}$)  CO map (Figure 14c), obtained at OVRO, indicates a very compact ($\sim$ 20$^{\prime\prime}$) north-south outflow centered on the position of RMS YSO G102.8051-00.7184A ($\sim$ 5$^{\prime\prime}$ northeast of G102.8051-00.7184B). G102.8051-00.7184A \& B are coincident with {\it WISE} Class I candidate, J221909.42+560501.2.

We identify three distinct jets along position angles $\sim$ 10, 40, \& -52$^{\circ}$ (Mol 143 Flow 1, 2, \& 3, respectively), which intersect at J221909.42+560501.2. G102.8051-00.7184A is the best candidate driving source for the jet along P.A. $\sim$ 10$^{\circ}$, which we identify with the compact N-S outflow of Fontani et al. (2004). G102.8051-00.7184A may also drive a jet along P.A. $\sim$ 40$^{\circ}$, but this is more difficult to distinguish due to the orientation of A \& B. G102.8051-00.7184B likely drives the jet along P.A. $\sim$ 128$^{\circ}$. We note that the the high-velocity CO emission contours seen in the single-antenna maps of ZHB05 and Fontani et al. (2004) are consistent with outflows along position angles of 40 \& -52$^{\circ}$.

\begin{figure}[htb!]
\figurenum{17}
\includegraphics[angle=0,scale=0.48]{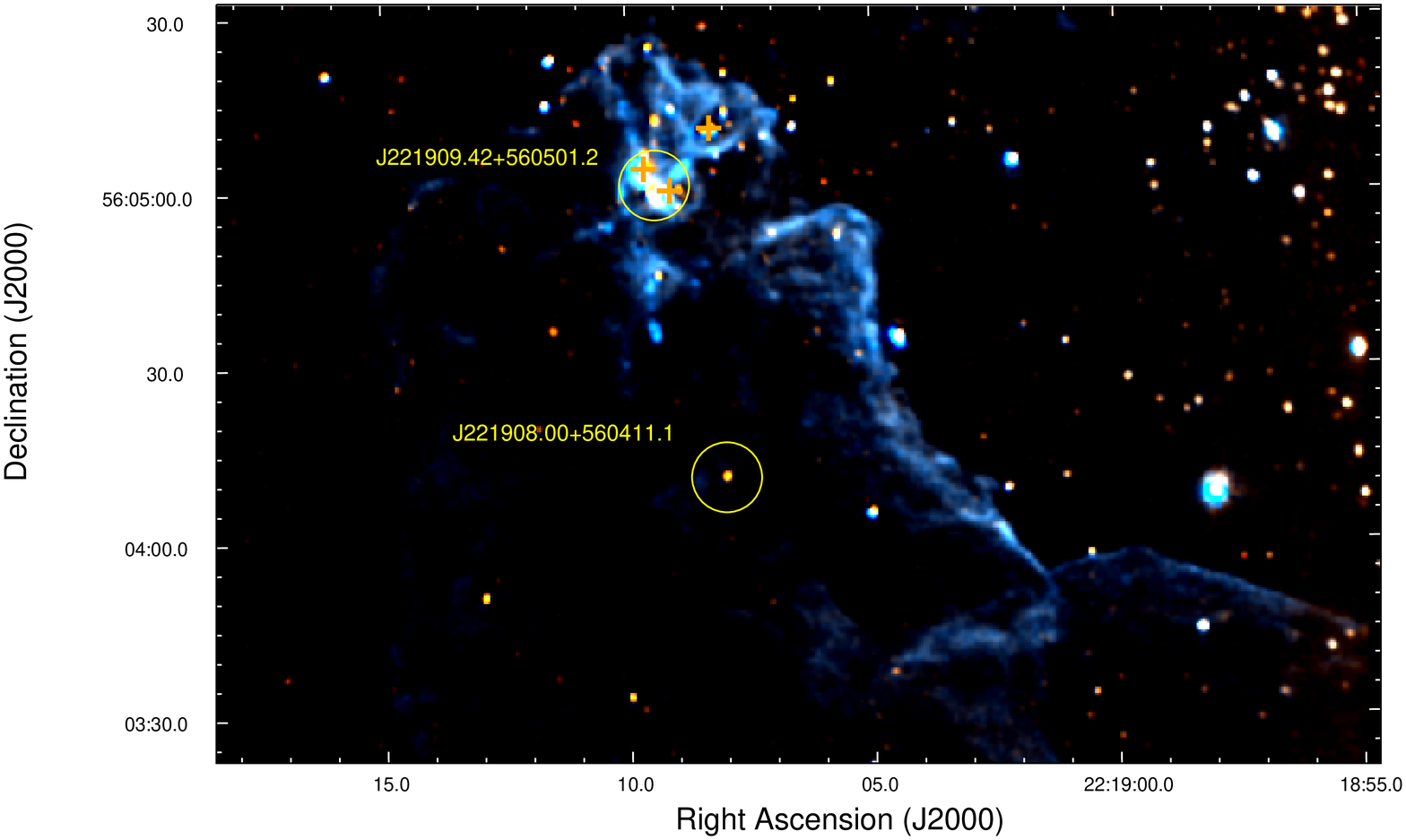}
\includegraphics[angle=0,scale=0.48]{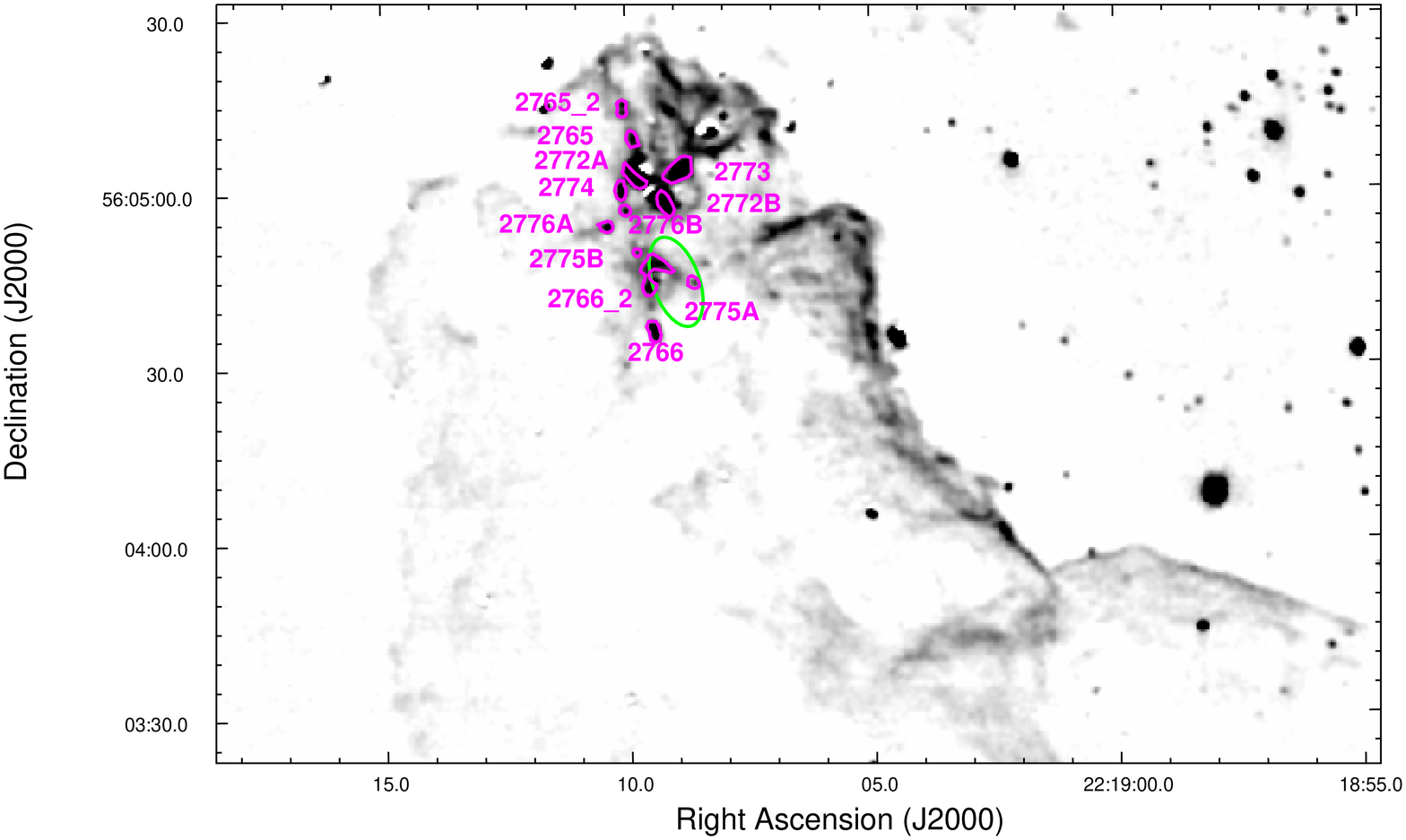}
\caption{Mol 143 (a) rgb image and (b) greyscale image, with labels as described in \S A.}
\end{figure}

\subsection{IRAS 22305+5803 (Mol 148)$^*$}

We identified eight Class I candidates in this region. VDR10 found three clumpy emission features directly north of the {\it IRAS} source position, which coincides with the object we identify as RMS YSO G105.5072+00.2294 and {\it WISE} Class I candidate J223223.87+581859.7 (Figures 18a \& b). The RMS YSO also coincides with the source VDR10 identify as `A'. The emission features are prominent in Figure 18b; however, they appear to be contaminated by fluorescent emission likely produced by a PDR associated with the Q-type object, G105.509+00.230. The objects VDR10 identify as `B' and `C' lie directly north of J223223.87+581859.7, towards the nebulous `white' emission in Figure 18a. The MHOs we discovered in this region lie well outside of the image presented in Figure A48 of VDR10. MHOs 2777D-G lie along a NE-SW P.A. of $\sim$ 22$^{\circ}$  from the RMS YSO, similar in direction to the CO outflow mapped by ZHB05 (Mol 148 Flow 1). MHOs 2777A-C lie along a P.A. of $\sim$ 73$^{\circ}$, centered on {\it WISE} Class I candidate J223219.41+581750.7 (Mol 148 Flow 2). 

\begin{figure}[htb!]
\figurenum{18}
\includegraphics[angle=0,scale=0.35]{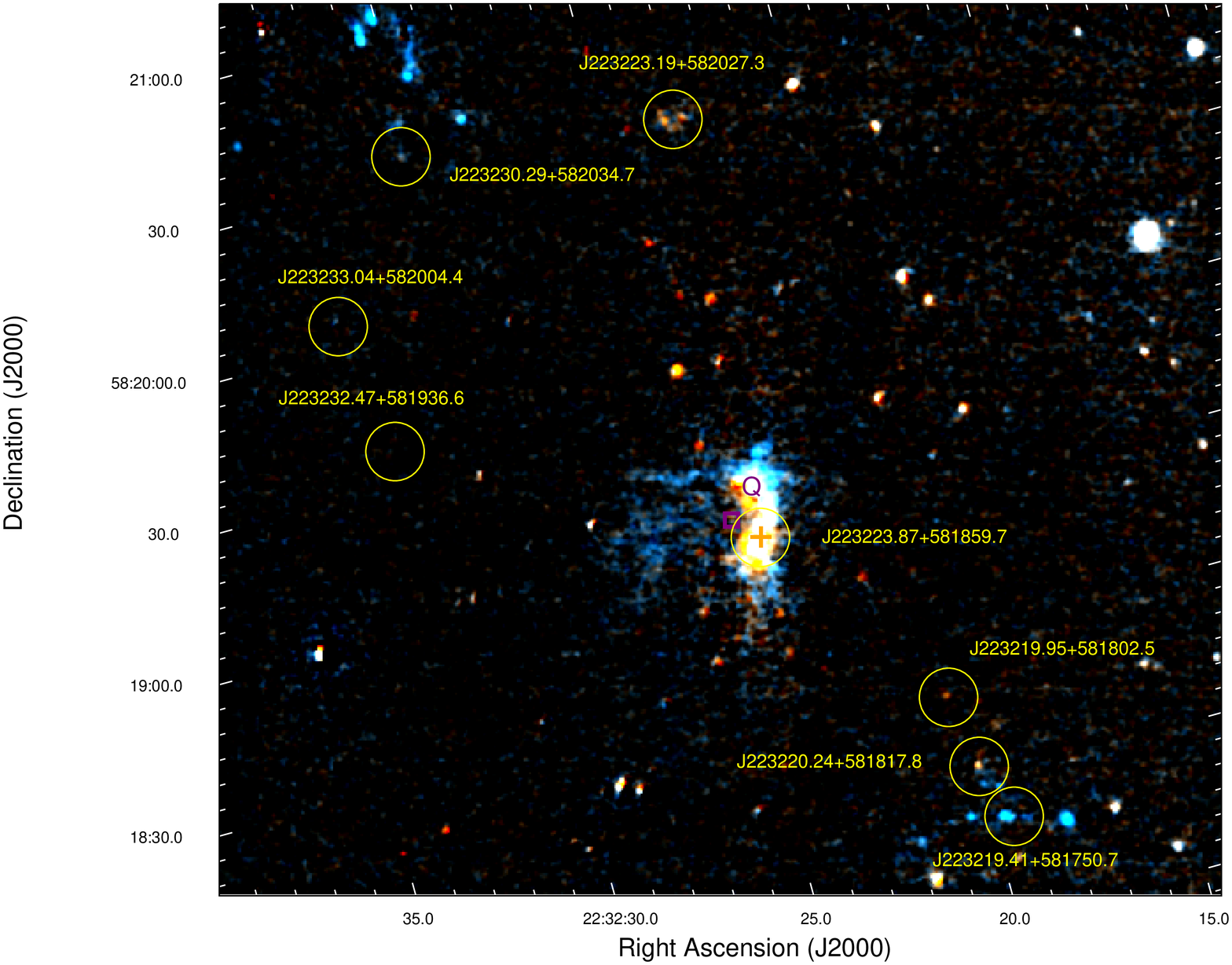}
\includegraphics[angle=0,scale=0.35]{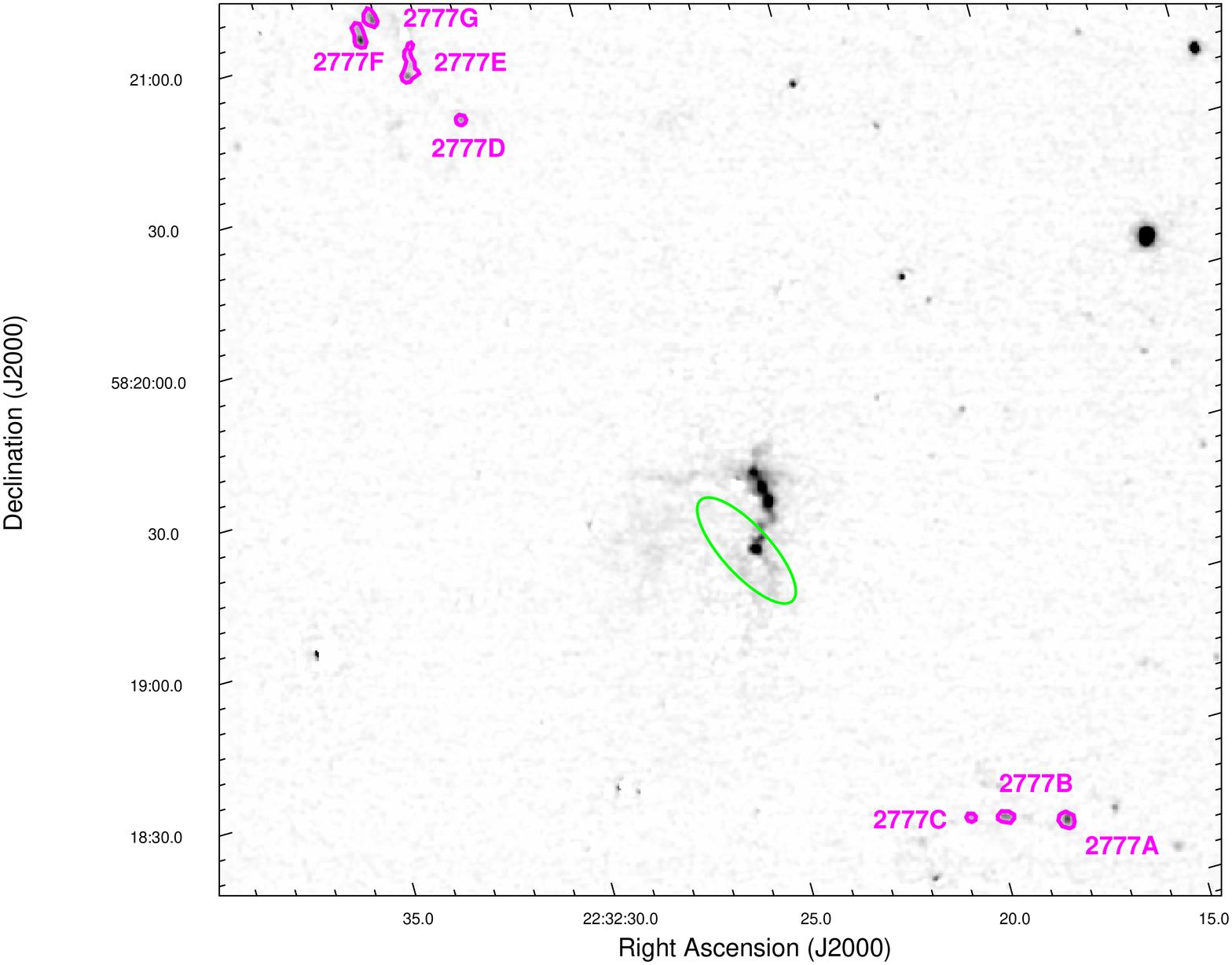}
\caption{Mol 148 (a) rgb image and (b) greyscale image, with labels as described in \S A.}
\end{figure}

\begin{figure}[htb!]
\figurenum{19}
\includegraphics[angle=0,scale=0.65]{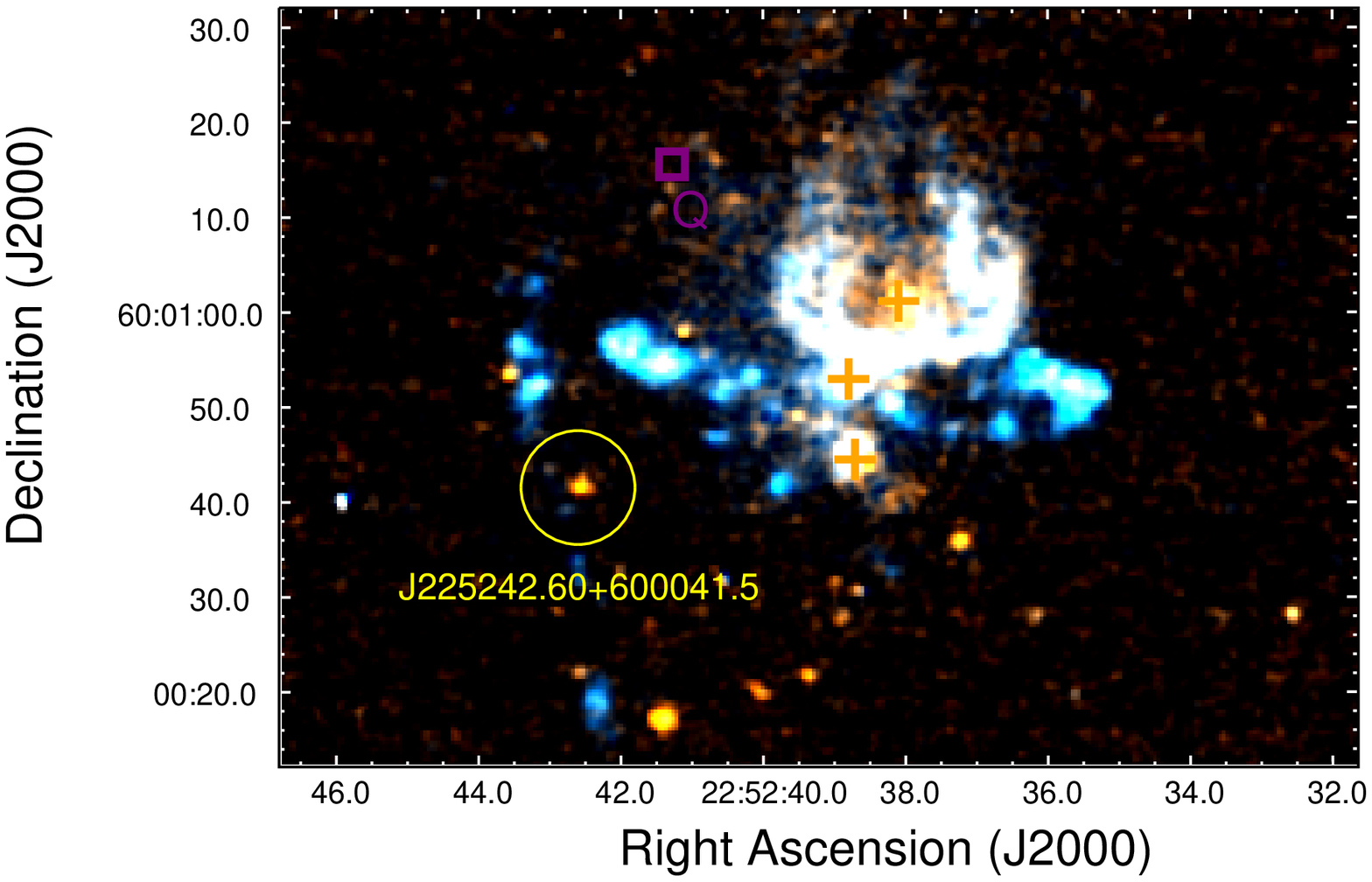}
\includegraphics[angle=0,scale=0.65]{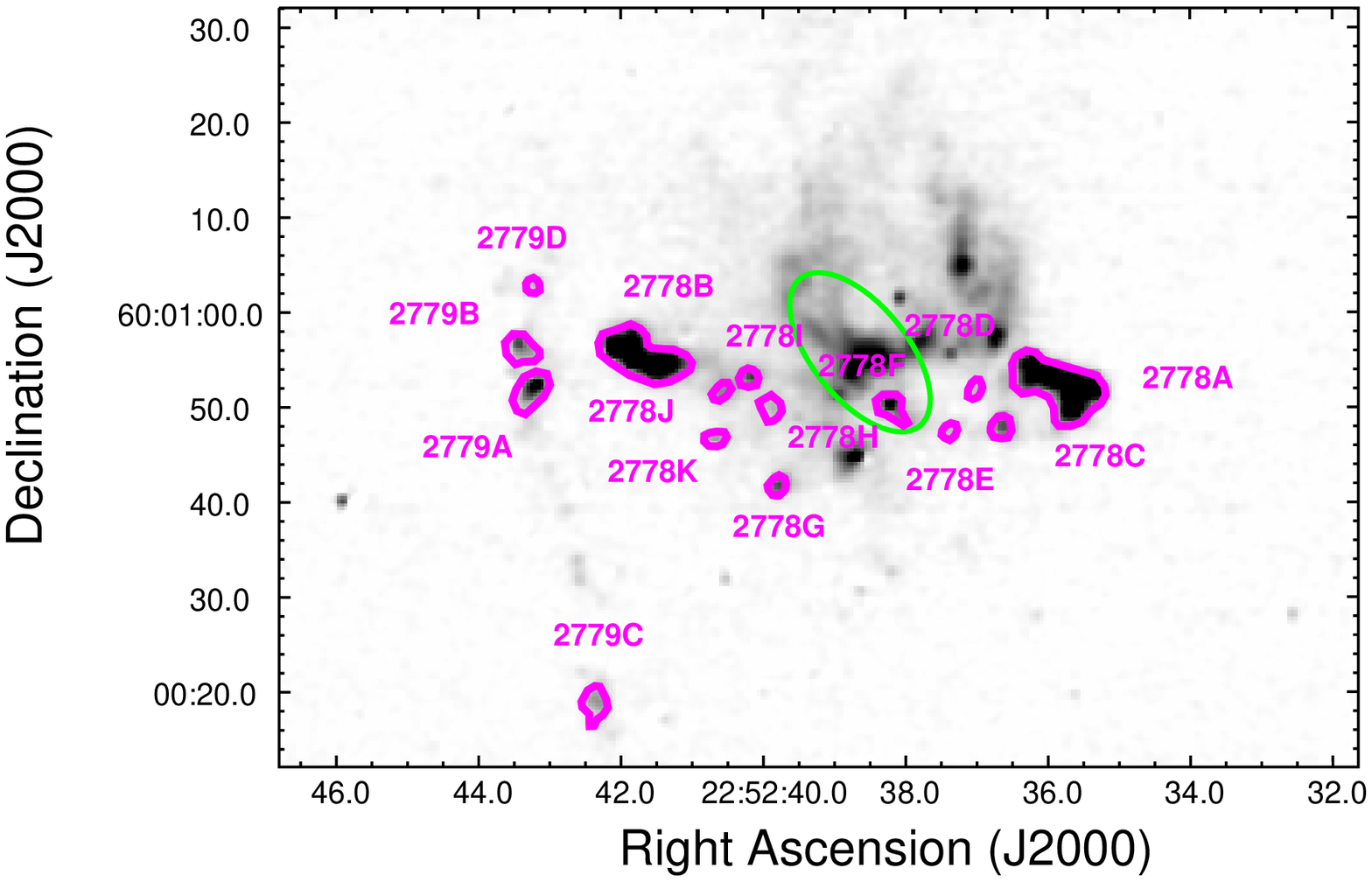}
\caption{Mol 151 (a) rgb image and (b) greyscale image, with labels as described in \S A.}
\end{figure}

\begin{figure}[htb!]
\figurenum{20}
\includegraphics[angle=0,scale=0.55]{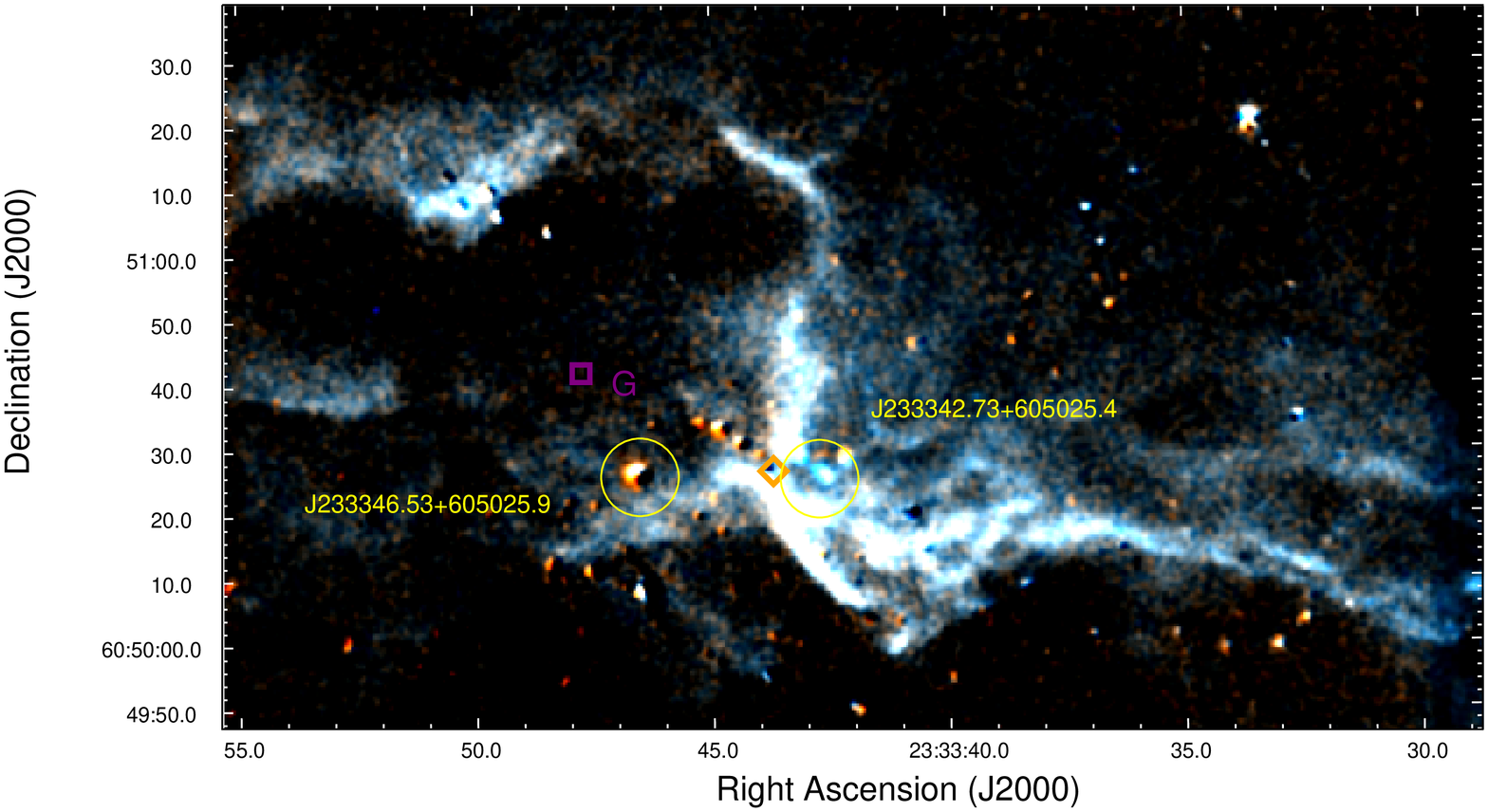}
\includegraphics[angle=0,scale=0.55]{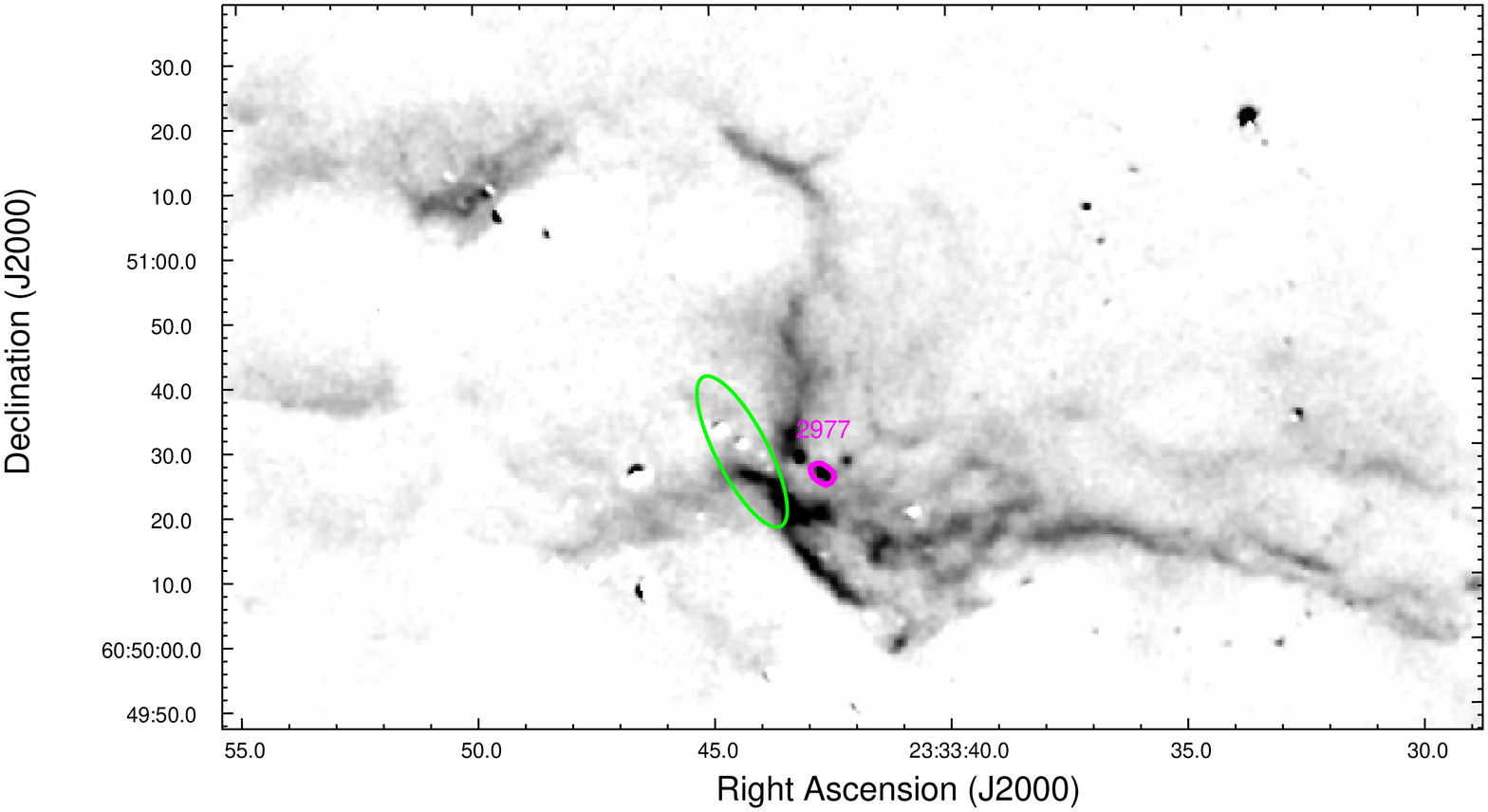}
\caption{Mol 158 (a) rgb image and (b) greyscale image, with labels as described in \S A.}
\end{figure}

\subsection{IRAS 22506+5944 (Mol 151)}

In this region, we identify a spectacular, compact jet that culminates in the bow-shaped MHOs 2278 A \& B (Figures 19a \& b). The jet is centered on RMS YSO, G108.5955+00.4935B (Mol 151 Flow 1). Another RMS YSO (G108.5955+00.4935A) is centered on the compact PDR just north of G108.5955+00.4935B and the jet. The location of G108.5955+00.4935B and its outflow on the periphery of the PDR is suggestive of triggered star formation, warranting further observations exploring the kinematics of this region. G\'omez-Ruiz et al. (2016) discovered six 44-GHz masers in this region, all of which lie south of the jet axis. One of these masers is coincident with MHO 2778G. The proximity of this MHO to another RMS YSO (G108.5955+00.4935C) is suggestive of a possible outflow associated with this source. The curved chain of MHOs 2779A-D is roughly centered on  {\it WISE} Class I candidate, J225242.60+600041.5, the likely driving source (Mol 151 Flow 2). We note that the extent of the outflows in this region are comparable in size to the resolution of the CO outflow map of ZHB05, precluding any meaningful comparison of morphology.

\subsection{IRAS 23314+6033 (Mol 158)}

The CO outflow mapped by ZHB05 in this region shows no hint of bipolarity. Both lobes peak at the western edge of the {\it IRAS} error ellipse, near the position of RMS \ion{H}{2} region G113.6041-00.6161 and {\it WISE} Class I candidate, J233342.73+605025.4 (Figures 20a \& b). These sources lie at the intersection of ridges of filamentary fluorescent emission. G113.6041-00.6161 is listed as a G-type (grouped) \ion{H}{2} region in the {\it WISE} Catalog of Galactic \ion{H}{2} Regions (Anderson et al. 2014).  We identify a single MHO (MHO 2977) toward J233342.73+605025.4, which would be consistent with, though hardly conclusive of, an outflow oriented along our line of sight.

\end{document}